\documentclass[prd,aps,letterpaper,floatfix,superscriptaddress,preprintnumbers,11pt,nofootinbib]{revtex4-1}
\usepackage{array}[=2016-10-06]
\usepackage{epsfig,epsf}
\usepackage{booktabs}
\usepackage{mathrsfs}
\usepackage{graphicx}
\usepackage{caption}
\usepackage{fontawesome5} 
\usepackage{pifont}
\usepackage{subcaption}
\usepackage{latexsym}
\usepackage{amsmath}
\usepackage{amssymb}
\usepackage{textcomp}
\usepackage{float}
\usepackage{cancel}
\usepackage{amsbsy}
\usepackage{tabularx}
\usepackage{placeins}
\usepackage{slashed}
\usepackage[numbered]{bookmark}
\usepackage[dvipsnames]{xcolor}

\usepackage{amsmath,amsfonts,amssymb}
\usepackage{bbold}
\usepackage{hyperref}
\DeclareGraphicsExtensions{.png,.eps,.pdf}
\graphicspath{{figures/}}
\usepackage{color}
\usepackage[normalem]{ulem}

\definecolor{darkgreen}{cmyk}{1,0,1,0.4}

\long\def\/*#1*/{}

\usepackage[utf8]{inputenc}
\usepackage{lineno}

\usepackage{multirow}
\usepackage{amsthm}
\usepackage{tikz}
\usetikzlibrary{arrows.meta,decorations.pathmorphing,decorations.markings}

\hypersetup{hidelinks,pdftitle={Can randomness lead to non-anarchical neutrino mixing?},pdfauthor={Aadarsh Singh and Sudhir K. Vempati}}
\begin{document}
\raggedbottom

\title{Can randomness lead to non-anarchical neutrino mixing?}
\author{Aadarsh Singh }
\affiliation{Centre for High Energy Physics, Indian Institute of Science, Bangalore 560012, India}
\author{Sudhir K. Vempati}
\affiliation{Centre for High Energy Physics, Indian Institute of Science, Bangalore 560012, India}
\date{September 8, 2026}

\begin{abstract}
We study how disorder and graph connectivity affect neutrino masses and mixing in fermionic theory spaces. Extending the construction of Craig and Sutherland, we compare a nearest-neighbour chain, a non-local graph with distance-suppressed links, and a sparse Petersen-type graph, for Dirac and Majorana neutrinos. Strong site disorder suppresses boundary propagation through localisation. When the heavy mass matrix instead has a dominant approximately uniform diagonal, light masses can arise from cancellations in its inverse. A path expansion makes the dependence on graph connectivity explicit. Our benchmark ensembles display broad mixing distributions for some flavour couplings and peaked distributions for others, particularly with weak site disorder and structured flavour input. These observations do not establish a Haar distribution at strong disorder or a statistical preference over neutrino anarchy at weak disorder. The results identify conditions under which disorder can preserve structured mixing, while leaving its quantitative explanatory advantage to a controlled ensemble comparison.

\end{abstract}

\maketitle

\section{Introduction}

Neutrino masses and lepton mixing pose distinct questions: why are neutrinos light, and why do their mixing angles differ so markedly from the quark angles? The measured lepton angles include two large angles and a smaller reactor angle, $\theta_{13}\simeq8.5^\circ$~\cite{esteban2025nufit}. Discrete and modular flavour symmetries provide one approach to these patterns~\cite{Altarelli:2010gt,King:2014nza,Feruglio:2019ybq}. Neutrino anarchy instead treats the mixing matrix statistically~\cite{hall2000neutrino,Haba:2000be}. These approaches motivate a separate question: what can a disordered theory space explain once its graph and flavour couplings have been specified?

Craig and Sutherland~\cite{craig2018exponential} showed that disorder in a chain of fermion masses can produce exponentially suppressed boundary couplings. The mechanism is related to Anderson localisation~\cite{PhysRev.109.1492,lagendijk2009fifty}: for a one-dimensional nearest-neighbour model with independent, nontrivial on-site disorder, eigenstates are localised in the infinite-system limit. A finite chain can appear extended when the localisation length exceeds its size. We therefore distinguish site disorder from hopping disorder and examine the finite geometries explicitly.

Fermionic clockwork also produces small effective couplings~\cite{giudice2017clockwork}. Its rectangular mass matrix has an unpaired chiral zero mode with a nonuniform profile. A discrete $\mathbb Z_N$ site-shift symmetry is not required. The square theory-space matrices used here have no generically protected zero mode. Their hierarchy is instead controlled by boundary propagation through the massive tower.

We compare a nearest-neighbour chain, a distance-suppressed non-local graph, and a sparse Petersen-type graph. Throughout, $N$ denotes the total number of sites in a displayed mass matrix and in the numerical captions, whereas $GP(n,k)$ has $2n$ vertices. The scan topology has $n=N/2$ and $k=n/2$ (hence $N$ divisible by four). Its inner links form a matching. This graph has degrees two and three and is not the standard Petersen graph $GP(5,2)$ with automorphism group $S_5$. Moreover, distance-dependent edge weights need not preserve the automorphisms of the unweighted graph.

The benchmarks distinguish two mechanisms. Strong site disorder can suppress endpoint overlaps by localisation. For an approximately uniform diagonal much larger than the hopping matrix, a convergent inverse-matrix expansion instead suppresses propagation between separated boundary sites. The mixing distributions depend additionally on where flavour violation is introduced. Strong-site benchmarks give broad distributions in the Hamiltonian
channel, while boundary flavour mixing produces endpoint
enhancement or, in some Dirac benchmarks, an interior peak.
Weak site disorder can preserve peaked distributions set by
the chosen flavour matrices. Neither observation alone establishes an improvement over anarchy.

Section~\ref{sec:recap} defines the geometries and disorder regimes. Section~\ref{sec:neutrino} develops the mass mechanism and flavour embedding. Section~\ref{sec:robust} discusses radiative corrections, the comparison with anarchy and constraints on active--heavy mixing. Sections~\ref{sec:strong} and \ref{sec:weak} present the benchmark results. Section~\ref{sec:conclusion} states their scope. Graph identities are given in Appendix~\ref{app:petersen}.

\section{Theory-space masses, disorder and localisation}
\label{sec:recap}

Anderson-like localization in four dimensions was demonstrated in a linear moose model~\cite{craig2018exponential}.  The fermionic action with link fields connecting left and right chiral fermions is given by
\begin{align}
 S={}&\int d^4x\,\Bigg\{\sum_{j=1}^{N}
 \left[\bar L_j i\slashed D L_j+\bar R_j i\slashed D R_j
 +\bar L_j M R_j+\mathrm{h.c.}\right]\nonumber\\
 &+\sum_{j=1}^{N-1}\left[\bar L_j\Phi_{j,j+1}R_{j+1}
 +\bar L_{j+1}\Phi_{j+1,j}R_j+\mathrm{h.c.}\right]\Bigg\}.
 \label{1}
\end{align} 
Here $\Phi_i$ represents the link fields and $L_i, R_i$ the chiral fermionic fields. When the link fields attain vacuum expectation values (vevs), the total Lagrangian including the kinetic terms is represented by Eq.\eqref{lag1}, where $\mathcal{H}$ represents mass terms that follow the underlying geometry in the theory space. 
\begin{align}
    \mathcal{L}  = \mathcal{L}_{kin}  - \sum_{i,j=1}^{N} \overline{L_{i}}\mathcal{H}_{i,j}R_j  + h.c. \label{lag1}
\end{align}
In a general manner, encompassing several models, $\mathcal{H}$ can be represented as follows, with $\kappa$ an integer taking values $\{0,1\}$: 
\begin{align}
    \mathcal{H}_{i,j} =&\epsilon_i \delta_{i,j} - t_i\delta_{i+1,j} - \kappa t_j\delta_{i,j+1} \label{Hmatrix}
\end{align}
We will call this the ACS (Anderson-Craig-Sutherland Lagrangian or model).
The local-chain link $t_i$ connects $i$ to $i+1$ in both directions when $\kappa=1$. In Eq.\eqref{Hmatrix}, on setting $\kappa=0$ and removing one left handed field $L_N$ we recover the well-known Clockwork model~\cite{giudice2017clockwork}, whose Lagrangian (in the notation where link fields have acquired vevs) reads
\begin{align}
\mathcal{L}_{CW} = \mathcal{L}_{kin} - \sum_{i=1}^{N-1} m\,\bar{L}_i R_i + \sum_{i=1}^{N-1} q m\,\bar{L}_i R_{i+1} + \text{h.c.}, \label{CW-Lagrangian}
\end{align}
with $\epsilon_i = m$ and $t_i = q m $. A crucial feature of the clockwork is the presence of a massless (zero) mode: the model as written has $N$ right-handed fields $R_1,\ldots,R_N$ and $N-1$ left-handed fields $L_1,\ldots,L_{N-1}$, so the mass matrix is rectangular ($(N-1) \times N$) and necessarily has a null vector---the celebrated clockwork zero mode. When $\kappa=1$ and the fields are in equal numbers (i.e.\ the mass matrix is square, $N\times N$), no zero mode is present in general, regardless of whether the parameters are uniform or random. When $\kappa=1$, we have the two-sided or double clockwork with similar assumptions on $\epsilon$ and $t$\cite{singh2025revisiting}\footnote{This limit is very similar to the deconstruction models.}. Interesting variations happen when $\epsilon_i$ and $t_i$ are made random when $\kappa=0$ \cite{de2020random} and $\kappa=1$ \cite{craig2018exponential}. The random clockwork model ($\kappa=0$) is when these parameters are chosen randomly in a range rather than being universal~\cite{de2020random}. The results for $\kappa=0$, the clockwork model can be found in Ref.~\cite{giudice2017clockwork,ibarra2018clockwork}. The particularly interesting case of $\kappa=1$ and random  $\epsilon_i$ has been studied in \cite{craig2018exponential}, which is also the topic of this work. 
It has been shown in \cite{craig2018exponential} that when $\epsilon_i$  are randomly varied in an interval such as $[2t, 2t + W]$, where $W$ is a parameter, the model exhibits Anderson-like localization of its wave functions. The localization is so effective that it can lead to exponential hierarchies in the couplings.

To set the notation and understand the model without randomness, let us consider the case when no parameter is random. The $\mathcal{H}$ in eq.\eqref{Hmatrix} leads to a mass matrix for the fermionic fields $\{L_i, R_i\}$ with $\kappa$ = 1, 
in the basis $(L_1, L_2, ... L_N,$ $R_1, R_2,...R_N)$  is a symmetric\footnote{The benchmarks use real entries. Physical masses are nonnegative singular or Takagi values, whereas real-symmetric tower eigenvalues may have either sign.} 
anti-diagonal block matrix
\begin{equation}
M_{mass} = \left[
\begin{array}{cc}
0 & M_A \\
M_A^T & 0
\end{array}
\right]
\label{eq:Mmass}
\end{equation}
where $M_A$ is the Dirac mass block, with entries of mass dimension one:
\begin{equation}
  M^{A} =
  \left[ {\begin{array}{ccccc}
    \epsilon_1 & -t & 0 & ... & 0 \\
    -t & \epsilon_2 & -t & ... & 0 \\
    0 & -t & \epsilon_3 & ... & 0\\
    ... & ... & ... & ... & ... \\
    0 & ... & ... & -t & \epsilon_N \\
  \end{array} } \right]
\end{equation}
Eigenvalues of matrix $M_A$ in the limiting case $\epsilon_i$ = $\epsilon$ $\forall$ i, which we will call the uniform case, are given by~\cite{kulkarni1999eigenvalues},\cite{gover1994eigenproblem},\cite{yueh2005eigenvalues}
\begin{align}
    \lambda_{k}=\epsilon-2t \cos \frac{k \pi}{N+1}, \label{5}
\end{align}
for $k \in\{1,2, \ldots, N\}$, and the corresponding elements of the eigenvectors $\xi_{j}^{(k)}$, are given by
\begin{align}
\xi_{j}^{(k)}=\rho^k \sin \frac{k j \pi}{N+1}, \hspace{1cm} j \in \{1,2, \ldots, N\} . \label{7}
\end{align}
Here $\xi_{j}^{(k)}$ is the $j^{\rm th}$ component of the $k^{\rm th}$ eigenvector (indexed by eigenvalue in ascending order), and $\rho^k$ is the normalisation constant for the $k^{\rm th}$ eigenvector, given by $\rho^k = \left[\sum_{j=1}^N \sin^2\!\frac{kj\pi}{N+1}\right]^{-1/2} = \sqrt{2/(N+1)}$.
\begin{figure}
    \begin{subfigure}{0.48\textwidth}
        \centering
        \includegraphics[width=\linewidth]{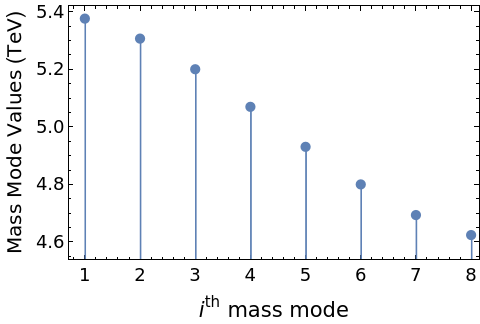}
        \caption{}
        \label{subfig:uniform-mass}
    \end{subfigure}
    \hfill
    \begin{subfigure}{0.48\textwidth}
        \centering
        \includegraphics[width=\linewidth]{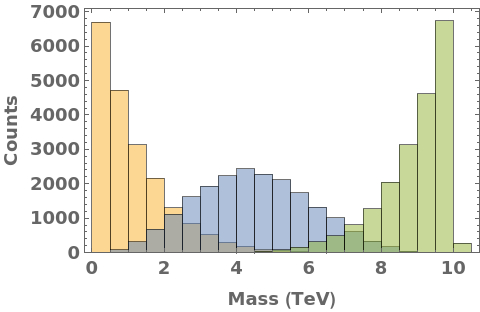}
        \caption{}
        \label{subfig:random-mass}
    \end{subfigure}
    
    \caption{{ Left: Mass mode distribution (eigenvalue spectrum) for the uniform case with $\epsilon_i = W$ and $t_i = t$, showing the cosine distribution of eq.~\eqref{5}. Right: Histogram of the three mass values (smallest, middle ($4^{th}$), largest) produced over 20000 random realisations of $\epsilon_i$ for the disordered case. The horizontal axis shows the mass value in TeV. The vertical axis (labelled ``counts'' in the figure) counts the number of random realisations in each bin. Parameters: $W = 5\,\text{TeV}$, $t = 0.2\,\text{TeV}$, $N = 8$, $\epsilon_i \in [-2W, 2 W]$.}}
    \label{anderson-local-eigenvalues}
\end{figure}

\subsection{Disorder \& Localization}

\begin{figure}
    \centering
    \begin{subfigure}{0.48\textwidth}
        \centering
        \includegraphics[width=\linewidth]{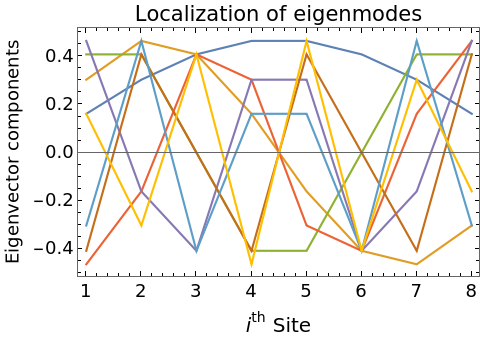}
        \caption{}
        \label{subfig:local-uniform}
    \end{subfigure}
    \hfill
    \begin{subfigure}{0.48\textwidth}
        \centering
        \includegraphics[width=\linewidth]{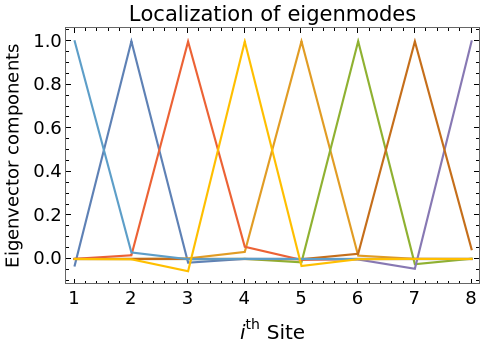}
        \caption{}
        \label{subfig:local-random}
    \end{subfigure}
    
    \caption{Eigenmodes $\xi_i$ of Local lattice with uniform sites $\epsilon_i$ = $W$ = 5 TeV $\&$ $ t_i = t = 0.2$ TeV (left) and random sites $ t_i = t = 0.2 $ TeV $\&$ $\epsilon_i \in$ $[-2W, 2W]$ (right) for $N$ = 8.}
    \label{firstanderson}
\end{figure}

As a simple illustration of site disorder, we take $\epsilon_i \in [-2W,2W]$ with universal hopping $t$ in Eq.~\eqref{Hmatrix} and $\kappa=1$. For concreteness we use $W=5~{\rm TeV}$, $t=0.2~{\rm TeV}$, and $N=8$, and compare the uniform case $\epsilon_i=W$ with the disordered case. The corresponding spectra are shown in Fig.~\ref{anderson-local-eigenvalues}: the uniform tridiagonal matrix gives the expected cosine spectrum, while disorder turns the fixed eigenvalues into distributions over random realisations. The associated eigenvectors, shown in Fig.~\ref{firstanderson}, demonstrate the key effect: the extended modes of the uniform chain become sharply localised on individual sites once random diagonal disorder is introduced. This is the Anderson-localisation mechanism relevant for the construction \cite{PhysRev.109.1492,tropper2021randomness,craig2018exponential}.

To compare the suppression of individual wavefunction components
in random clockwork (RCW) and ACS-type models, we define
\begin{align}
\xi_0^{\min}=\min_i |\xi_0^i|, \qquad i=1,\ldots,N .
\label{xl0}
\end{align}
Here $\xi_0^i$ denotes the $i$th site component of the zero mode
in clockwork, or of the lightest eigenmode in the ACS case.
A smaller $\xi_0^{\min}$ indicates stronger suppression of at
least one component. This can accompany localisation, but a
small component can also arise from a node and does not by
itself determine the spatial extent of the mode. The distinction between the two setups is important: fermionic clockwork models with $\kappa=0$ have a rectangular mass matrix, with one extra right-handed field, and therefore a protected zero mode, while the ACS model with $\kappa=1$ is generically square and has no zero mode. In the RCW benchmark \cite{de2020random}, both site and hopping parameters are random, $\epsilon_i\in[t,t+W]$ and $t_i\in[0,W]$, whereas in the ACS variants below we allow disorder in $\epsilon_i$, $t_i$, or both.
\begin{figure}
    \centering
    \begin{subfigure}{0.48\textwidth}
        \centering
        \includegraphics[width=\linewidth]{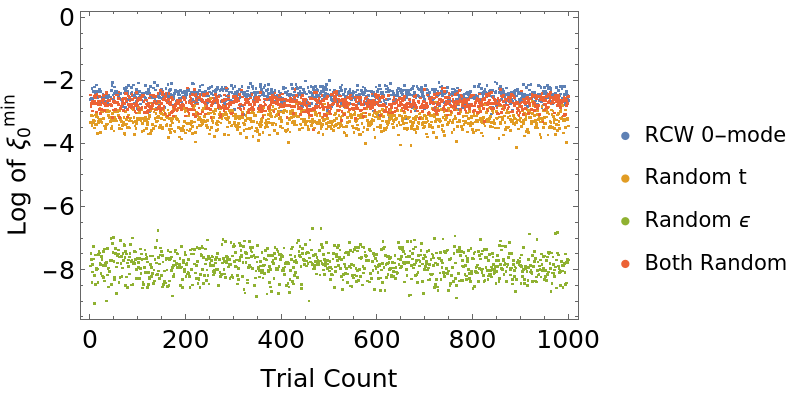}
        \caption{}
        \label{subfig:loc-median-50runs}
    \end{subfigure}
    \hfill
    \begin{subfigure}{0.48\textwidth}
        \centering
        \includegraphics[width=\linewidth]{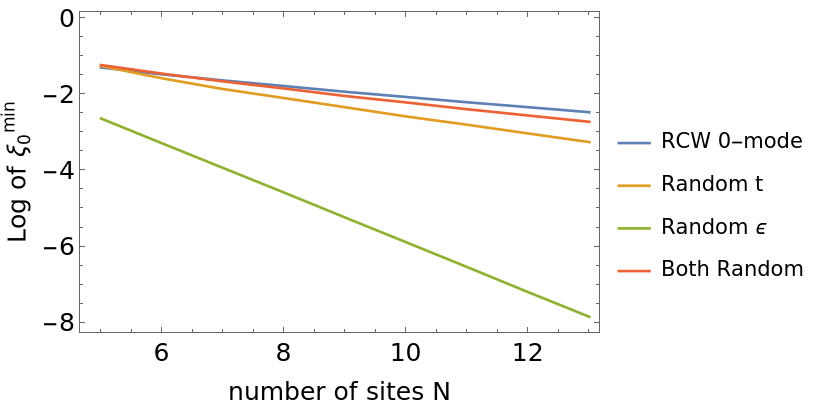}
        \caption{}
        \label{subfig:loc-varying-N}
    \end{subfigure}
    \caption{{ Median of $\log_{10}\xi_0^{\min}$ (eq.~\eqref{xl0}) for the random clockwork (RCW) and various ACS disorder models, with $N = 14$ sites (left) and varying $N \in [5,14]$ (right). Each data point is the median over 50 independent random realisations, each comprising 1000 disorder draws, with $W = 10\,\text{TeV}$ and $t = 1\,\text{TeV}$. Model parameters are listed in Table~\ref{tablecomparision}.}}
    \label{compare}
\end{figure}

Fig.~\ref{compare} compares $\xi_0^{\min}$ for RCW and the ACS variants, with the parameter choices listed in Table~\ref{tablecomparision}. For these ranges, the ACS model with random diagonal entries
gives the strongest suppression of the minimum component.
This is consistent with hopping being small compared with
the site disorder. The other ensembles give comparable
minimum-component suppression, although this quantity alone
does not establish comparable localisation lengths.

\begin{table}[tbp]
\centering
\small
\setlength{\arrayrulewidth}{0.4pt}
\setlength{\tabcolsep}{6pt}
\renewcommand{\arraystretch}{1.25}
\caption{Parameters considered for Clockwork and both-sided Hamiltonian with $W$ = 10 TeV and $t$ = 1 TeV.}
\label{tablecomparision}
\begin{tabularx}{\linewidth}{@{}>{\raggedright\arraybackslash}p{0.30\linewidth} *{3}{>{\centering\arraybackslash}X}@{}}
\noalign{\hrule height 0.7pt\vskip 5pt}
\textbf{Scenario} & \textbf{$\kappa$} & \textbf{$\epsilon_i$ (TeV)} & \textbf{$t_i$ (TeV)} \\
\noalign{\vskip 3pt\hrule height 0.4pt\vskip 5pt}
Random Clockwork & 0 & [t, t+W] & [0, W] \\
\addlinespace[5pt]
Random $\epsilon_i$ & 1 & [t, t+W] & $t$ \\
\addlinespace[5pt]
Random $t_i$ & 1 & W & [0, W] \\
\addlinespace[5pt]
Random $\epsilon_i$ $\&$ $t_i$ & 1 & [t, t+W] & [0, W] \\
\noalign{\vskip 4pt\hrule height 0.7pt}
\end{tabularx}
\end{table}

\subsection{Strong, Weak localizations and Geometry}
\subsubsection{Strong Disorder}
For strong independent site disorder, with width large compared with the hopping scale, localisation suppresses propagation along the chain~\cite{craig2018exponential}. The estimate below is a parametric strong-disorder expression, not an exact localisation length for every energy or distribution. For the non-local and Petersen-type graphs, the relevant path lengths, weighted connectivity and finite size must also be considered. Random hopping has a distinct band-centre behaviour on bipartite graphs and is not covered by the on-site-disorder argument~\cite{izrailev2012anomalous,cheraghchi2005localization}.
\begin{equation}
L_{loc}^{-1}  \sim \ln\!\left(\frac{W}{2t} - 1\right).
\label{eq:Lloc-strong}
\end{equation}
Also in strong localization conditions, the eigenvectors are exponentially localized at a certain site $i_j$ ($i^{th}$ mode localized at $i_j^{th}$ site), leading the eigenvector matrix to have $k^{th}$ components proportional to, $\xi^k_{i}$ $\propto$ $exp[(-|k-i_j|/L_{loc})]$ \cite{craig2018exponential}.

We now turn our attention to the role that underlying geometries might play in the strong localization regime. We compare three cases, all of which show very similar results except for the difference in the magnitude of $L_{loc}$. 
We consider three particular geometries, which are sort of extreme cases in terms of the links of the ``hopping" terms. The three cases we consider are (i) Only Nearest Neighbour Links/Mass terms ($L_{NN}$), (ii) All possible links between sites, including nearest neighbour, next to nearest, etc. ($L_{ALL}$), (iii) Links based on a specific geometry
($L_G$). The case of $L_{NN}$ is the one where the hopping terms or $t$ terms are restricted to be only nearest neighbour ones as in Ref.~\cite{craig2018exponential}. The case of $L_{ALL}$ is considered in Ref.~\cite{tropper2021randomness}, where, in addition to nearest neighbour mass terms, mass terms with all other possible sites are also considered with reducing weight depending on the distance from the sites.  Among the many possible choices of $L_G$, we focus on the Petersen graph, whose graph-theoretic properties are reviewed in Appendix~\ref{app:petersen}. It provides a sparse intermediate connectivity between the local and fully-connected extremes. To the best of our knowledge, the Petersen graph has not previously been studied in this context.

\begin{figure}
    \includegraphics[width=0.8\linewidth]{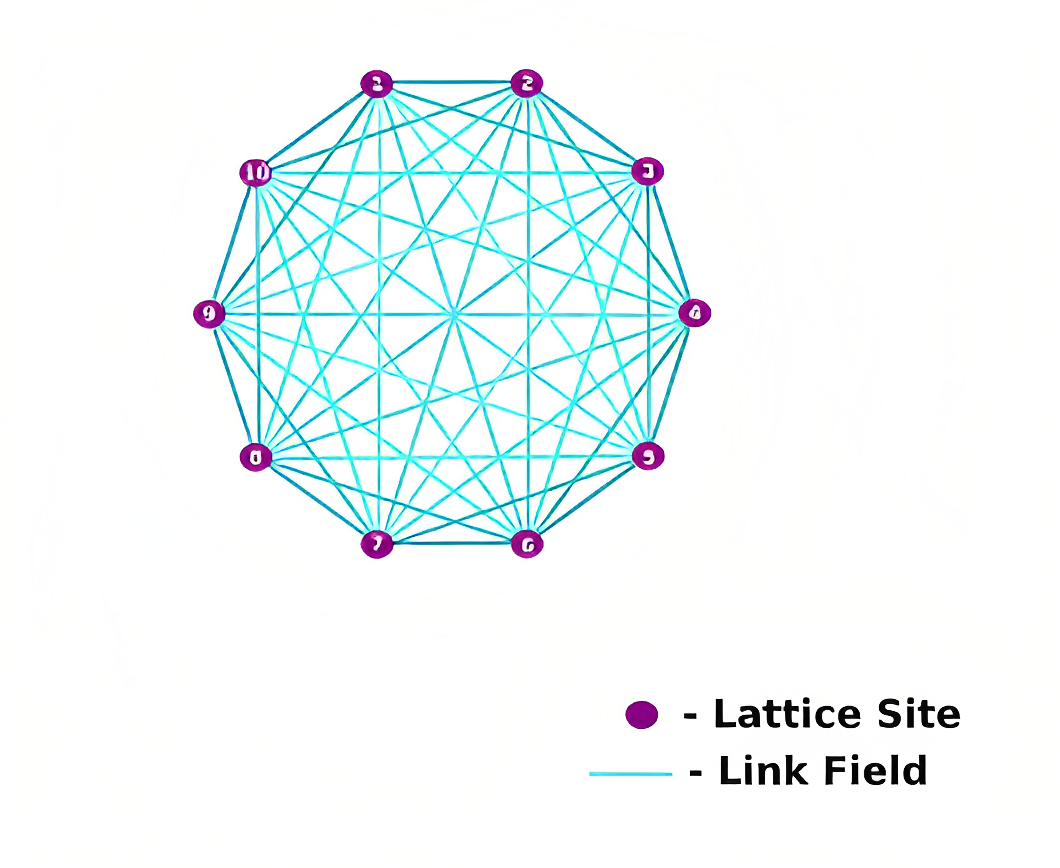}\\
   \caption{ Long-range non-local lattice representation for $N$ = 10 sites. The pictorial length of the link does not correlate with the strength of the coupling between the fields of different sites.} \label{Nonlocal}
\end{figure}

The Hamiltonian for the completely local case coincides with Eq.~\eqref{Hmatrix} for $\kappa = 1$, as discussed in the previous section. 
The corresponding completely non-local construction was introduced in Ref.~\cite{tropper2021randomness}, building on the framework of Ref.~\cite{nosov2019correlation}. 
In this work, motivated by the non-local Hamiltonian considered for scalar fields \cite{tropper2021randomness} :
\begin{align}
\mathcal{L}_{+}
= \frac{1}{2} \sum_{i=1}^{N}\left(\partial_{\mu} \pi_{i}\right)^{2}
- \frac{1}{2} \sum_{j=1}^{N} \epsilon_{j} \pi_{j}^{2}
- \frac{1}{2} \sum_{i=1}^{N-1} \sum_{j=i+1}^{N} \frac{t}{b^{\,j-i}}\left(\pi_{i} + \pi_{j}\right)^{2},
\end{align}
where $b$ is the decay parameter, $\pi_i$ are scalar fields and $i,j$ label the sites of the theory space. 
In the decaying regime ($b>1$), this class of long-range Hamiltonians exhibits exponential localization, and the construction can be straightforwardly extended to fermionic fields. 
The corresponding graph of link fields is shown in Fig.~\ref{Nonlocal}.
The corresponding Lagrangian with fermions is given by 
\begin{align}
\label{nonlocalDirac}
    \mathcal{L}_{non-local} &= \mathcal{L}_{Kin} - \sum_{i,j=1}^{N} \overline{L_{i}}\epsilon_{i,j}R_j - \sum_{i,j=1}^{N} \overline{L_{i}}\frac{t}{b^{|i-j|}}\left(1-\delta_{i, j}\right)R_j  + h.c. 
\end{align}
where $\epsilon_{ij} = \epsilon_i \delta_{i,j}$ with $\epsilon_i \in$ $[-2W, 2W]$. The  Dirac mass matrix for non-local Hamiltonian in basis $\{ L_i , R_j\}$, assuming $t_i$ = $t$, is
$ \mathcal{L}_{mass} = \Bar{L}_i \mathcal{M}_{non-local}R_j + H.c.$ with
\begin{equation}
\mathcal{M}_{non-local} =
\left[ {\begin{array}{ccccc}
\displaystyle\epsilon_1 & \displaystyle\frac{t}{b} & \displaystyle\frac{t}{b^2} & \cdots & \displaystyle\frac{t}{b^{N-1}} \\[10pt]
\displaystyle\frac{t}{b} & \displaystyle\epsilon_2 & \displaystyle\frac{t}{b} & \cdots & \displaystyle\frac{t}{b^{N-2}} \\[10pt]
\displaystyle\frac{t}{b^2} & \displaystyle\frac{t}{b} & \displaystyle\epsilon_3 & \cdots & \displaystyle\frac{t}{b^{N-3}} \\[10pt]
\vdots & \vdots & \vdots & \ddots & \vdots \\[10pt]
\displaystyle\frac{t}{b^{N-1}} & \cdots & \cdots & \displaystyle\frac{t}{b} & \displaystyle\epsilon_N
\end{array} } \right]
\end{equation}
\\
It is instructive to look at the eigenvalues and corresponding eigenvectors for matrix $M_{non-local}$ in the uniform case where $b \rightarrow 1$ and $ \epsilon_i \rightarrow \epsilon$. They are given by:
\begin{align}
\lambda_1 &= \epsilon + (N-1)t \\
\lambda_i &= \epsilon - t, \quad \text{with } i \in \{2, 3, \ldots, N\}
\end{align}
The corresponding eigenvectors in the uniform limit are
\begin{equation}
\xi = \begin{pmatrix}
\xi_1 \\ \xi_2 \\ \xi_3 \\ \vdots \\ \xi_N
\end{pmatrix}
= \begin{pmatrix}
\frac{1}{\sqrt{N}} & \frac{1}{\sqrt{N}} & \frac{1}{\sqrt{N}} & \cdots & \frac{1}{\sqrt{N}} \\[6pt]
-\frac{1}{\sqrt{2}} & \frac{1}{\sqrt{2}} & 0 & \cdots & 0 \\[6pt]
-\frac{1}{\sqrt{2}} & 0 & \frac{1}{\sqrt{2}} & \cdots & 0 \\[6pt]
\vdots & \vdots & \vdots & \ddots & \vdots \\[6pt]
-\frac{1}{\sqrt{2}} & 0 & 0 & \cdots & \frac{1}{\sqrt{2}}
\end{pmatrix}
\end{equation}

\begin{figure}
\centering
\begin{subfigure}{0.32\textwidth}
    \includegraphics[width=\linewidth]{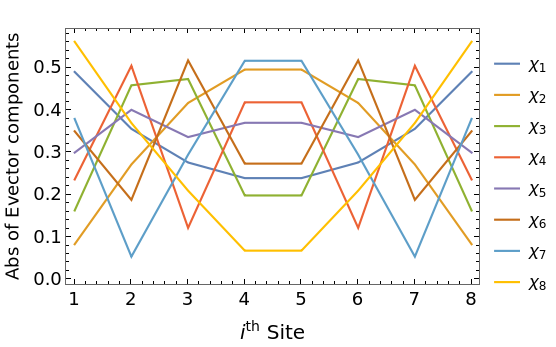}
    \caption{}
    \label{fig:nl-b0}
\end{subfigure}
\hfill
\begin{subfigure}{0.32\textwidth}
    \includegraphics[width=\linewidth]{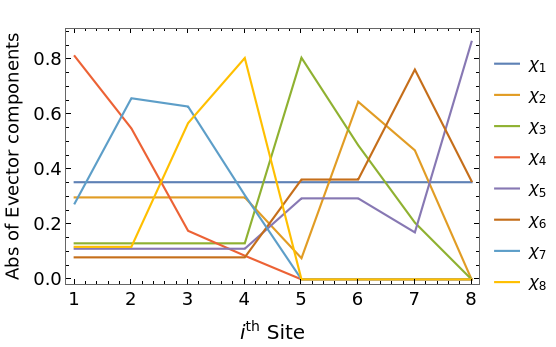}
    \caption{}
    \label{fig:nl-b1}
\end{subfigure}
\hfill
\begin{subfigure}{0.32\textwidth}
    \includegraphics[width=\linewidth]{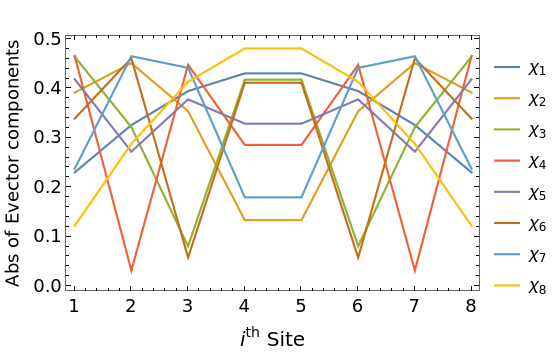}
    \caption{}
    \label{fig:nl-b2}
\end{subfigure}
\caption{Mass modes $\xi_i$ of Non-Local lattice having uniform sites $\epsilon_i = 2W$, $W = 5$ TeV, $t = 1$ TeV, $N = 8$ and increasing (left), constant (middle) and decreasing (right) non-neighbouring couplings with distance for $b = 0.7, 1$ and $2$ respectively.}
\label{nl-modes}
\end{figure}

For $b\ne1$ the eigenvectors are obtained numerically. Fig.~\ref{nl-modes} shows numerical eigenvectors for uniform site parameters at the three stated weights. As can be seen, there is no significant localization of modes in these cases. The plots are shown for three representative values of the decay parameter $b$, namely $b = 0.7$, $1$, and $2$, as illustrated in Fig.~\ref{nl-modes} (a), (b) and (c) respectively.

\begin{figure}
    \begin{subfigure}{0.48\textwidth}
        \centering
        \includegraphics[width=\linewidth]{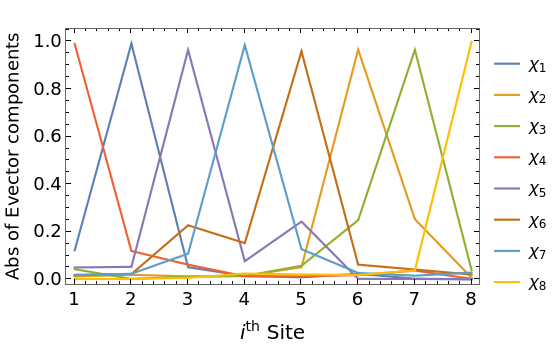}
        \caption{}
        \label{subfig:nl-mass-random}
    \end{subfigure}
    \hfill
    \begin{subfigure}{0.48\textwidth}
        \centering
        \includegraphics[width=\linewidth]{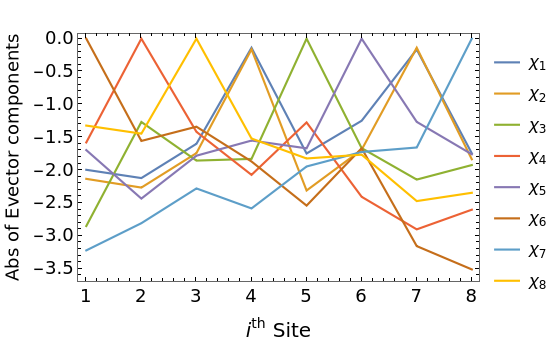}
        \caption{}
        \label{subfig:nl-mass-log}
    \end{subfigure}
    
    \caption{Mass modes $\xi_i$ of Non-Local lattice with random site terms (left) and Log of components of mass modes (right) for $\epsilon_i \in$ $[-2W, 2W]$ with $W$ = 5 TeV, $t$ = 1/4 TeV, $b$ = 2 and $N$ = 8.}
    \label{disorder-nonlocal-modes}
\end{figure}

However, the situation changes once disorder is introduced into the parameters of the Hamiltonian $\mathcal{H}$/ mass matrix. 
Assuming disorder in the diagonal terms, the resulting eigenmodes for the same geometry are shown in Fig.~\ref{disorder-nonlocal-modes}. The parameters used are given in the caption of the figure: $\epsilon_i \in [-2W, 2W]$ with $W$ = 5 TeV, $t$ = 1/4 TeV, $b$ =2 and $N$ = 8.
Large diagonal disorder leads to Anderson-like localization even in non-local geometries, provided the hopping strengths decay with distance.
This demonstrates that localization is a robust feature of disordered theory-space constructions and is not restricted to purely local geometries.

The Petersen-type construction gives an intermediate sparse connectivity. Let $N=2n$ with even $n\ge4$. Label the inner sites $v_i=i+1$ and the outer sites $u_i=n+i+1$, for $i=0,\ldots,n-1$. The unordered edge set is
\begin{equation}
 E=\{\{u_i,u_{i+1\bmod n}\},\{u_i,v_i\}:0\le i<n\}
 \cup\{\{v_i,v_{i+n/2}\}:0\le i<n/2\}.
 \label{eq:petersen-edges}
\end{equation}
There are $N$ vertices and $5N/4$ edges. For $I,J=1,\ldots,N$, define
\begin{equation}
 \mathcal H^{\rm pet}_{IJ}=\epsilon_I\delta_{IJ}
 +(1-\delta_{IJ})A_{IJ}\frac{t_{IJ}}{b^{|I-J|}},
 \qquad A_{IJ}=\mathbb 1[\{I,J\}\in E].
 \label{petersen-Hamiltonian}
\end{equation}
Here $t_{IJ}=t_{JI}$. The uniform-hopping illustration has $t_{IJ}=t$. The distance in the weight is the difference of the assigned integer labels, including the closing outer edge. This convention reproduces the explicit matrix below and is not the shortest cyclic distance. The Lagrangian is
\begin{equation}
 \mathcal L_{\rm Petersen}=\mathcal L_{\rm kin}
 -\sum_{I,J=1}^N\bar L_I\mathcal H^{\rm pet}_{IJ}R_J+\mathrm{h.c.}
 \label{peter_lag}
\end{equation}

For illustration, the Dirac mass matrix for the generalized Petersen
geometry used in our scans, specialised to $N=8$, can be written as
\begin{equation}
\mathcal{L}_{\rm mass}=\bar L_i M_{\rm Petersen} R_j+\text{h.c.},
\label{eq:Lmass-petersen}
\end{equation}
with
\begin{equation}
\mathcal{M}_{Petersen} =
\left[
\begin{array}{cccccccc}
\epsilon_1 & 0 & \dfrac{t}{b^2} & 0 & \dfrac{t}{b^4} & 0 & 0 & 0 \\[2pt]
0 & \epsilon_2 & 0 & \dfrac{t}{b^2} & 0 & \dfrac{t}{b^4} & 0 & 0 \\[2pt]
\dfrac{t}{b^2} & 0 & \epsilon_3 & 0 & 0 & 0 & \dfrac{t}{b^4} & 0 \\[2pt]
0 & \dfrac{t}{b^2} & 0 & \epsilon_4 & 0 & 0 & 0 & \dfrac{t}{b^4} \\[2pt]
\dfrac{t}{b^4} & 0 & 0 & 0 & \epsilon_5 & \dfrac{t}{b} & 0 & \dfrac{t}{b^3} \\[2pt]
0 & \dfrac{t}{b^4} & 0 & 0 & \dfrac{t}{b} & \epsilon_6 & \dfrac{t}{b} & 0 \\[2pt]
0 & 0 & \dfrac{t}{b^4} & 0 & 0 & \dfrac{t}{b} & \epsilon_7 & \dfrac{t}{b} \\[2pt]
0 & 0 & 0 & \dfrac{t}{b^4} & \dfrac{t}{b^3} & 0 & \dfrac{t}{b} & \epsilon_8
\end{array}
\right]
\end{equation}

\begin{figure}
    \includegraphics[width=0.28\linewidth]{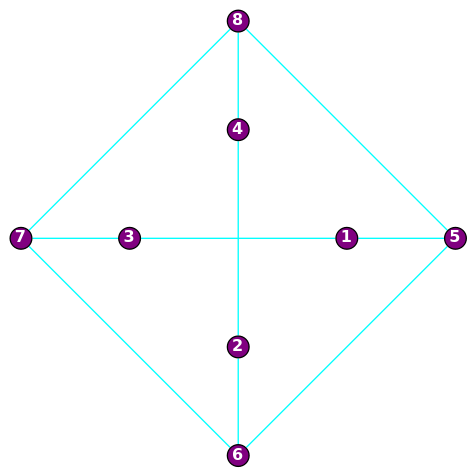} \hspace{1cm} \includegraphics[width=0.5\linewidth]{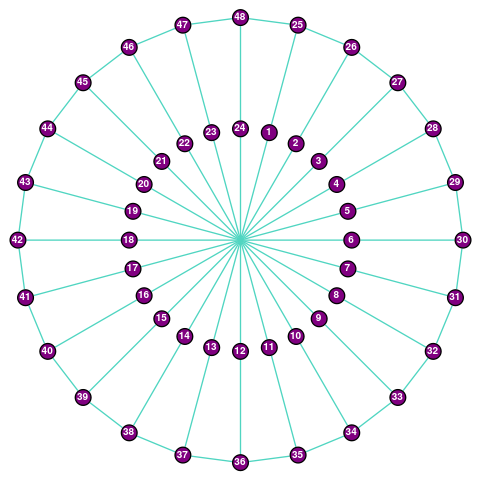} \\
    \caption{Petersen-type matching graphs $GP(n,n/2)$ with $n=4$ (left) and $n=24$ (right), giving $N=8$ and $N=48$ total sites, respectively.} \label{petersen}
\end{figure}
For the eight-site matrix at $b=1$ and $\epsilon_i=\epsilon$, the eigenvalues can be stated compactly as $\lambda=\epsilon+t a$, with adjacency eigenvalues
\begin{equation}
 a\in\left\{\frac{-1\pm\sqrt5}{2}\ \text{(each with twice multiplicity)},\quad
 \frac{3\pm\sqrt5}{2},\quad\frac{-1\pm\sqrt{13}}{2}\right\}.
 \label{petersen-evalue}
\end{equation}

\begin{figure}
    \begin{subfigure}{0.48\textwidth}
        \centering
        \includegraphics[width=\linewidth]{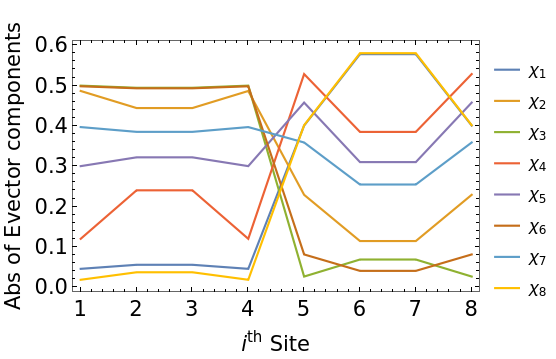}
        \caption{}
        \label{subfig:pet-uniform}
    \end{subfigure}
    \hfill
    \begin{subfigure}{0.48\textwidth}
        \centering
        \includegraphics[width=\linewidth]{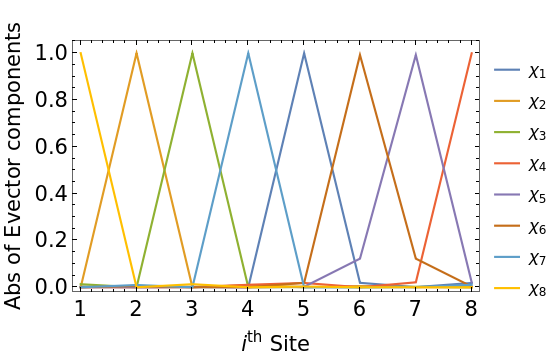}
        \caption{}
        \label{subfig:pet-random}
    \end{subfigure}
    
    \caption{Mass modes $\xi_i$ of Petersen graph with uniform sites $\epsilon_i$ = $W$ (left) and random sites $\epsilon_i \in [-2W, 2W]$ (right) for $N$ = 8, $W$ = 5 TeV, $t$ = 1/4 TeV and $b$ = 2.}
    \label{Petermode}
\end{figure}

\begin{figure}[ht]
    \centering
    \begin{subfigure}{0.32\textwidth}
        \centering
        \includegraphics[width=1\textwidth]{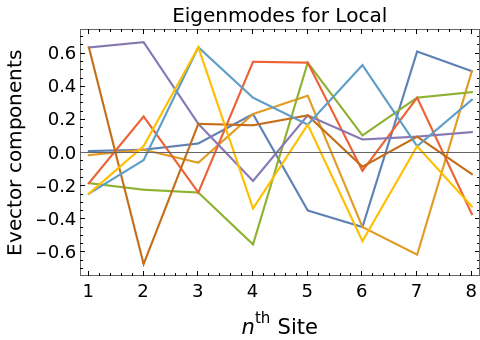}
        \caption{}
        \label{subfig:local-wave}
    \end{subfigure}
    \hfill
    \begin{subfigure}{0.32\textwidth}
        \centering
        \includegraphics[width=1\textwidth]{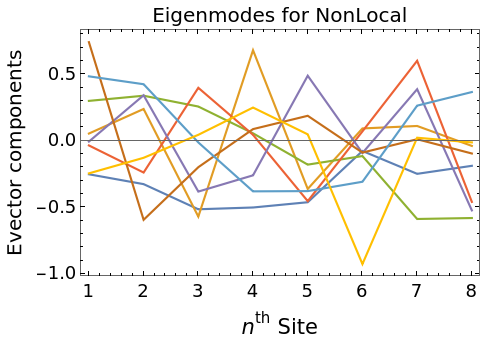}
        \caption{}
        \label{subfig:nonlocal-wave}
    \end{subfigure}
    \hfill
    \begin{subfigure}{0.32\textwidth}
        \centering
        \includegraphics[width=1\textwidth]{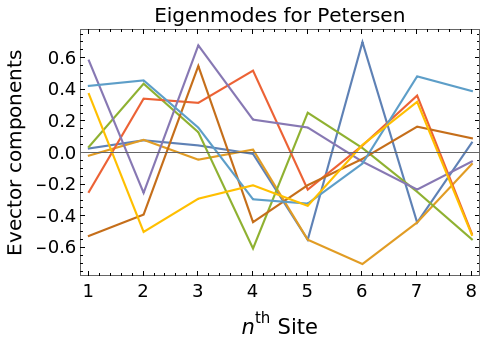}
        \caption{}
        \label{subfig:pet-wave}
    \end{subfigure}
    
    \caption{Eigenmodes for the three geometries under strong hopping disorder: local (left), non-local (middle), and Petersen (right). Parameters: $\epsilon_i = W = 0.25\,\text{TeV}$, $b = 2$, $t_i \in [-t,\,t]$, $t = 10\,\text{TeV}$, $N = 8$.}
    \label{wavefunction-strong-coupling-2}
\end{figure}

In general, this mass matrix will not have a 0-mode though one can produce a 0-mode by carefully choosing the site term $\epsilon_i$ in a uniform limiting case. In Fig.~\ref{Petermode}, we present the normalized eigenvectors $\chi_i$ for two cases, (i) uniform case where $\epsilon_i$ = W, and (ii) $\epsilon_i \in [-2W, 2W]$ strong disorder.
The left panel in the plot shows the uniform case ($\epsilon_i = W$ and $b>1$), while the right panel corresponds to the random case with $\epsilon_i \in [-2W,\,2W]$. Several modes have much larger components on one half of the
sites than on the other. This can be understood from the distance
weights, which suppress the spokes connecting the inner and outer
vertices. The two halves are therefore weakly coupled, allowing
a mode to reside predominantly on one half even without disorder.
Its components on the other half remain nonzero.
\begin{figure}
     \includegraphics[width=\linewidth]{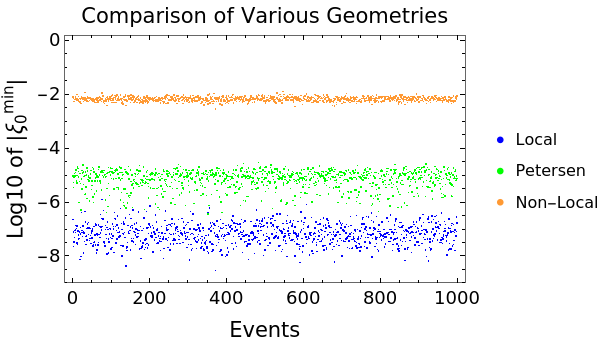} \\
   \caption{ \label{lvsnlvspet} Medians over 1,000 groups of 50 trials each of $\log_{10}|\xi_0^{\min}|$ for the lightest mode in the three geometries under strong site disorder ($\epsilon_i \in [0.5W, 1.5W]$), with $W = 5\,\text{TeV}$, $t = 0.25\,\text{TeV}$, $b = 2$, $N = 12$.}
\end{figure}
\begin{table}[tbp]
\centering
\small
\setlength{\arrayrulewidth}{0.4pt}
\setlength{\tabcolsep}{6pt}
\renewcommand{\arraystretch}{1.25}
\caption{Parameters considered for Local, Non-local and Petersen Hamiltonian with $W$ = 5 TeV, $t$ = 0.25 TeV and $b$ = 2.}
\label{disorder_lvsnlvspet_parameter}
\begin{tabularx}{\linewidth}{@{}>{\raggedright\arraybackslash}p{0.30\linewidth} *{3}{>{\centering\arraybackslash}X}@{}}
\noalign{\hrule height 0.7pt\vskip 5pt}
\textbf{Scenario} & \textbf{N (Sites)} & \textbf{$\epsilon_i$ (Strong Disorder)} & \textbf{$\epsilon_i$ (Weak Disorder)} \\
\noalign{\vskip 3pt\hrule height 0.4pt\vskip 5pt}
Local & 12 & [W - $\frac{W}{2}$, W + $\frac{W}{2}$] & [W - $\frac{t}{2}$, W + $\frac{t}{2}$] \\
\addlinespace[5pt]
Non-local & 12 & [W - $\frac{W}{2}$, W + $\frac{W}{2}$] & [W - $\frac{t}{2}$, W + $\frac{t}{2}$] \\
\addlinespace[5pt]
Petersen & 12 & [W - $\frac{W}{2}$, W + $\frac{W}{2}$] & [W - $\frac{t}{2}$, W + $\frac{t}{2}$] \\
\noalign{\vskip 4pt\hrule height 0.7pt}
\end{tabularx}
\end{table}\\
Now, in the case of strong disorder where $\epsilon_i \in [-2W, 2W]$, we see that all the modes are localized. However, the relative localization can be different for different geometries. To see this, we can compare the $\xi_0^{min}$ (introduced in eq.\eqref{xl0}) parameter for the three types of lattices we have studied with strong disorder in the diagonal elements of their Hamiltonian $\epsilon_i$s$\in [W/2, 3W/2]$. For the parameters in Table~\ref{disorder_lvsnlvspet_parameter}, Fig.~\ref{lvsnlvspet} shows the smallest values
of $\xi_0^{\min}$ in the local geometry, followed by Petersen and
non-local geometries. The local chain permits propagation only
through successive nearest-neighbour links, whereas additional
links provide alternative routes that can weaken component
suppression. The observed ordering is therefore physically
plausible for these benchmarks, although it also depends on the
link weights and disorder. The plotted quantity compares minimum
components rather than localisation lengths.
\begin{center}
\it{Strong Hopping Disorder}
\end{center}
For the strong disorder scenario, we can also have the large disorders in the hopping terms $t_i$'s. This will correspond to having uniform diagonal elements $\epsilon_i$'s, i.e, $\epsilon_i = \epsilon$ $\forall$ $i$ in the Hamiltonian matrices $\mathcal{H}_{i,j}$ for all geometries with randomly chosen off-diagonal couplings $t_i \in [-t, t]$ for $t \gg \epsilon$. The question then remains whether strong hopping disorder could lead to strong/weak localization. But such is not necessarily the case with large random off-diagonal or hopping terms. Fig.~\ref{wavefunction-strong-coupling-2} demonstrates the eigenmodes for the three lattices: 1) local (left), 2) Non-local (middle), and 3) Petersen (right) with large randomness in the coupling parameters, respectively. The parameters considered for the plots are $\epsilon_i$ = $W$ = 0.25 TeV, $b$ = 2 and $t_i \in$  [-t, t], $t$ = 10 TeV, $N$ = 8. These finite-size examples do not show the uniform, strong endpoint suppression seen with site disorder.

The other mechanism that can produce the hierarchical scale without needing localization of modes is the GIM-like cancellation mechanism \cite{singh2025revisiting}, which works on the unitarity property of eigenmodes of the matrix. For one Dirac generation, taking both boundary mass entries
equal to $v$, the leading physical light mass is
\begin{align}
m_0 \simeq v^2
\left|
\sum_{i=1}^{N}
\frac{\xi_1^i\xi_N^i}{\lambda_i}
\right|.
\end{align}
Here $v$ denotes the chosen boundary mass scale, taken equal
to the Higgs vev, and the $\lambda_i$ are signed tower
eigenvalues. This approximation requires an invertible heavy
block and perturbatively small boundary mixing. The examined strong-hopping benchmarks do not have a dominant uniform diagonal or a quasi-degenerate spectrum. They do not realise the controlled mass-suppression limit developed below.
\subsubsection{Weak Disorder}
\begin{figure}[ht]
\centering
\begin{subfigure}{0.32\textwidth}
    \includegraphics[width=1\textwidth]{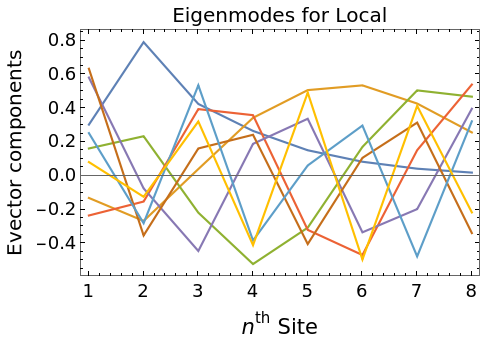}
    \caption{}
    \label{fig:local}
\end{subfigure}
\hfill
\begin{subfigure}{0.32\textwidth}
    \includegraphics[width=1\textwidth]{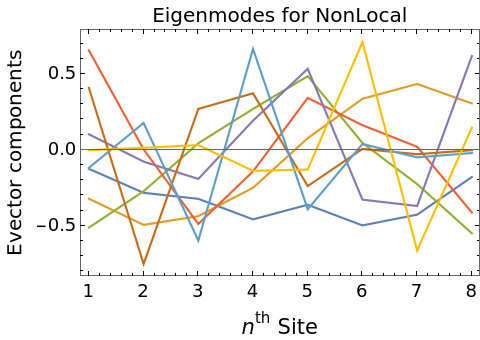}
    \caption{}
    \label{fig:nonlocal}
\end{subfigure}
\hfill
\begin{subfigure}{0.32\textwidth}
    \includegraphics[width=1\textwidth]{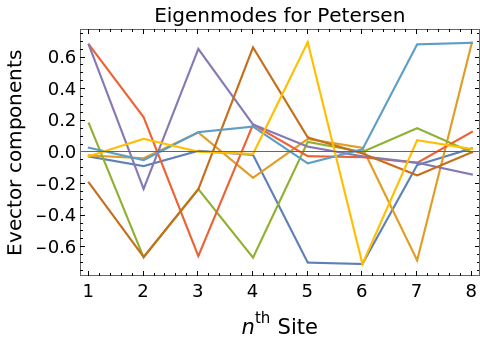}
    \caption{}
    \label{fig:pet}
\end{subfigure}
\caption{Wavefunctions for three geometries: 1) local (left), 2) nonlocal (middle) and 3) Petersen (right) with $\epsilon_i \in [W-t, W+t]$, $W = 5$ TeV, $b = 2$ and $t_i = t = 0.2$ TeV, $N = 8$.}
\label{wavefunctions}
\end{figure}

\begin{center}
    \textit{{ Weak Site Disorder}}
\end{center} 
The disorders in the site $\epsilon_i$'s (diagonal) and hopping $t_i$'s (off-diagonal) couplings can also be weak in nature as compared to other Hamiltonian matrix elements. This scenario is also widely studied in the literature, particularly in condensed matter systems \cite{tessieri2012anomalous,bernardet2002destruction,hrahsheh2012disordered}. Reducing the site disorder relative to the hopping generally allows
the modes to spread over more sites, making their boundary components
less strongly suppressed in a finite lattice. However, small disorder
relative to a common diagonal mass does not by itself imply extended
modes. The common diagonal shifts all eigenvalues equally without
changing the eigenvectors, while their spatial profiles depend on
the disorder and hopping strengths.
Fig.~\ref{wavefunctions} shows the wavefunctions for three lattices: 1) local (left), 2) Non-local (middle), and 3) Petersen (right) with small randomness in the parameters. The parameters chosen for the plots are $\epsilon_i \in [W-t, W+t]$, $W$ = 5 TeV, $b$ = 2 and $t_i =t$   = 0.2 TeV, $N$ = 8. The effect of the lattice on the shape of wavefunctions is evident from the figure. Apart from slight localization, the lattice structure also provides some qualitative features to the wavefunction, as is evident specially in Fig.~\ref{wavefunctions} (c).
Even when the individual wavefunctions do not provide sufficient
boundary suppression, a small mass can arise from cancellations
among their contributions. When the diagonal terms are close to a
common value that dominates both their fluctuations and the hopping
matrix, the eigenvalues have small relative splittings. Contributions
from different modes then carry nearly equal inverse-mass weights.
For couplings at distinct sites, orthogonality cancels the common
part of these contributions, leaving a smaller remainder controlled
by the splittings and the paths connecting the two sites. This is
the GIM-like suppression mechanism developed below and does not
require localisation.

A convenient way to understand the origin of quasi–degeneracies in the weak–disorder regime is to
note that adding a uniform diagonal shift does not alter the eigenvectors of the Hamiltonian. For any
real symmetric mass matrix $A$ with eigenvalues $\{\lambda_i\}$ and orthonormal eigenvectors
$\{\xi^{i}\}$, the shifted matrix
\begin{equation}
B = A + n\,\mathbb{1}
\end{equation}
shares the same eigenvectors, while its eigenvalues are uniformly translated,
\begin{equation}
\lambda_i \;\longrightarrow\; \lambda_i + n .
\end{equation}
where $n \in \Re$ represents the shift. The uniform shift leaves the eigenvectors and absolute
eigenvalue splittings unchanged. However, for
$n\gg\max_i|\lambda_i|$, the relative splittings
$|\lambda_i-\lambda_j|/n$ become small, so the heavy states
are quasi-degenerate relative to their common mass scale. The spread in inverse eigenvalues, which controls the efficiency
of the GIM-like cancellation mechanism discussed below, is then suppressed:
\begin{align}
\delta &\equiv 
\frac{1}{n+\lambda_1}-\frac{1}{n+\lambda_N} \\
&= \frac{\lambda_N-\lambda_1}{(n+\lambda_1)(n+\lambda_N)}
\;\xrightarrow{n\gg\max_i|\lambda_i|}\; 0 ,
\end{align}
showing that a large uniform contribution to the diagonal entries compresses the inverse spectrum even
if the original eigenvalues were moderately split. This behaviour is central to the appearance of
quasi–degenerate heavy states in the weak–disorder regime.

To make this mechanism explicit, we define the boundary
propagator $\zeta$, which has inverse-mass dimension and
controls the induced light mass once the boundary couplings
are included:
\begin{equation}
\zeta \;\equiv\;
\sum_{i=1}^{N}
\frac{\xi^{i}_{1}\, \xi^{i}_{N}}
{\lambda_i + n} .
\end{equation}
Here $\xi^{i}_{1}$ and $\xi^{i}_{N}$ denote the overlaps of the $i$th eigenvector with the first and $N$th
sites (or left/right endpoints of the chain), respectively. 
The dependence of the suppression on geometry and boundary placement
can be understood directly from the inverse matrix. Let
\(O=(\xi^1,\ldots,\xi^N)\) be the orthogonal matrix whose columns are
the eigenvectors of the real symmetric matrix \(A\). Then
\[
A=O\,\mathrm{diag}(\lambda_i)\,O^T,
\qquad
B=A+nI=O\,\mathrm{diag}(\lambda_i+n)\,O^T.
\]
For two distinct attachment sites \(p\) and \(q\), this gives
\begin{equation}
\zeta_{pq}(n)
\equiv
\sum_{i=1}^{N}
\frac{\xi_p^i\xi_q^i}{n+\lambda_i}
=
(B^{-1})_{pq},
\label{eq:zeta-inverse-element}
\end{equation}
provided \(B\) is invertible. The quantity defined above corresponds
to \(p=1\) and \(q=N\).

For \(n>\max_i|\lambda_i|\), each denominator admits the convergent
geometric expansion
\[
\frac{1}{n+\lambda_i}
=
\frac{1}{n}
\sum_{r=0}^{\infty}
\left(-\frac{\lambda_i}{n}\right)^r.
\]
Using
\[
\sum_i \lambda_i^r \xi_p^i\xi_q^i=(A^r)_{pq},
\]
we therefore obtain
\begin{align}
\zeta_{pq}(n)
&=
\sum_{r=0}^{\infty}
\frac{(-1)^r(A^r)_{pq}}{n^{r+1}}
\nonumber\\
&=
-\frac{A_{pq}}{n^2}
+\frac{(A^2)_{pq}}{n^3}
-\frac{(A^3)_{pq}}{n^4}
+\cdots .
\label{eq:zeta-inverse-expansion}
\end{align}
The \(1/n\) term vanishes because \(p\neq q\) and
\(\sum_i\xi_p^i\xi_q^i=\delta_{pq}\). This is the leading
orthogonality cancellation.

The higher powers have a simple interpretation in terms of paths.
For example,
\[
(A^2)_{pq}=\sum_j A_{pj}A_{jq}
\]
sums the contributions from all two-step walks between the attachment
sites. More generally, \((A^r)_{pq}\) sums products of matrix entries
along walks with \(r\) steps. Diagonal entries correspond to remaining
at the same site and cannot shorten the number of links required to
connect \(p\) and \(q\).

If the shortest connecting path contains \(d\) links, then
\((A^r)_{pq}=0\) for \(r<d\). Hence
\begin{equation}
\zeta_{pq}(n)
=
\frac{(-1)^d(A^d)_{pq}}{n^{d+1}}
+\mathcal{O}\!\left(n^{-d-2}\right).
\label{eq:zeta-shortest-path}
\end{equation}
The coefficient \((A^d)_{pq}\) is the sum of the hopping products
along the shortest paths. It is nonzero for the positive hopping
couplings considered here. Thus each additional required hop
introduces another inverse power of the diagonal offset.

\begin{figure}[ht]
     \centering
\includegraphics[width=0.62\textwidth]{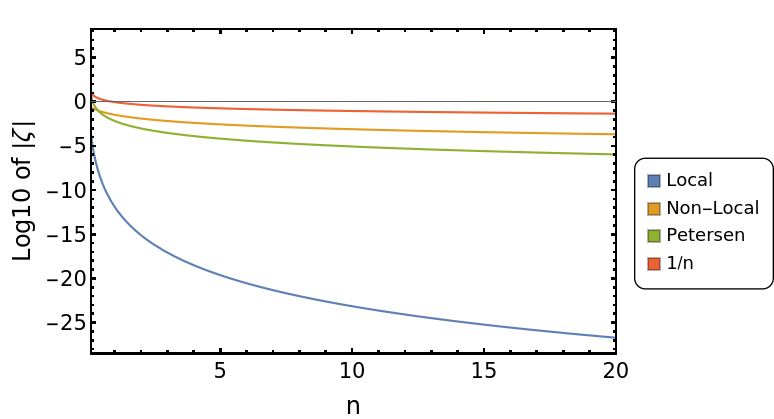} 
   \caption{ The GIM-like cancellation parameter $\zeta$ (in TeV$^{-1}$, $\log_{10}$ scale) as a function of the uniform diagonal offset $n$ (in TeV), for the three geometries, together with the $1/n$ reference curve. Parameters: $\epsilon_i \in [W-t, W+t]$, $W = 0.2\,\text{TeV}$, $b = 1$, $t_i = t = 0.1\,\text{TeV}$, $N = 12$.} \label{cancellation}
\end{figure}

For a local chain with uniform nearest-neighbour hopping \(t\),
the unique shortest path between \(p<q\) contains \(q-p\) links.
Consequently,
\begin{equation}
\zeta_{pq}(n)
\sim
\frac{(-1)^{q-p}t^{q-p}}{n^{q-p+1}}.
\label{eq:zeta-local-attachments}
\end{equation}
Neighbouring attachments therefore give
\(\zeta_{p,p+1}\sim-t/n^2\), whereas the endpoints of an
\(N\)-site chain give
\(\zeta_{1N}\sim(-t)^{N-1}/n^N\).
For the displayed \(N=12\), \(b=1\) benchmarks, the shortest boundary
paths contain eleven, one and two links in the local, non-local
and Petersen geometries, respectively. Their leading magnitudes are
therefore
\[
|\zeta_{\mathrm{local}}|\sim\frac{t^{11}}{n^{12}},
\qquad
|\zeta_{\mathrm{nonlocal}}|\sim\frac{t}{n^2},
\qquad
|\zeta_{\mathrm{Petersen}}|\sim\frac{t^2}{n^3}.
\]
This explains the different rates of decrease. No localisation is required for this result. The uniform shift leaves
the eigenvectors unchanged, and the suppression also occurs for
uniform site masses.
Thus $\zeta=(B^{-1})_{1N} = [(A+nI)^{-1}]_{1N}$, and Fig.~\ref{cancellation} shows its magnitude together with the (1/n) reference curve. The constant shift by $n$ in the denominator
illustrates how a uniform enhancement of the diagonal masses drives the inverse–eigenvalue
differences toward zero, thereby strengthening the cancellation and reducing the smallest effective
mass scale.
The decrease of $|\zeta|$ with increasing $n$ therefore reflects
stronger suppression of boundary propagation and, for fixed boundary
couplings, a smaller induced light-mass scale. The different rates
follow from the shortest connecting paths, as derived above.
A dominant uniform diagonal with weak disorder can thus generate hierarchies through
GIM-like cancellations without requiring localisation \cite{singh2025revisiting}.
\begin{center}
    \textit{{ Weak Hopping Disorder}}
\end{center} 
The weak disorder of the couplings in the Hamiltonian can also be considered in the off-diagonal or hopping terms $t_i$'s instead of the diagonal (site) terms $\epsilon$'s. Thus, in the Hamiltonian $H_{i,j}$ of the theory space, the off-diagonal terms are randomised weakly, i.e., the magnitude of off-diagonal couplings is smaller than the diagonal entries. The large common diagonal makes the relative spectral splittings
small, allowing cancellations to suppress propagation between
distinct boundary sites. It does not determine the spatial extent
of the eigenmodes. Indeed, rescaling all hopping terms by a common
factor at fixed uniform site mass changes the eigenvalue splittings
but leaves the eigenvectors unchanged.
\begin{figure}[ht]
    \centering
    \begin{subfigure}{0.32\textwidth}
        \centering
        \includegraphics[width=1\textwidth]{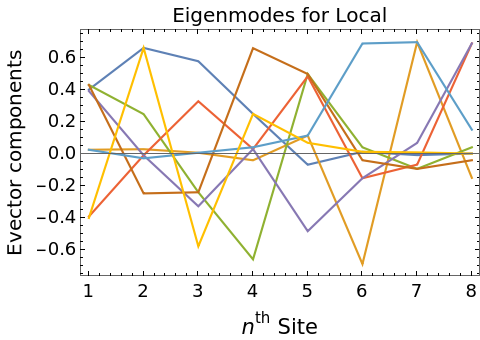}
        \caption{}
        \label{subfig:local-wavef}
    \end{subfigure}
    \hfill
    \begin{subfigure}{0.32\textwidth}
        \centering
        \includegraphics[width=1\textwidth]{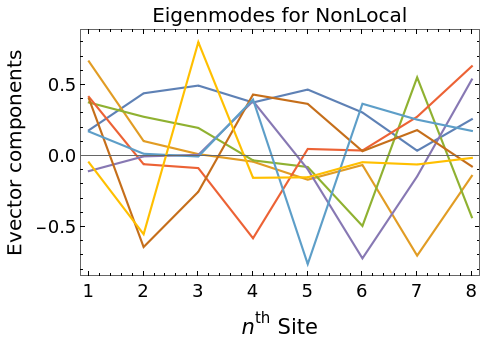}
        \caption{}
        \label{subfig:nonlocal-wavef}
    \end{subfigure}
    \hfill
    \begin{subfigure}{0.32\textwidth}
        \centering
        \includegraphics[width=1\textwidth]{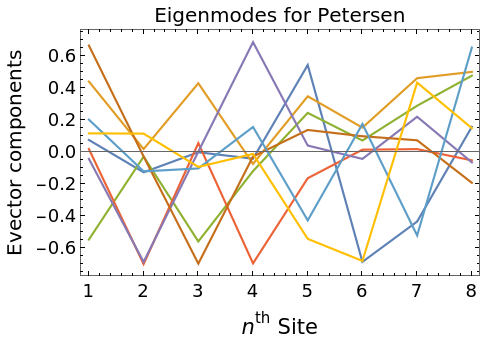}
        \caption{}
        \label{subfig:peter-wavef}
    \end{subfigure}
    
    \caption{Eigenmodes for the three geometries under weak hopping disorder: local (left), non-local (middle), and Petersen (right). Parameters: $\epsilon_i = W = 10\,\text{TeV}$, $b = 1.2$, $t_i \in [-t,\,t]$, $t = 0.1\,\text{TeV}$, $N = 8$. The dominant diagonal permits GIM-like suppression of
boundary propagation without requiring localisation.}
    \label{wavefunction-strong-coupling}
\end{figure}

Fig.~\ref{wavefunction-strong-coupling} shows the wavefunctions for three lattices: 1) local (left), 2) Non-local (middle), and 3) Petersen (right) with small randomness in the coupling parameters, respectively. The numbers chosen for the plot are $\epsilon_i$ = $W$ = 10 TeV, $b$ = 1.2 and $t_i \in$  [-t, t], $t$ = 0.1 TeV, $N$ = 8.
The orthonormality of eigenmodes is guaranteed by the real and symmetric mass matrix structure. Light masses can therefore again be generated through the GIM-like cancellation mechanism, making this setting viable for explaining mass hierarchies in the SM.

\begin{figure}[ht]
\centering
\begin{subfigure}{0.329\textwidth}
    \includegraphics[width=1\textwidth]{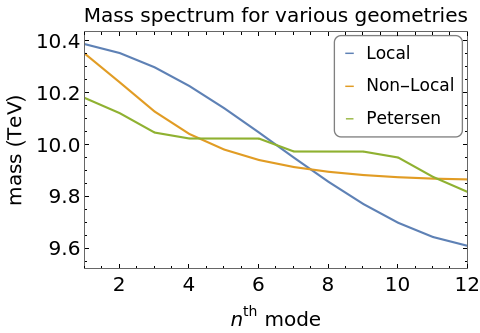}
    \caption{}
    \label{fig:uniform}
\end{subfigure}
\hfill
\begin{subfigure}{0.329\textwidth}
    \includegraphics[width=1\textwidth]{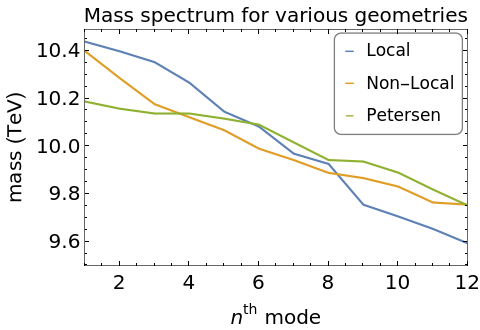}
    \caption{}
    \label{fig:weak}
\end{subfigure}
\hfill
\begin{subfigure}{0.329\textwidth}
    \includegraphics[width=1\textwidth]{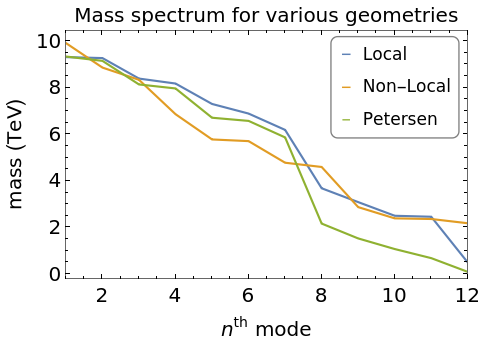}
    \caption{}
    \label{fig:strong}
\end{subfigure}
\caption{Mass spectrum of heavy modes (in TeV on the vertical axis, mode index on the horizontal axis) for three theory spaces: local, non-local and Petersen. (a) Uniform scenario ($\epsilon_i = 2W$), (b) weak site disorder ($\epsilon_i \in [2W-t, 2W+t]$), (c) large site disorder ($\epsilon_i \in [-2W, 2W]$) with $t_i = t$, $W = 5$ TeV, $t = 0.2$ TeV, $b = 2$, $N = 12$.}
\label{mass_uniform_weak_site}
\end{figure}

The heavy spectrum depends on the site masses, link strengths
and graph connectivity. When the site masses are nearly uniform,
the links determine the splittings around their common value,
making differences between geometries easier to see. Strong site
disorder instead makes the random diagonal entries the main
source of spectral variation, which can obscure these geometric
features without eliminating the effect of the links. For the
real symmetric matrices considered here, the physical heavy
masses are the absolute values of the eigenvalues. This geometric effect on the mass spectrum is preserved in weak disorder case and is numerically demonstrated in Fig.~\ref{mass_uniform_weak_site} with parameters used for the Hamiltonian as $W$ = 5 TeV, $t$ = 0.2 TeV, $b$ = 2 and $N$ = 12 for site (diagonal) disorder in both weak and strong cases. The other parameters and range of randomness used are mentioned in the caption of the figure. For each run, weak disorder induces only small perturbations in the spectrum, whereas strong disorder leads to large spectral variations. The plots shown are for one representative run.

\section{Neutrino Masses}
\label{sec:neutrino}

\subsection{Localization Models}
In the Majorana scenario, tiny neutrino masses are generated by the localisation mechanism of the ACS construction, supplemented by additional Majorana mass terms as in Ref.~\cite{craig2018exponential}. The same framework also applies to Dirac neutrinos \footnote{In the Dirac construction we impose lepton-number conservation
to forbid Majorana masses. This assumption is not imposed in
the Majorana construction.}. The generalized ACS Lagrangian is given by 
\begin{align}
    \mathcal{L}_{ACS}  = \mathcal{L}_{kin}  - \sum_{i,j=1}^{N} \overline{L_{i}}\mathcal{H}_{i,j}R_j  + h.c. \label{ACS}
\end{align}
where $\mathcal{H}_{ij}$ is the Hamiltonian for one of the three geometries: (i) the local chain eq.~\eqref{local-hamiltonian}, (ii) the non-local graph eq.~\eqref{nonlocal-hamiltonian}, or (iii) the Petersen graph eq.~\eqref{petersen-Hamiltonian}.
This leads to localized modes in the ACS model. Now we need to connect it to neutrino masses. Similar to extra-dimensional and clockwork frameworks, we take $L_L$ (left handed leptonic doublets) and $\nu_R$ (right handed neutrino singlets) to contain the Standard Model left-handed neutrino and an
additional right-handed neutrino singlet, respectively and couple them to the theory-space fermions at the two endpoints of the lattice so that the overlap between them is minimal.

For one generation, write the mass matrix after symmetry breaking in the block form
\begin{equation}
 \mathcal M_D=\begin{pmatrix}0&a\,e_p^T\\ b\,e_q&\mathcal H\end{pmatrix},
 \qquad m_D^{\rm eff}=-ab(\mathcal H^{-1})_{pq},
 \label{interaction}
\end{equation}
where $a,b$ are boundary masses, $e_p,e_q$ select the coupled sites, and the effective expression is valid when boundary mixing is perturbative and the heavy block is invertible. For a real symmetric $\mathcal H$, let
$O^T\mathcal H O=\operatorname{diag}(\lambda_k)$ with signed eigenvalues. Then
\begin{equation}
 (\mathcal H^{-1})_{pq}=\sum_k\frac{O_{pk}O_{qk}}{\lambda_k}.
 \label{eq:boundary-resolvent}
\end{equation}
Here $\lambda_k$ are signed eigenvalues, while the physical Dirac masses are singular values. With modes centred at $i_k$ and a common localisation-length estimate,
\begin{equation}
 |O_{pk}O_{qk}|\propto
 \exp[-(|p-i_k|+|q-i_k|)/L_{\rm loc}].
 \label{eq:xi-overlap}
\end{equation}
For chain endpoints the exponent is $-(N-1)/L_{\rm loc}$. Equation~\eqref{eq:boundary-resolvent} also shows why a near-zero heavy eigenvalue or a resonant path requires separate treatment. To quantify the spatial concentration of a normalised heavy mode,
we introduce its participation number,
\begin{equation}
    P_k \equiv
    \left(\sum_{i=1}^{N}|O_{ik}|^4\right)^{-1},
    \qquad 1 \leq P_k \leq N .
    \label{eq:participation-number}
\end{equation}
A mode confined to a single site has $P_k=1$, whereas a mode
distributed equally over all sites has $P_k=N$. More generally,
$P_k$ measures the effective number of occupied sites.
Thus, $P_k\ll N$ indicates spatial concentration, although
suppression of the light mass additionally depends on the
endpoint overlaps and inverse eigenvalues entering the boundary
propagator. For exactly degenerate eigenvalues, individual
participation numbers depend on the choice of orthonormal
basis within the degenerate eigenspace.

\subsection{GIM-like cancellation and the graph-path expansion}
\label{sec:gim-mechanism}
Write $\mathcal H=\bar\epsilon\mathbb 1+T$, where $T$ includes hopping and diagonal fluctuations. If $\|T\|_2<|\bar\epsilon|$, the inverse has the convergent expansion
\begin{equation}
 (\mathcal H^{-1})_{pq}=\frac{\delta_{pq}}{\bar\epsilon}
 +\frac1{\bar\epsilon}\sum_{r=1}^{\infty}
 \frac{(-1)^r(T^r)_{pq}}{\bar\epsilon^r}.
 \label{eq:resolvent-paths}
\end{equation}
For distinct boundary sites, the zeroth-order term vanishes because
$\sum_k O_{pk}O_{qk}=0$. This identity is exact in every disorder realisation. Disorder does not break orthogonality. The residual comes from unequal inverse eigenvalues. This is the sense in which the suppression is GIM-like~\cite{singh2025revisiting}.
The path expansion gives a stronger explanation than spectral near-degeneracy alone. If the shortest nonzero hopping path from $p$ to $q$ has $d$ edges, then $(T^r)_{pq}=0$ for $r<d$. Diagonal disorder cannot shorten that path. Thus the leading boundary propagator is of order $t^d/\bar\epsilon^{d+1}$, multiplied by the weighted sum of shortest paths. For a uniform open chain with endpoints $p=1,q=N$,
\begin{equation}
 |m_D^{\rm eff}|\simeq\frac{|ab|}{|\bar\epsilon|}
 \left|\frac{t}{\bar\epsilon}\right|^{N-1}.
 \label{eq:gim-chain-scale}
\end{equation}
Longer paths give the higher-order corrections. Extra edges can shorten the leading path and weaken suppression, even for comparable spectral widths. Graph weights and boundary placement are therefore physical inputs, not incidental details of a mass-matrix example.
To quantify destructive cancellation in the boundary propagator,
we define the cancellation ratio
\begin{equation}
    \mathcal{C}_{pq}
    \equiv
    \frac{\left|\sum_k a_k^{(pq)}\right|}
         {\sum_k\left|a_k^{(pq)}\right|},
    \qquad
    a_k^{(pq)} \equiv \frac{O_{pk}O_{qk}}{\lambda_k},
    \label{eq:cancellation-ratio}
\end{equation}
where the signed eigenvalues $\lambda_k$ are retained and the
denominator is assumed nonzero. By the triangle inequality,
$0\leq\mathcal{C}_{pq}\leq1$: values close to unity indicate
little destructive cancellation, whereas
$\mathcal{C}_{pq}\ll1$ indicates that the net propagator is
much smaller than the sum of the absolute mode contributions.
Indeed,
\begin{equation}
    \left|(H^{-1})_{pq}\right|
    = \mathcal{C}_{pq}\sum_k\left|a_k^{(pq)}\right|.
    \label{eq:propagator-suppression-decomposition}
\end{equation}
Together with the participation numbers, this separates
spatial concentration of the contributing modes from
cancellation among their contributions. Both effects can
coexist in a single realisation. A small cancellation ratio
does not by itself imply fine-tuning, since orthogonality
and graph structure can enforce the suppression.

For $a=b=0.174$~TeV, $\bar\epsilon=10$~TeV, $t=0.5$~TeV and $N=8$, direct inversion gives $|m_D^{\rm eff}|\simeq2.41$~eV. With $t=0.3$~TeV and the other parameters unchanged, the value is approximately $0.0666$~eV. These are one-generation examples.
We define $\bar\epsilon=\langle\epsilon_i\rangle$, $\sigma_\epsilon^2=\operatorname{Var}(\epsilon_i)$ and $\sigma_t^2=\operatorname{Var}(t_{ij})$. For $\epsilon_i\sim\mathcal U(-2W,2W)$, $\sigma_\epsilon=2W/\sqrt3$. For $\epsilon_i\sim\mathcal U(W-t,W+t)$, $\bar\epsilon=W$ and $\sigma_\epsilon=t/\sqrt3$. Similarly, zero-mean hoppings of width $t\ll\bar\epsilon$ can give a quasi-degenerate tower.

To compare the mechanisms across geometries and disorder
regimes, Table~\ref{tab:tower-suppression-diagnostics}
summarises the single-generation tower diagnostics.
In addition to the participation numbers defined above,
we use
\begin{equation}
    \overline P = \frac{1}{N}\sum_k P_k,
    \qquad
    P_{\rm end}
    = \frac{\sum_k |a_k^{(pq)}|P_k}
           {\sum_k |a_k^{(pq)}|}.
    \label{eq:averaged-participation}
\end{equation}
The first averages over all tower modes, whereas the second
weights modes by the magnitudes of their contributions to
the endpoint propagator before cancellation.

\begin{table*}[t]
\centering
\caption{
Single-generation tower diagnostics for $N=8$ and coupled
sites $(p,q)=(1,8)$. Entries give the median and the
16th--84th percentile interval in brackets, over $10^3$
realisations per case.
The site and raw hopping distributions, in TeV, are:
strong site, $\epsilon_i\sim U[-10,10]$, $t_{ij}=0.2$.
Weak site, $\epsilon_i\sim U[9.8,10.2]$, $t_{ij}=0.2$.
Strong hopping, $\epsilon_i=0.25$, $t_{ij}\sim U[-10,10]$.
Weak hopping, $\epsilon_i=10$, $t_{ij}\sim U[-0.1,0.1]$.
Local geometry retains nearest-neighbour links only.
For Non-local and Petersen-type geometries, allowed links
carry the factor $2^{-|i-j|}$ in the specified site labelling.
The same random draws are used for all three geometries.
}
\label{tab:tower-suppression-diagnostics}

\renewcommand{\arraystretch}{1.15}
\setlength{\tabcolsep}{4pt}
\resizebox{\textwidth}{!}{%
\begin{tabular}{@{}llcccc@{}}
\hline
\textbf{Regime}
& \textbf{Geometry}
& $\boldsymbol{\overline P}$
& $\boldsymbol{P_{\rm end}}$
& $\boldsymbol{\mathcal C_{pq}}$
& $\boldsymbol{|G_{pq}|}$ \textbf{(TeV$^{-1}$)} \\
\hline

Strong site
& Local
& $1.034\,[1.008,1.178]$
& $1.032\,[1.007,1.241]$
& $0.122\,[0.00721,0.532]$
& $(2.56\,[0.230,68.4])\times10^{-10}$ \\

& Non-local
& $1.023\,[1.006,1.142]$
& $1.012\,[1.003,1.093]$
& $0.426\,[0.0894,0.856]$
& $(8.88\,[2.87,46.8])\times10^{-5}$ \\

& Petersen-type
& $1.003\,[1.001,1.034]$
& $1.002\,[1.001,1.027]$
& $0.365\,[0.0579,0.791]$
& $(4.56\,[1.28,27.7])\times10^{-6}$ \\
\hline

Weak site
& Local
& $4.543\,[4.151,5.010]$
& $4.848\,[4.352,5.262]$
& $(1.69\,[1.47,1.97])\times10^{-12}$
& $(1.29\,[1.24,1.32])\times10^{-13}$ \\

& Non-local
& $3.150\,[2.811,3.566]$
& $4.036\,[3.470,4.514]$
& $(4.61\,[2.23,9.90])\times10^{-4}$
& $(1.38\,[1.36,1.41])\times10^{-5}$ \\

& Petersen-type
& $2.160\,[1.849,2.509]$
& $2.324\,[1.878,2.857]$
& $(1.71\,[0.843,3.74])\times10^{-5}$
& $(3.07\,[3.00,3.13])\times10^{-7}$ \\
\hline

Strong hopping
& Local
& $3.318\,[2.715,4.187]$
& $3.261\,[2.612,4.356]$
& $0.656\,[0.208,0.949]$
& $0.187\,[0.0211,1.31]$ \\

& Non-local
& $3.650\,[3.215,4.136]$
& $3.313\,[2.413,4.213]$
& $0.589\,[0.204,0.886]$
& $0.183\,[0.0379,0.800]$ \\

& Petersen-type
& $2.770\,[2.395,3.304]$
& $2.806\,[2.220,3.674]$
& $0.500\,[0.147,0.866]$
& $0.114\,[0.0275,0.677]$ \\
\hline

Weak hopping
& Local
& $3.318\,[2.715,4.187]$
& $3.323\,[2.671,4.331]$
& $(4.58\,[0.397,27.0])\times10^{-17}$
& $(1.36\,[0.0605,10.9])\times10^{-18}$ \\

& Non-local
& $3.650\,[3.215,4.136]$
& $3.714\,[3.028,4.412]$
& $(8.70\,[2.81,16.4])\times10^{-5}$
& $(4.01\,[1.34,6.33])\times10^{-6}$ \\

& Petersen-type
& $2.770\,[2.395,3.304]$
& $2.835\,[2.362,3.638]$
& $(5.31\,[1.07,15.4])\times10^{-7}$
& $(1.43\,[0.287,3.88])\times10^{-8}$ \\
\hline
\end{tabular}%
}
\end{table*}

Table~\ref{tab:tower-suppression-diagnostics} distinguishes
spatial concentration from cancellation in endpoint propagation.
For the strong-site benchmarks, both participation measures
are close to unity, indicating strongly concentrated modes.
Cancellation also contributes, particularly in Local geometry.
The weak-site and weak-hopping benchmarks have larger
participation numbers but much smaller cancellation ratios.
Strong hopping instead gives median cancellation ratios
between $0.50$ and $0.66$ and substantially larger propagators.
These results illustrate two complementary suppression mechanisms.

The identical $\overline P$ statistics in the strong- and
weak-hopping ensembles follow from the matched random draws.
For each geometry and realisation,
\begin{equation}
H_{\rm weak\ hopping}
=0.01\,H_{\rm strong\ hopping}
 +(9.9975\,{\rm TeV})\,I.
\end{equation}
The eigenvectors and individual participation numbers are
therefore unchanged, while the eigenvalue denominators alter
the endpoint weights and their cancellation.
This provides an explicit example of comparable spatial
participation coexisting with very different propagator
suppression.

\subsection{Extension to three generations} \label{sec:gauge}
\subsubsection{Dirac Case}
Let $\mathbb H$ be the full $3N\times3N$ Dirac tower mass matrix, including the graph and all flavour couplings. Its entries may be written
\begin{equation}
 \mathbb H_{(i\alpha)(j\beta)}
 =Y_{\rm Ham}^{\alpha\beta}\mathcal H_{ij}^{\alpha\beta},
 \label{Dirac-Lagrangian-3Flavour}
\end{equation}
With three light families the complete broken-phase mass matrix is
\begin{equation}
 \mathcal M_D=\begin{pmatrix}0_{3\times3}&B_L\\B_R&\mathbb H\end{pmatrix},
 \qquad m_D^{\rm eff}=-B_L\mathbb H^{-1}B_R.
 \label{DiracInteraction-Lagrangian-3Flavour}
\end{equation}
Here $B_L$ is $3\times3N$, $B_R$ is $3N\times3$, and both have mass dimension one. For endpoint couplings, $(B_L)_{a,(i\alpha)}=a_{a\alpha}\delta_{i1}$ and $(B_R)_{(j\beta),b}=b_{\beta b}\delta_{jN}$. In a charged-lepton mass basis, the left singular vectors of $m_D^{\rm eff}$ determine the leading light mixing matrix. Otherwise $U_{\rm PMNS}=U_{eL}^\dagger U_{\nu L}$.

A real matrix need not be symmetric: the displayed $Y_{\rm Ham}$ benchmarks are nonsymmetric. Consequently the general Dirac calculation requires a singular-value decomposition of the full matrix. The single orthogonal eigenvector formula in Eq.~\eqref{eq:boundary-resolvent} applies only to a symmetric one-generation block. If $\mathbb H=H_G\otimes F$ and the endpoints factorise in flavour, the effective matrix is proportional to $Y_LF^{-1}Y_R$, with proportionality factor $(H_G^{-1})_{pq}$. In that special case geometry changes the overall mass scale but not the mixing angles. Nonfactorising flavour disorder or different boundary supports are necessary for an independent geometric effect on mixing.

Gauge invariance must also be checked before interpreting the broken-phase blocks as a complete model. A singlet tower admits $\bar L_{\rm SM}\widetilde H S_R$ at one end and a singlet mass mixing $\bar S_L\nu_R$ at the other. Exact lepton number forbids Majorana masses in the Dirac case. If both boundary masses are instead taken to arise from the Standard Model Higgs, one end requires a singlet and the other a doublet. Two such towers do not communicate merely by assigning these representations: an explicit gauge-invariant Higgs-mediated connector is needed, and it changes the full heavy block and its matching. The two-Higgs-boundary benchmarks are therefore broken-phase mass-matrix examples until that connector and its parameters are specified. A singlet completion with a gauge-singlet scalar generating the second boundary mass is another possibility, but its scalar sector is an additional model input.

\subsubsection{Majorana Case}

For the Majorana scenario, we consider the Dirac-like Theory space with Majorana neutrinos $\Psi$ motivated from the Lagrangian considered by Craig et. al. in \cite{craig2018exponential}. 
\begin{align}
\mathcal{L}_{} = \mathcal{L}_{kin} -g_\Psi\,\bar L_1\Psi - \sum_{i,j=1}^{N} \overline{L_{i}}\mathcal{H}_{i,j}R_j - \tfrac12 W \Psi\Psi + h.c. \label{craig}
\end{align}
Here $g_\Psi$ denotes the coupling to the Majorana field $\Psi$.
It is independent of the hopping parameters entering $H_{i,j}$, although in some benchmarks it is set equal to the hopping prefactor. We generalise their Lagrangian with a theory space Hamiltonian $\mathcal{H}_{i,j}$ for various geometries as before.
In this scenario, too, the Hamiltonian captures the geometry of theory space. The fermionic fields of the theory space $L_i, R_i$ still have the Dirac nature. The Majorana nature is produced from the couplings with the $\Psi$ fields.
The model can be trivially extended to the three-flavour case to account for three hierarchical masses corresponding to three generations of the SM. The full Lagrangian for the three-flavour case is given by
\begin{align}
    \mathcal{L}_{} = \mathcal{L}_{kin} -t^{\alpha,\beta} \Bar{L_{1}^{\alpha}}\Psi^{\beta} - \sum_{i,j=1}^{N} Y^{\alpha,\beta}_{Ham}\overline{L_{i}^{\alpha}}\mathcal{H}_{i,j}^{\alpha,\beta}R_j^{\beta} - \tfrac12 W^{\alpha \beta} \Psi^{\alpha}\Psi^{\beta} + h.c. \label{Majorana-Lagrangian-3flavour}
\end{align}
{ with $W^{\alpha,\beta}$ denoting a symmetric Majorana mass matrix.\footnote{A Majorana bilinear depends only on its symmetric part. If an unsymmetrised benchmark is used, its effective coefficient is $(W+W^T)/2$, with the conventional factor $1/2$ in the Majorana mass Lagrangian.}} The matrix $t^{\alpha,\beta}$ describes the coupling to the
Majorana fields. In the benchmarks considered here it is flavour
diagonal, $t^{\alpha,\beta}=g_\Psi\delta_{\alpha\beta}$. Here, the Majorana field $\Psi^{\beta}$ is coupled to the first mode of left-handed fermion of the $\alpha$ flavour $L_1^{\alpha}$. With
 \begin{align}
     \mathcal{H}_{i,j}^{\alpha,\beta} = \epsilon_i^{\alpha,\beta}\delta_{i,j}
+
h_i^{\alpha,\beta}\delta_{i+1,j}
+
h_j^{\alpha,\beta}\delta_{i,j+1} \label{local-hamiltonian}
 \end{align}
for the nearest-neighbour theory space, where
$h_i^{\alpha,\beta}$ denotes the hopping on the link connecting
sites $i$ and $i+1$ within the corresponding tower block.
These graph couplings are distinct from the boundary couplings
$t^{\alpha,\beta}$. For the site-disorder benchmarks,
$h_i^{\alpha,\beta}=t$, whereas for the hopping-disorder
benchmarks they are sampled from the stated interval $[-t,t]$. The interaction term between SM neutrinos of different flavours and new right-handed chiral fermions is given by:
\begin{align}
     \mathcal{L}_{Int.} = -  \Bar{L_L^{\alpha}}\tilde{H}R_N^{\alpha} + h.c. \label{Maj-interaction}
\end{align}
$L_L^{\alpha}$ is the ${\alpha}^{th}$ generation SM lepton doublet. Here, neutrinos are coupled to the last mode of right-handed fermions of each flavour. Depending on different sets of assumptions on the parameters $t^{\alpha,\beta}$, $\epsilon_i^{\alpha,\beta}$,
$h_i^{\alpha,\beta}$ and $W^{\alpha,\beta}$, one can find different sets of parameters leading to neutrinos of eV masses. 
Furthermore, we can introduce mixing among flavours by incorporating non-zero flavour-violating $W^{\alpha, \beta}$ couplings among Majorana neutrinos or by introducing a non-diagonal flavour space Hamiltonian mixing $Y^{\alpha,\beta}_{Ham}$ similar to the Dirac scenario. 

 The Lagrangian in 
Eqs.~\eqref{Majorana-Lagrangian-3flavour} and 
\eqref{Maj-interaction} can be promoted to a form invariant under 
the Standard Model (SM) gauge symmetry 
\(
SU(3)_c \times SU(2)_L \times U(1)_Y
\)
by assigning appropriate transformation properties to the fields. 
We take the vector-like fields $L_i$ and $R_i$ to transform as singlets 
under the SM gauge group. The Majorana field $\Psi$ is also assumed 
to be a complete SM singlet. With these charge assignments, the mass 
terms in Eq.~\eqref{Majorana-Lagrangian-3flavour} are trivially 
gauge invariant.
The interaction term in Eq.~\eqref{Maj-interaction} is constructed 
to respect SM gauge invariance. The SM lepton doublet $L_L$ and the 
SM Higgs doublet transform as $SU(2)_L$ doublets, while the right-handed 
field $R_N$ is taken to be an $SU(2)_L$ singlet. The contraction of 
$SU(2)_L$ indices therefore proceeds in the standard manner, ensuring 
invariance under the full SM gauge group.

Consequently, the resulting Majorana Lagrangian is fully consistent 
with SM gauge symmetry and remains valid for arbitrary choices of 
the underlying theory-space graph or Hamiltonian structure.

\enlargethispage{\baselineskip}
For numerical matching it is useful to write all neutral fields as left-handed Weyl fields. If $\mathbb M_H$ is their symmetric heavy Majorana block and $B$ their coupling to active neutrinos, then
\begin{equation}
 \mathcal M_M=\begin{pmatrix}0&B\\B^T&\mathbb M_H\end{pmatrix},
 \qquad m_M^{\rm eff}=-B\mathbb M_H^{-1}B^T.
 \label{eq:majorana-schur}
\end{equation}
Physical masses and mixing follow from a Takagi factorisation, or from the real-symmetric limit with the necessary sign phases. The heavy block includes the $\Psi$ fields.
The shared tower propagator also permits a direct comparison
of the one-generation Dirac and Majorana suppression.
Writing the dimensionful boundary mass entries as
$g_L$, $g_R$ and $g_\Psi$, the leading effective masses are
\begin{equation}
m_D^{\rm eff}=|g_Lg_R G_{pq}|,\qquad
m_M^{\rm eff}
=\frac{|g_Lg_\Psi G_{pq}|^2}{|M_\Psi|}.
\end{equation}
For matched entries
$g_L=g_R=g_\Psi=0.173\,{\rm TeV}$ and
$M_\Psi=10\,{\rm TeV}$, this gives
$m_M^{\rm eff}=(m_D^{\rm eff})^2/|M_\Psi|$
for each realisation.
For example, the weak-site Non-local benchmark gives median
effective masses of $4.14\times10^5\,{\rm eV}$ and
$1.72\times10^{-2}\,{\rm eV}$, respectively.
These illustrate the additional Majorana suppression.

\section{Robustness, statistical interpretation and limitations}
\label{sec:robust}
The light masses are set by strongly suppressed elements of the inverse heavy mass matrix, and the mixing results depend on the chosen flavour input. We therefore discuss whether radiative corrections preserve the suppression, how the mixing distributions should be compared with neutrino anarchy, and which constraints apply to the active--heavy mixing.\\
A correction $\delta\mathbb H$ to the heavy block changes its inverse as
\begin{equation}
 \delta(\mathbb H^{-1})=-\mathbb H^{-1}\delta\mathbb H\,\mathbb H^{-1}
 +O(\delta\mathbb H^2).
 \label{eq:rob:resolvent}
\end{equation}
In the large-diagonal regime described by
Eq.~\eqref{eq:resolvent-paths}, the leading boundary propagator
depends on the shortest connecting paths and their weights.
An additional connection can therefore change a small
$G_{pq}$ appreciably, even when the correction is small
compared with the heavy masses. Symmetries constrain the allowed corrections, but identifying
which terms are generated requires the interactions of the
underlying theory. For uniform site and link parameters, graph
symmetries respected by the full theory and its vacuum relate
corrections to symmetry-equivalent links. Generic disorder can
break these relations, while radiative corrections remain
loop-suppressed for perturbative couplings and moderate
logarithms of scale ratios.

A common multiplicative renormalisation
$\mathbb H\mapsto z\mathbb H$ preserves the heavy eigenvectors
and the ratios $t/\bar\epsilon$, while rescaling the boundary
propagator as $G_{pq}\mapsto G_{pq}/z$. Such a correction
therefore preserves the pattern of suppression.
Non-universal corrections can instead change the mode profiles
and the cancellations between their contributions. Their
effect on a small boundary propagator can consequently be
larger than their relative effect on the heavy masses.
The form and magnitude of these corrections depend on the
interactions of the link fields and the boundary sector,
which are not fully specified by the mass matrices studied
here. A quantitative assessment of radiative stability
therefore requires an explicit completion and is beyond
the present analysis. Below the heavy thresholds, Standard Model running from
the TeV scale to the electroweak scale produces small
changes in the mixing angles for hierarchical light-neutrino
spectra, owing to the small charged-lepton Yukawa couplings.
Flavour-independent running can rescale the overall mass
scale without changing the mixing angles.
Near-degenerate neutrino masses can enhance the
flavour-dependent effects~\cite{Antuschetal2005,ChankowskiPokorski2002}.\\
The site and link distributions are inputs of the effective theory. Landscape ensembles motivate a statistical treatment of couplings~\cite{douglas2003statistics,ashok2004counting,Balasubramanian:2008tz} but do not fix these distributions. The suppression itself does not rely on tuning independent parameters, and instead follows from the graph structure together with localisation or a power of the modest ratio $t/\bar\epsilon$. Establishing technical naturalness~\cite{tHooft:1979rat,Barbieri:1987fn} and computing the corrections above requires an explicit ultraviolet completion of the link-field sector, including the corrections to the boundary couplings.

Neutrino anarchy uses the Haar measure on $U(3)$ as an
invariant reference for leptonic mixing~\cite{hall2000neutrino,Haba:2000be}.
For the real mass matrices considered here, the corresponding
reference is $O(3)$. Random site or hopping parameters do not
by themselves imply Haar-distributed mixing, and a broad
angle histogram need not follow this measure.
The distributions in Secs.~\ref{sec:strong} and~\ref{sec:weak}
show how the mixing angles respond to disorder for fixed
benchmark flavour matrices. Concentrated distributions
indicate that the angles remain close across disorder
realisations. Whether those angles agree with observations
also depends on the chosen flavour structure.
In particular, the weak-site flavour matrices were fitted
to the observed angles, so their agreement does not provide
an independent statistical test. A quantitative comparison
with anarchy would require specified flavour priors and
consistent sampling. We do not perform that comparison here.

Finally, viable light masses do not guarantee viable active--heavy mixing. In the Dirac block convention the mixing is $\Theta=B_L\mathbb H^{-1}$, and the leading nonunitarity of the light mixing matrix is $\eta=\Theta\Theta^\dagger/2$. Unlike the endpoint propagator $G_{pq}$, active--heavy
mixing need not inherit suppression between distant sites.
When the inverse heavy block is dominated by its diagonal,
$|\Theta|\simeq |a/\epsilon_p|$. For $a=174\,\mathrm{GeV}$,
this gives $\eta\simeq1.5\times10^{-4}$ at
$|\epsilon_p|=10\,\mathrm{TeV}$ and
$\eta\simeq1.5\times10^{-2}$ at
$|\epsilon_p|=1\,\mathrm{TeV}$.
For $\epsilon_p$ uniform on $[-10,10]\,\mathrm{TeV}$,
$10\%$ of draws satisfy $|\epsilon_p|<1\,\mathrm{TeV}$.
Near small heavy eigenvalues, the full mass matrix must
be used to assess the mixing. The bounds on $\eta$ are flavour dependent~\cite{Antusch:2006vwa}, and lepton-flavour-violating decays such as $\mu\to e\gamma$, whose amplitude contains $\sum_k\Theta_{\mu k}\Theta_{ek}^*F(M_k^2/M_W^2)$, constrain the fourth power of the mixing. These constraints must be imposed on the same samples used in any quantitative comparison of mixing distributions.

\section{Numerical Analysis and Results: Strong Disorder}
\label{sec:strong}
In the present section, we focus on the numerical analysis and results where we examine illustrative mass scales and mixing distributions. We do not perform a simultaneous likelihood fit to oscillation data. In particular, we consider two cases: (a) Disorder in the site mass terms $\epsilon_i$s (diagonal terms), and (b) disorder in the hopping terms $t_i$s (off-diagonal mass terms). We present a comprehensive analysis of both these cases. The neutrino masses and mixing numbers we consider are mentioned in Appendix \ref{neutrino_numbers}.
Throughout the numerical discussion, ``anarchical'' describes a broad distribution without a stable preferred peak. The term does not imply equality to a Haar measure. Comparisons among geometries use the benchmark parameters, which are not always matched in site number or weighted hopping norm. 

\subsection{Site Disorder (Randomness in diagonal terms)}
The case of site disorder is an interesting one, as it could lead to exponential localization of each mode of the Hamiltonian. Localisation can generate effective neutrino Yukawa couplings
of order $10^{-12}$ from TeV-scale fermions. Stronger
suppression per link can, in principle, reduce the number
of sites needed to reach a given hierarchy compared with
a clockwork chain at a fixed hopping ratio. The advantage
depends on the chosen disorder and coupling parameters. Since the nature of neutrinos, being Dirac or Majorana, is still uncertain from the experiment, in the following sections we have considered both scenarios and studied their impact on the masses and mixing angles produced.
We now present the results for the Petersen geometry. The corresponding results for the other two geometries are given in Appendix~\ref{app:strong_site}.

\subsubsection{Dirac}

\begin{figure}
    \begin{subfigure}{0.48\textwidth}
        \centering
        \includegraphics[width=\linewidth]{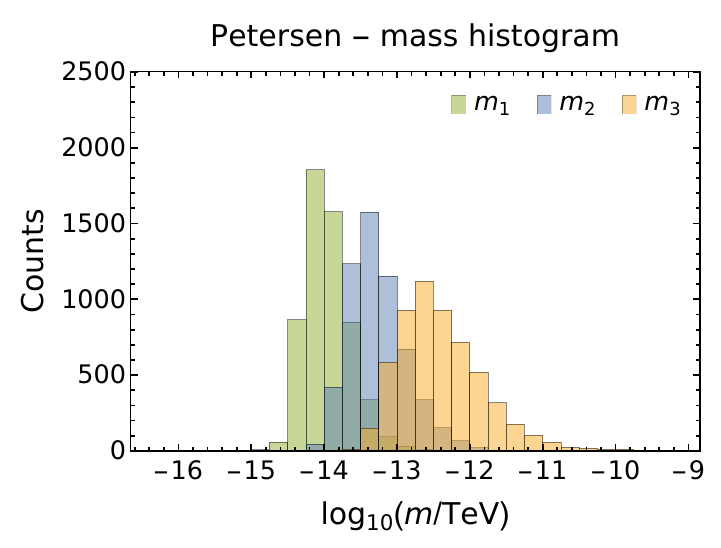}
        \caption{}
        \label{subfig:pet-logmass}
    \end{subfigure}
    \hfill
    \begin{subfigure}{0.48\textwidth}
        \centering
        \includegraphics[width=\linewidth]{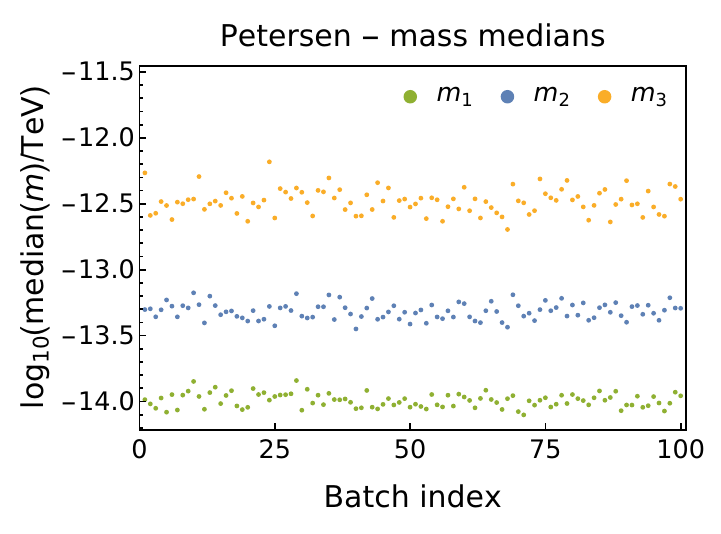}
        \caption{}
        \label{subfig:pet-mass}
    \end{subfigure}
    
    \caption{Histograms of $\log_{10}(m/\mathrm{TeV})$ from $10^4$
random realisations (left) and batch medians of the three
lightest physical Dirac masses, shown on a logarithmic scale,
for 100 batches of 100 realisations (right).
Results are for the boundary-Yukawa mixing channel in the
Petersen geometry with $W=5\,\mathrm{TeV}$,
$b=5$, $N=12$, and $t=0.1\,\mathrm{TeV}$.}
    \label{petersen_masses}
\end{figure}

\begin{figure}
    \begin{subfigure}{0.48\textwidth}
        \centering
        \includegraphics[width=\linewidth]{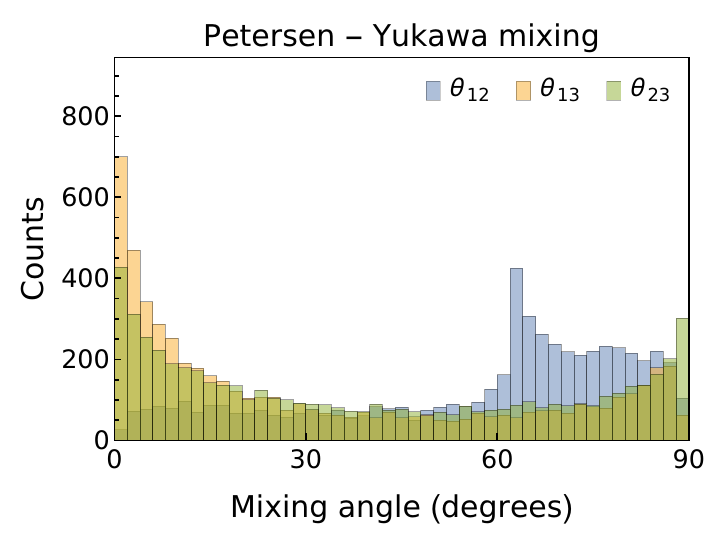}
        \caption{}
        \label{subfig:pet-yukangle}
    \end{subfigure}
    \hfill
    \begin{subfigure}{0.48\textwidth}
        \centering
        \includegraphics[width=\linewidth]{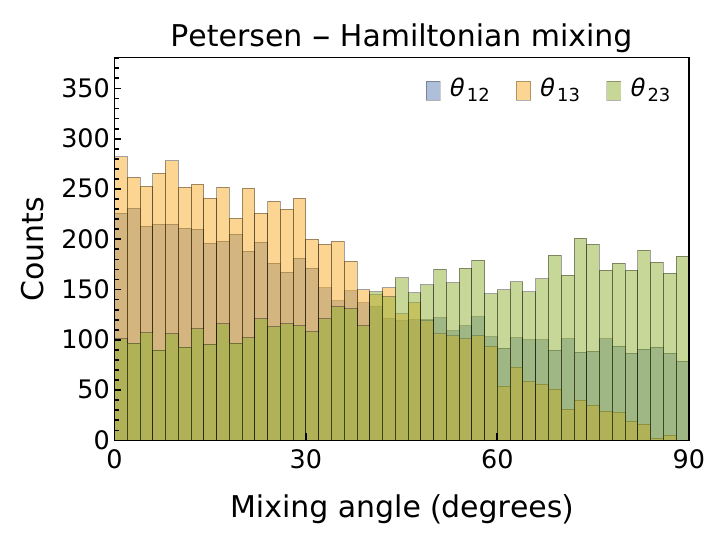}
        \caption{}
        \label{subfig:pet-hamangle}
    \end{subfigure}
    
    \caption{Histogram of PMNS mixing angles over several runs in the Petersen geometry: Yukawa mixing $Y^{\alpha,\beta}_{yuk}$ (left) and Hamiltonian mixing $Y^{\alpha,\beta}_{Ham}$ (right), with matrices as in eq.~\eqref{18}. Parameters: $W = 5\,\text{TeV}$, $b = 5$, $N = 12$, $t = 0.1\,\text{TeV}$.}
    \label{dirac-strong-site-peter}
\end{figure}

\paragraph{\textbf{Petersen}}
The Hamiltonian for this geometry is given in section \ref{sec:recap}.
Strong site disorder suppresses propagation through the Petersen
geometry. Its additional weighted links provide routes absent
from the local chain, allowing a different degree of boundary
suppression. The resulting light-mass scales depend on these
paths and their weights, without requiring a universal ordering
of localisation lengths among the three geometries. We use this
Hamiltonian in the three-generation Dirac construction below. For three flavours, we analyze a scenario with $N=12$ and a wide randomness interval, $\epsilon_i \in [-2W, 2W]$, alongside fixed $t_i = t$ where $W = 5\,\text{TeV}$, $b = 5$, and $t = 0.1\,\text{TeV}$. The resultant neutrino masses are shown in Fig.~\ref{petersen_masses}. We employ both the flavour-diagonal left-handed SM neutrino Yukawa couplings and non-diagonal right-handed neutrino Yukawa couplings, $Y_{yuk}^{\alpha,\beta}$. Specifically, $Y_{yuk}^{\alpha,\beta}$ is constructed as a unique realization of a random $O(1)$ $3 \times 3$ matrix, and its induced mixing is visualised in the left panel of Fig.~\ref{dirac-strong-site-peter}. Conversely, the right panel of Fig.~\ref{dirac-strong-site-peter} displays the scenario where diagonal $Y_{yuk}^{\alpha,\beta}$ couplings are used with off-diagonal flavour matrices $Y^{\alpha,\beta}_{Ham}$. It should be noted that we do not consider either $Y_{yuk}^{\alpha,\beta}$ or $Y_{Ham}^{\alpha,\beta}$ to be anarchical in Nature. For numerical concreteness, we chose 
\begin{align}
Y_{yuk} &= 
\begin{bmatrix}
  1 & 0.4 & 0.6 \\
  0.3 & 1 & 0.8 \\
  0.9 & 0.3 & 1 \\
\end{bmatrix},
\hspace{2cm}
Y_{Ham} = 
\begin{bmatrix}
  1 & 0.4 & 0.6 \\
  0.3 & 1 & 0.8 \\
  0.9 & 0.3 & 1 \\
\end{bmatrix} \label{18}
\end{align}

Table~\ref{tab:strong-site-dirac-parameters} lists the parameters
used for the strong-site Dirac benchmarks in all three geometries.
The same parameters apply to the boundary-Yukawa and Hamiltonian
mixing channels.
\begin{table}[tbp]
\centering
\small
\setlength{\arrayrulewidth}{0.4pt}
\setlength{\tabcolsep}{6pt}
\renewcommand{\arraystretch}{1.25}
\caption{Parameters for the strong-site Dirac benchmarks in both
flavour-mixing channels. Site masses are sampled uniformly from
the listed interval, while the raw hopping is fixed.
The decay base $b$ applies only to the non-local and Petersen
geometries.}
\label{tab:strong-site-dirac-parameters}
\begin{tabularx}{\linewidth}{
@{}l c c >{\centering\arraybackslash}X c@{}
}
\noalign{\hrule height 0.7pt\vskip 5pt}
\textbf{Geometry}
& $\boldsymbol{N}$
& $\boldsymbol{b}$
& $\boldsymbol{\epsilon_i}$ \textbf{(TeV)}
& $\boldsymbol{t}$ \textbf{(TeV)} \\
\noalign{\vskip 3pt\hrule height 0.4pt\vskip 5pt}
Local     & 12 & --- & $[-10,10]$ & $0.1$ \\
Non-local & 12 & 5   & $[-10,10]$ & $0.1$ \\
Petersen  & 12 & 5   & $[-10,10]$ & $0.1$ \\
\noalign{\vskip 4pt\hrule height 0.7pt}
\end{tabularx}
\end{table}
For the Petersen geometry, the additional links provide more paths
between sites than in the nearest-neighbour chain, although strong
site disorder still suppresses propagation between the boundaries.
The connection to neutrino mixing is captured by
$m_D^{\mathrm{eff}}=-B_LH^{-1}B_R$: the light mass matrix combines
propagation through the heavy sector with the flavour structure
of the boundary couplings. Suppression of the propagation amplitudes
can therefore generate small masses without necessarily making
the mixing angles small.

In the boundary-Yukawa channel, $Y_{Ham}^{\alpha,\beta}$ is diagonal,
so different flavours do not mix within the heavy sector.
Nevertheless, the non-diagonal $Y_{yuk}^{\alpha,\beta}$ couples
these otherwise separate flavour sectors at the boundary.
The relative propagation amplitudes of the flavour sectors vary
across disorder realisations, changing their contributions to the
effective mass matrix. The mixing pattern therefore reflects
both these relative amplitudes and the chosen boundary Yukawa
matrix. For the benchmark in
Fig.~\ref{dirac-strong-site-peter} (left), this produces an interior
peak near $63^\circ$ in $\theta_{12}$, strong enhancement near
$0^\circ$ in $\theta_{13}$, and enhancement near both endpoints
in $\theta_{23}$, together with broad tails. The interior peak
is sensitive to the input Yukawa matrix and is not a universal
prediction of the Petersen geometry.

For non-diagonal $Y_{Ham}^{\alpha,\beta}$, flavours can instead
mix during propagation through the heavy sector, even when the
boundary couplings are flavour diagonal. Disorder then changes
both the propagation amplitudes and their flavour composition.
For the chosen benchmark, the resulting mixing-angle distributions
are broad, as shown in
Fig.~\ref{dirac-strong-site-peter} (right), with
$\theta_{12}$ and $\theta_{13}$ slightly favouring smaller angles and
$\theta_{23}$ slightly favouring larger angles. This broadness does not
by itself establish a Haar-distributed mixing ensemble.
The results of all three geometries are summarised in
Table~\ref{table-strong-site}.
\begin{table}[tbp]
\centering
\small
\setlength{\arrayrulewidth}{0.4pt}
\setlength{\tabcolsep}{6pt}
\renewcommand{\arraystretch}{1.25}
\caption{Strong-site Dirac mixing patterns for the chosen
benchmarks and flavour matrices. Geometry dependence is
visible in the boundary-Yukawa channel and less pronounced
in the Hamiltonian channel. Peak features are input-dependent.
Broad distributions do not imply Haar-distributed mixing.}
\label{table-strong-site}
\begin{tabularx}{\linewidth}{@{}>{\raggedright\arraybackslash}p{0.27\linewidth} *{3}{>{\centering\arraybackslash}X}@{}}
\noalign{\hrule height 0.7pt\vskip 5pt}
\textbf{Flavour channel} & \textbf{Local} & \textbf{Non-local} & \textbf{Petersen-type} \\
\noalign{\vskip 3pt\hrule height 0.4pt\vskip 5pt}
Boundary Yukawa & Strong endpoint enhancement & Broad with an interior peak & Broad with an interior peak \\
\addlinespace[5pt]
Hamiltonian & Broad & Broad & Broad \\
\noalign{\vskip 4pt\hrule height 0.7pt}
\end{tabularx}
\end{table}

\subsubsection{Majorana}

Now we will study the scale of the masses and the mixing angles produced for the same three geometries, for the case of Majorana neutrinos as per the Lagrangian \eqref{craig}. The results of Petersen geometry are described here, and the rest are mentioned in Appendix \ref{app:strong_site}.

\begin{figure}[ht]
    \begin{subfigure}{0.48\textwidth}
        \centering
        \includegraphics[width=\linewidth]{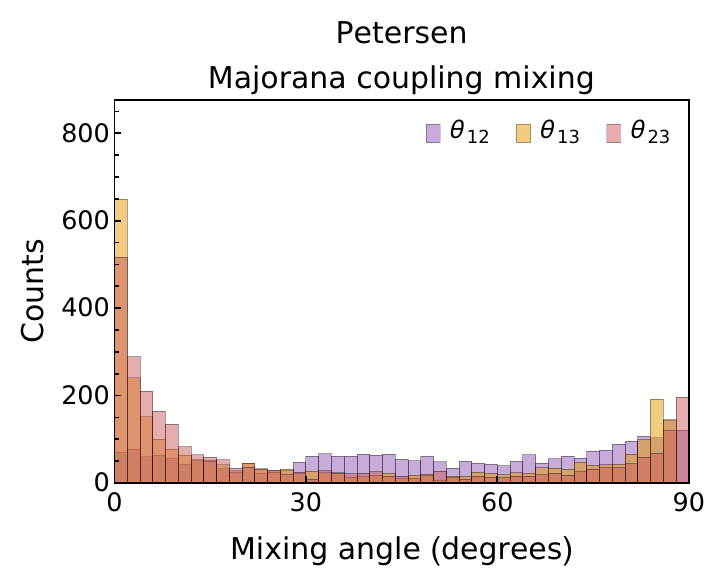}
        \caption{}
        \label{subfig:pet-yukangle-maj}
    \end{subfigure}
    \hfill
    \begin{subfigure}{0.48\textwidth}
        \centering
        \includegraphics[width=\linewidth]{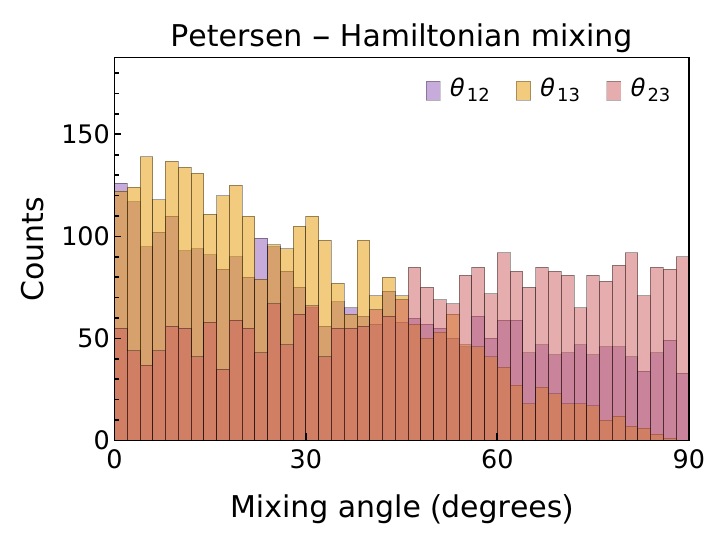}
        \caption{}
        \label{subfig:pet-hamangle-maj}
    \end{subfigure}
    
    \caption{Histogram of PMNS mixing angles in the Petersen geometry (Majorana, strong site disorder): Majorana coupling $W^{\alpha,\beta}
=\frac{W}{2}
\left(Y_{\rm yuk}+Y_{\rm yuk}^{T}\right)_{\alpha\beta}$ (left) and Hamiltonian mixing $Y^{\alpha,\beta}_{Ham}$ (right), with matrices as in eq.~\eqref{18}. Parameters: $\epsilon_i\in[-5,5]\,\mathrm{TeV}$,
$b=5$, $N=8$, and $t=1/3\,\mathrm{TeV}$.}
    \label{maj-strong-site-pet}
\end{figure}

\paragraph{\textbf{Petersen}}
The Lagrangian for this geometry in the Majorana scenario is given by eq.(\ref{craig}) with Hamiltonian eq.(\ref{petersen-Hamiltonian}).
For the Petersen geometry, we use $N=8$, a randomness
interval $\epsilon_i\in[-5,5]\,\mathrm{TeV}$, decay base
$b=5$, and fixed hopping $t=1/3\,\mathrm{TeV}$. The coupling to the Majorana fields in Eq.~\eqref{Majorana-Lagrangian-3flavour} is
$t^{\alpha,\beta}=0.2\,\mathrm{TeV}\,\delta_{\alpha\beta}$. We apply the same two mixing approaches: first, cases with non-diagonal right-handed neutrino Majorana couplings ($W^{\alpha,\beta}$), set at O(1) values, alongside flavour-diagonal left-handed SM couplings Fig.~\ref{maj-strong-site-pet} (left), and second, scenarios featuring diagonal $W^{\alpha,\beta}$ couplings with off-block Hamiltonian mixing ($Y^{\alpha,\beta}_{Ham}$) Fig.~\ref{maj-strong-site-pet} (right).
\begin{figure}[ht]
    \begin{subfigure}{0.48\textwidth}
        \centering
        \includegraphics[width=\linewidth]{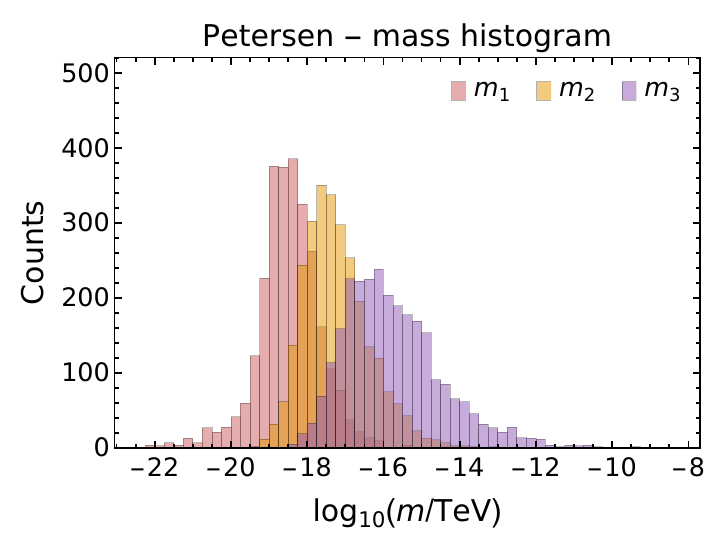}
        \caption{}
        \label{subfig:pet-logmass-maj}
    \end{subfigure}
    \hfill
    \begin{subfigure}{0.48\textwidth}
        \centering
        \includegraphics[width=\linewidth]{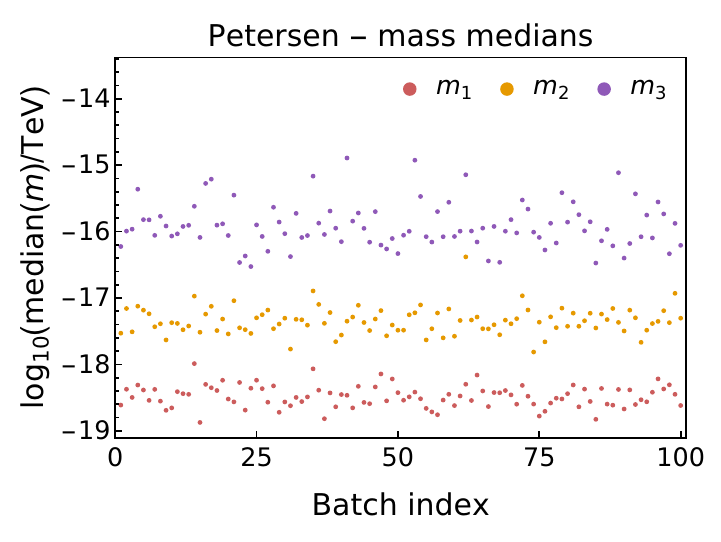}
        \caption{}
        \label{subfig:pet-mass-maj}
    \end{subfigure}
    
    \caption{Histograms of $\log_{10}$(mass/TeV) over various runs (left) and median of the three lightest masses over 100 runs (right) in the Petersen geometry (Majorana case). Parameters: $\epsilon_i\in[-5,5]\,\mathrm{TeV}$,
$b=5$, $N=8$, and $t=1/3\,\mathrm{TeV}$.} \label{neutrino_mass_strongsite_maj}
\end{figure}
In the Majorana-coupling channel, diagonal
$Y_{Ham}^{\alpha,\beta}$ keeps the heavy tower flavours
separate, while the non-diagonal $W^{\alpha,\beta}$
connects them through the Majorana fields.
The effective light mass matrix contains propagation
from each external neutrino to this Majorana sector.
Consequently, the relative propagation amplitudes weight
the Majorana flavour structure on both sides.
Strong disorder can make these amplitudes very unequal,
favouring alignment with particular flavour directions.
Their overall suppression produces small masses, but
does not by itself require small mixing angles. The Majorana construction also introduces seesaw suppression.
The light masses depend quadratically on the suppressed effective
Dirac mass connecting the SM neutrinos to the Majorana sector,
and inversely on the Majorana mass scale.
For the chosen Petersen benchmark, Fig.~\ref{maj-strong-site-pet}(a) shows
endpoint enhancement together with a broad population
of intermediate angles. 
Compared with the local benchmark, the endpoint
concentration is weaker, showing that the geometry
affects the distributions through the propagation
amplitudes. Their detailed shapes also depend on the
chosen Majorana coupling matrix.

In the Hamiltonian channel, non-diagonal
$Y_{Ham}^{\alpha,\beta}$ allows flavours to mix within
the heavy tower itself. Disorder then changes both
the strength and flavour composition of propagation,
even when $W^{\alpha,\beta}$ is diagonal.
For this benchmark, the resulting distributions in
Fig.~\ref{maj-strong-site-pet}(b) are broad, with $\theta_{12}$ and $\theta_{13}$
favouring smaller angles and $\theta_{23}$ favouring
larger angles. Their qualitative similarity across
the three geometries indicates less pronounced
geometry dependence in this channel. Fig.~\ref{neutrino_mass_strongsite_maj} shows the neutrino masses generated in this scenario.

Strong site disorder suppresses propagation between the
boundaries and generates small neutrino masses in these
Majorana benchmarks. However, small propagation amplitudes
do not by themselves imply small mixing angles, which
depend on the relative flavour structure of the effective
mass matrix. The Majorana-coupling channel produces
endpoint-enhanced distributions with non-zero
intermediate-angle populations. Endpoint concentration
is strongest in the local geometry and weaker in the
non-local and Petersen benchmarks. The Hamiltonian
channel produces broad distributions that are
qualitatively similar across the three geometries. Table~\ref{table-maj-site-strong-param} mentions all the parameters used for the numerical analysis of three geometries. The results obtained for the three geometries are summarised in Table~\ref{table-maj-site-strong}.

\begin{table}[tbp]
\centering
\small
\setlength{\arrayrulewidth}{0.4pt}
\setlength{\tabcolsep}{6pt}
\renewcommand{\arraystretch}{1.25}
\caption{Parameters for the strong-site Majorana benchmarks
with $b=5$ and fixed raw hopping $t=1/3\,\mathrm{TeV}$.
The Majorana mass prefactor is $5\,\mathrm{TeV}$, and the
coupling to $\Psi$ is $0.2\,\mathrm{TeV}$.}
\label{table-maj-site-strong-param}
\begin{tabularx}{\linewidth}{@{}>{\raggedright\arraybackslash}p{0.30\linewidth} *{3}{>{\centering\arraybackslash}X}@{}}
\noalign{\hrule height 0.7pt\vskip 5pt}
\textbf{Scenario} & \textbf{N}
& \textbf{$\epsilon_i$ (TeV)}
& \textbf{$t$ (TeV)} \\
\noalign{\vskip 3pt\hrule height 0.4pt\vskip 5pt}
Local & 8 & $[-5,5]$ & $1/3$ \\
\addlinespace[5pt]
Non-local & 8 & $[-5,5]$ & $1/3$ \\
\addlinespace[5pt]
Petersen & 8 & $[-5,5]$ & $1/3$ \\
\noalign{\vskip 4pt\hrule height 0.7pt}
\end{tabularx}
\end{table}
\begin{table}[tbp]
\centering
\small
\setlength{\arrayrulewidth}{0.4pt}
\setlength{\tabcolsep}{6pt}
\renewcommand{\arraystretch}{1.25}
\caption{Strong-site Majorana mixing patterns for the chosen
benchmarks and flavour matrices. Endpoint enhancement is
strongest in the local Majorana-coupling channel.
Geometry dependence is less pronounced in the Hamiltonian
channel. Broad distributions do not imply Haar-distributed mixing.}
\label{table-maj-site-strong}
\begin{tabularx}{\linewidth}{@{}>{\raggedright\arraybackslash}p{0.27\linewidth} *{3}{>{\centering\arraybackslash}X}@{}}
\noalign{\hrule height 0.7pt\vskip 5pt}
\textbf{Mixing Type} & \textbf{Local} & \textbf{Non-local} & \textbf{Petersen} \\
\noalign{\vskip 3pt\hrule height 0.4pt\vskip 5pt}
$W^{\alpha,\beta}$ Majorana Mixing & Strong endpoint enhancement & Endpoint enhancement with broad tails & Endpoint enhancement with broad tails \\
\addlinespace[5pt]
$Y^{\alpha,\beta}_{Ham}$ Hamiltonian Mixing & Broad & Broad & Broad \\
\noalign{\vskip 4pt\hrule height 0.7pt}
\end{tabularx}
\end{table}

\subsection{ Hopping Disorder (Randomness in off-diagonal terms)}
In the strong hopping-disorder regime, the off-diagonal couplings $t_i$ are drawn randomly from the interval $[-t,\,t]$, while the site mass terms satisfy $\epsilon_i \ll t$. 
As discussed above, the displayed finite-system modes do not exhibit sufficient endpoint suppression, and the spectrum is not quasi-degenerate. 
For these benchmarks neither mechanism provides the required controlled suppression. 
These parameter choices do not generate a sufficient hierarchy.

\section{Weak disorder: mass suppression and structured mixing}
\label{sec:weak}

 In the following section, we will consider two scenarios with weak disorder in the diagonal mass terms, i.e. site disorder and weak disorder in the off-diagonal terms, i.e. hopping disorder. The aim is to show explicit examples where weak disorder can lead to structured flavoured mixing angles in the neutrino sector.

\subsection{Site Disorder}

We will again consider two scenarios for the nature of right-handed neutrinos, as Dirac and Majorana, with various geometries and different types of flavour mixings. We analyse the role of three distinct geometries—local, Petersen, and non-local—in shaping the neutrino mass hierarchy and mixing structure. 
In what follows, we focus on the Petersen geometry, while the corresponding results for the local and non-local cases are presented in Appendix~\ref{app:weak_site}.

\subsubsection{Dirac}
In the Dirac scenario, similar to the earlier scenario of large disorder, the Lagrangian and Hamiltonian are kept the same as in eq.(\ref{ACS}), depending on the geometry considered. The three-flavour Lagrangian is again given by eq.(\ref{Dirac-Lagrangian-3Flavour}). For the mixings among the three generations, the flavour Yukawa couplings  $Y^{\alpha,\beta}_{yuk}$ and the flavour Hamiltonian $Y^{\alpha,\beta}_{Ham}$ cases are considered. The mixing angles depend on the flavour structure of the
effective light mass matrix. Weak site fluctuations can preserve
a concentrated mixing pattern when they produce only small
changes in this structure. The distributions below test this
behaviour for the chosen flavour matrices.

\paragraph{\textbf{Petersen}}
In the Petersen geometry with weak site disorder, the Hamiltonian used in the one-flavour Lagrangian eq.(\ref{ACS}) is eq.(\ref{petersen-Hamiltonian}). The distribution of mass spectrum in this scenario is that of the Petersen lattice with a tiny perturbation. This can be clearly distinguished from the local scenario mass spectrum. The uniform-site mode profiles are illustrated in Fig.~\ref{Petersen_spectrum}
of Appendix~\ref{app:petersen}. The distance weights used in the present
benchmark can lift degeneracies of the unweighted graph. The parameters used for the numerical results in this scenario are mentioned in Table~\ref{table-weak-site-dirac}. The site masses are sampled from
$\epsilon_i\in[9.8,10.2]\,\mathrm{TeV}$, with the
site-disorder half-width held fixed at $0.2\,\mathrm{TeV}$.
The hopping is uniform on the allowed links before
geometric weighting, with $t\simeq0.087\,\mathrm{TeV}$
for Yukawa mixing and $t\simeq0.086\,\mathrm{TeV}$
for Hamiltonian mixing. Both benchmarks use $N=16$
and $b=3$.
\begin{figure}[ht]
    \begin{subfigure}{0.34\textwidth}
        \centering
        \includegraphics[width=1\textwidth]{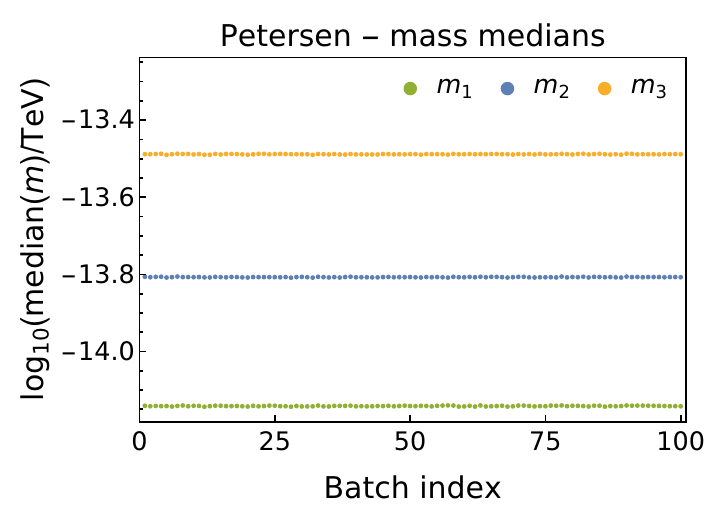}
        \caption{}
        \label{subfig:peter-mass}
    \end{subfigure}
    \hfill
    \begin{subfigure}{0.325\textwidth}
        \centering
        \includegraphics[width=1\textwidth]{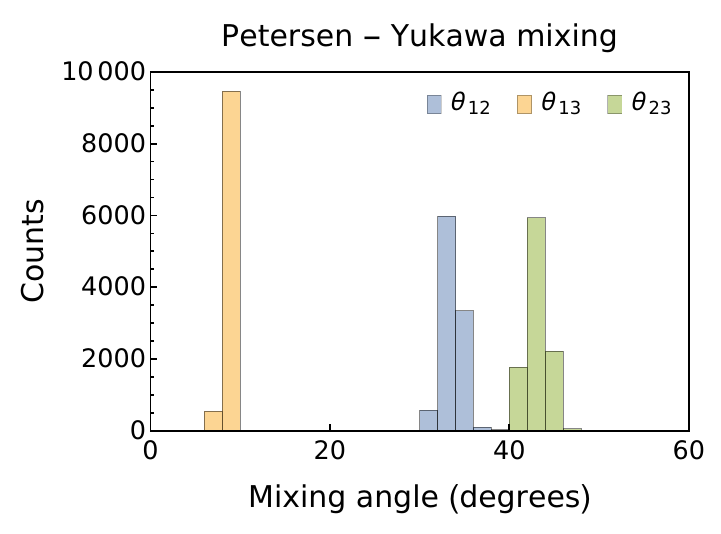}
        \caption{}
        \label{subfig:peter-yukangle}
    \end{subfigure}
    \hfill
    \begin{subfigure}{0.32\textwidth}
        \centering
        \includegraphics[width=1\textwidth]{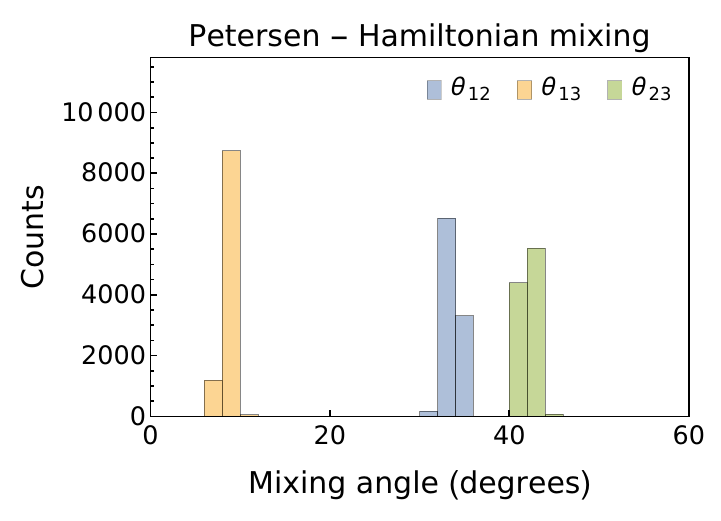}
        \caption{}
        \label{subfig:peter-hamangle}
    \end{subfigure}
    
    \caption{Median of the three lightest masses (in TeV, $\log_{10}$ scale) over 100 runs (left) and histograms of PMNS mixing angles for Yukawa mixing $Y^{\alpha,\beta}_{yuk}$ (middle) and Hamiltonian mixing $Y^{\alpha,\beta}_{Ham}$ (right), with matrices as in eq.~\eqref{31}. Petersen geometry, weak site disorder:
$\epsilon_i\in[9.8,10.2]\,\mathrm{TeV}$,
$b=3$, and $N=16$.
The hopping is $t\simeq0.087\,\mathrm{TeV}$ for Yukawa
mixing and $t\simeq0.086\,\mathrm{TeV}$ for Hamiltonian mixing.}
    \label{petersen-weak-site-dirac}
\end{figure}
Fig.~\ref{petersen-weak-site-dirac} (a) demonstrates the scale of mass produced (left) and the mixing angles generated for both Yukawa $Y_{yuk}^{\alpha,\beta}$ (middle) and Hamiltonian mixing $Y^{\alpha,\beta}_{Ham}$ (right).\\
The narrow mixing-angle distributions can be understood
by considering how weak site disorder perturbs the
effective mass matrix,
$m_D^{\mathrm{eff}}=-B_LH^{-1}B_R$.
The flavour matrices are held fixed while the site
masses fluctuate over a small interval. For these
benchmarks, the fluctuations change the relative
flavour structure only slightly, so the directions
that diagonalise the light mass matrix remain close
across realisations. Both the boundary-Yukawa and
Hamiltonian channels therefore produce concentrated
mixing-angle distributions. Their peak locations
depend on the chosen flavour matrices and propagation
through the heavy sector, rather than being fixed
by weak disorder alone. For concreteness, benchmark $Y_{yuk}$ values and $Y_{Ham}$, we chose are
\begin{align}
Y_{yuk} &\simeq 
\begin{bmatrix}
 0.39 & -0.21 & 0.12 \\
0.53 & 0.56 & -0.60 \\
0.30 & 1.4 & 1.5 \\
\end{bmatrix},
\hspace{2cm}
Y_{Ham} \simeq 
\begin{bmatrix}
1.6 & -1.0 & 0.58 \\
0.62 & 0.52 & -0.59 \\
0.054 & 0.30 & 0.33 \\
\end{bmatrix} \label{31}
\end{align}

These matrices specify the Petersen Dirac benchmarks.
The fitted entries and hopping parameters are displayed
to two significant figures, and the unrounded values are
released with the code.
For the chosen Petersen benchmark,
Fig.~\ref{petersen-weak-site-dirac}(a) shows the small
neutrino mass scales and their stability across batches.
The boundary-Yukawa and Hamiltonian channels produce
narrowly peaked mixing-angle distributions, as shown
in Fig.~\ref{petersen-weak-site-dirac}(b) and
Fig.~\ref{petersen-weak-site-dirac}(c), respectively.
The fixed flavour matrices set the central mixing
pattern together with propagation through the heavy
sector. Weak site disorder introduces only a limited
variation in this pattern for these benchmarks.
This concentration of mixing angles should not be
confused with spatial localisation of the heavy modes.

Table~\ref{table-weak-site-dirac} lists the parameters
used for all three geometries.
Table~\ref{Table-weak-site-summary} summarises their
mixing distributions, showing that narrowly peaked
patterns can be obtained in both flavour channels
for suitable inputs in each geometry.
The peak locations are therefore benchmark-dependent,
rather than universal predictions of weak site disorder.
\begin{table}[tbp]
\centering
\small
\setlength{\arrayrulewidth}{0.4pt}
\setlength{\tabcolsep}{6pt}
\renewcommand{\arraystretch}{1.25}
\caption{Parameters for the weak-site Dirac benchmarks.
The site-disorder half-width is fixed at
$0.2\,\mathrm{TeV}$ independently of the hopping.
The decay base $b$ applies only to the non-local
and Petersen geometries. Hopping values are displayed
to two significant figures.}
\label{table-weak-site-dirac}
\begin{tabularx}{\linewidth}
{@{}llcc>{\centering\arraybackslash}Xc@{}}
\noalign{\hrule height 0.7pt\vskip 5pt}
\textbf{Geometry} & \textbf{Mixing}
& \textbf{N} & \textbf{b}
& \textbf{$\epsilon_i$ (TeV)}
& \textbf{$t$ (TeV)} \\
\noalign{\vskip 3pt\hrule height 0.4pt\vskip 5pt}
Local & Yukawa & 6 & --- & $[9.8,10.2]$ & $\simeq0.057$ \\
\addlinespace[5pt]
Local & Hamiltonian & 9 & --- & $[9.8,10.2]$ & $\simeq0.36$ \\
\addlinespace[5pt]
Non-local & Yukawa & 16 & 4 & $[9.8,10.2]$ & $\simeq0.063$ \\
\addlinespace[5pt]
Non-local & Hamiltonian & 16 & 4 & $[9.8,10.2]$ & $\simeq0.064$ \\
\addlinespace[5pt]
Petersen & Yukawa & 16 & 3 & $[9.8,10.2]$ & $\simeq0.087$ \\
\addlinespace[5pt]
Petersen & Hamiltonian & 16 & 3 & $[9.8,10.2]$ & $\simeq0.086$ \\
\noalign{\vskip 4pt\hrule height 0.7pt}
\end{tabularx}
\end{table}

\begin{table}[tbp]
\centering
\small
\setlength{\arrayrulewidth}{0.4pt}
\setlength{\tabcolsep}{6pt}
\renewcommand{\arraystretch}{1.25}
\caption{Weak-site Dirac mixing patterns for the chosen
benchmarks. Both flavour channels produce narrowly peaked
distributions in all three geometries. Peak locations
depend on the input flavour matrices and model parameters.}
\label{Table-weak-site-summary}
\begin{tabularx}{\linewidth}{@{}>{\raggedright\arraybackslash}p{0.27\linewidth} *{3}{>{\centering\arraybackslash}X}@{}}
\noalign{\hrule height 0.7pt\vskip 5pt}
\textbf{Mixing Type} & \textbf{Local} & \textbf{Non-local} & \textbf{Petersen} \\
\noalign{\vskip 3pt\hrule height 0.4pt\vskip 5pt}
$Y_{yuk}^{\alpha,\beta}$ Yukawa Mixing & Narrowly peaked & Narrowly peaked & Narrowly peaked \\
\addlinespace[5pt]
$Y^{\alpha,\beta}_{Ham}$ Hamiltonian Mixing & Narrowly peaked & Narrowly peaked & Narrowly peaked \\
\noalign{\vskip 4pt\hrule height 0.7pt}
\end{tabularx}
\end{table}

\subsubsection{Majorana}
In the Majorana neutrino scenario, the Lagrangian is given by eq.(\ref{craig}) with different Hamiltonians for geometries depending on the case being studied. The light masses arise from suppressed boundary propagation
together with the Majorana mass scale. This suppression does
not require localisation. The Majorana neutrinos $\Psi$ are assumed to have masses at the fundamental scale.
The Lagrangian for the three-generation scenario is given as before Eq.~(\ref{Majorana-Lagrangian-3flavour}). The flavour-violating couplings in this scenario are also the same as in the Majorana scenario of strong site disorder, i.e, via $W^{\alpha,\beta}$ and Hamiltonian $Y^{\alpha,\beta}_{Ham}$.

\paragraph{\textbf{Petersen}}
Weak site disorder perturbs propagation through the
Petersen graph. The light Majorana masses combine
the suppressed boundary propagators with the
Majorana-sector inverse mass matrix. Their suppression
therefore depends on both propagation through the
graph and the Majorana mass scale. The Lagrangian and Hamiltonian for Petersen with Majorana for the flavour case are given by eq.(\ref{craig}) and eq.(\ref{petersen-Hamiltonian}) respectively. Both Petersen Majorana benchmarks use
$\epsilon_i\in[9.8,10.2]\,\mathrm{TeV}$ and a Majorana
mass prefactor of $10\,\mathrm{TeV}$.
For Majorana-coupling mixing we use $N=16$, $b=1.5$,
and $t=1.0\,\mathrm{TeV}$.
For Hamiltonian mixing we use $N=8$, $b=2$,
and $t\simeq0.64\,\mathrm{TeV}$.
The coupling to $\Psi$ is flavour diagonal and equals
the hopping prefactor in each benchmark,
$t^{\alpha,\beta}=t\,\delta_{\alpha\beta}$.
The respective flavour matrices are 
\begin{equation}
\begin{aligned}
W &\simeq
(10\,\mathrm{TeV})
\begin{pmatrix}
1.6 & -0.36 & 0.11 \\
-0.36 & 0.95 & -0.42 \\
0.11 & -0.42 & 0.80
\end{pmatrix},
\hspace{2cm}
Y_{\rm Ham} &\simeq
\begin{pmatrix}
1.6 & -0.98 & 0.58 \\
0.55 & 0.55 & -0.63 \\
0.075 & 0.34 & 0.36
\end{pmatrix}.
\end{aligned}
\label{eq:petersen-weak-site-majorana-flavour}
\end{equation}
In the Majorana-coupling channel, $Y_{\rm Ham}=I_3$.
In the Hamiltonian channel,
$W^{\alpha,\beta}=10\,\mathrm{TeV}\,\delta_{\alpha\beta}$.
\begin{figure}[ht]
    \centering
    \begin{subfigure}{0.32\textwidth}
        \centering
        \includegraphics[width=1\textwidth]{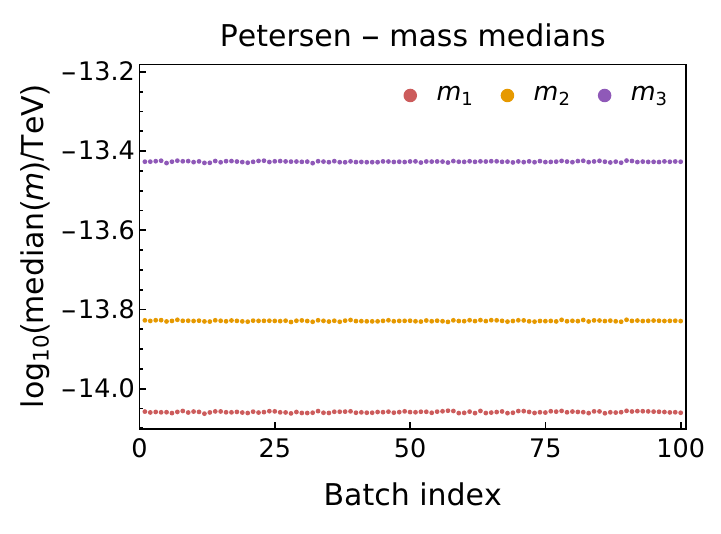}
        \caption{}
        \label{subfig:pet-hammass-maj}
    \end{subfigure}
    \hfill
    \begin{subfigure}{0.32\textwidth}
        \centering
        \includegraphics[width=1\textwidth]{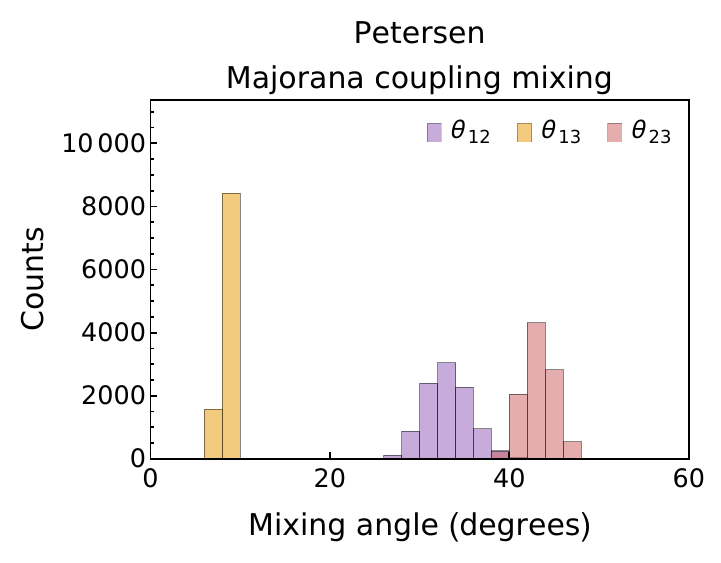}
        \caption{}
        \label{subfig:pet-yukangle-maj-weak}
    \end{subfigure}
    \hfill
    \begin{subfigure}{0.32\textwidth}
        \centering
        \includegraphics[width=1\textwidth]{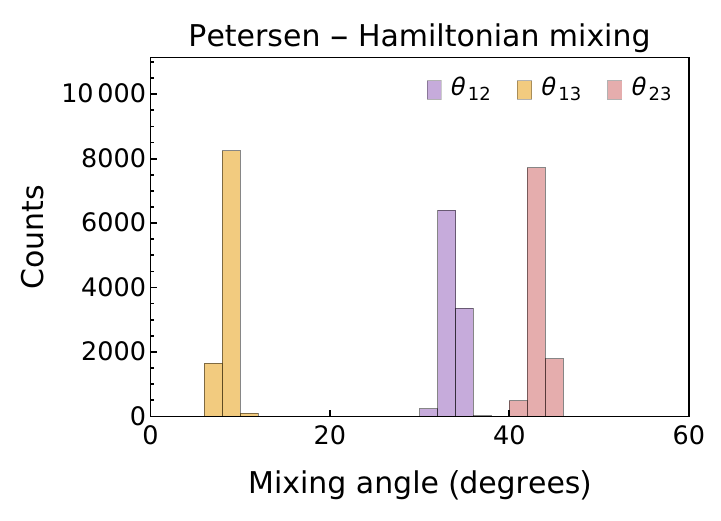}
        \caption{}
        \label{subfig:pet-hamangle-maj-weak}
    \end{subfigure}
    
    \caption{Median of the three lightest masses (in TeV, $\log_{10}$ scale) over 100 runs (left) and histograms of PMNS mixing angles for Majorana coupling $W^{\alpha,\beta}$ (middle) and Hamiltonian mixing $Y^{\alpha,\beta}_{Ham}$ (right). The flavour matrices are given in
Eq.~\eqref{eq:petersen-weak-site-majorana-flavour}.
Both benchmarks have
$\epsilon_i\in[9.8,10.2]\,\mathrm{TeV}$.
The Majorana-coupling benchmark uses
$N=16$, $b=1.5$, and $t=1.0\,\mathrm{TeV}$.
The Hamiltonian benchmark uses
$N=8$, $b=2$, and $t\simeq0.64\,\mathrm{TeV}$.}
    \label{maj-weak-site-pet}
\end{figure}

Fig.~\ref{maj-weak-site-pet} demonstrates the median of three masses produced (left), the mixing angles due to Majorana mixing $W^{\alpha,\beta}$ (middle) and Hamiltonian mixing $Y^{\alpha,\beta}_{Ham}$ (right).
Both the Majorana-coupling and Hamiltonian channels
produce peaked mixing-angle distributions.
In the former, the propagation amplitudes weight
the flavour structure supplied by $W^{\alpha,\beta}$.
In the latter, flavours also mix within the heavy
tower through $Y_{Ham}^{\alpha,\beta}$.
For these benchmarks, weak site fluctuations
produce limited changes in the resulting light
mixing matrix, preserving concentration around
the values set by the chosen inputs.
The Majorana-coupling distributions are somewhat
wider, particularly for $\theta_{12}$, but the
mixing is neither absent nor restricted to the
Hamiltonian channel.

\begin{table}[tbp]
\centering
\small
\setlength{\arrayrulewidth}{0.4pt}
\setlength{\tabcolsep}{5pt}
\renewcommand{\arraystretch}{1.25}
\caption{Parameters for the weak-site Majorana benchmarks.
The Majorana mass prefactor is $10\,\mathrm{TeV}$,
and the flavour-diagonal coupling to $\Psi$ equals $t$.
The decay base $b$ applies only to non-local and Petersen
geometries. Hopping values are displayed to two significant
figures.}
\label{table-maj-site-weak-param}
\begin{tabular*}{\linewidth}
{@{\extracolsep{\fill}}llcccc@{}}
\noalign{\hrule height 0.7pt\vskip 5pt}
\textbf{Geometry} & \textbf{Mixing}
& \textbf{N} & \textbf{b}
& \textbf{$\epsilon_i$ (TeV)}
& \textbf{$t$ (TeV)} \\
\noalign{\vskip 3pt\hrule height 0.4pt\vskip 5pt}
Local & Majorana coupling & 5 & --- & $[9.8,10.2]$ & $\simeq0.69$ \\
\addlinespace[5pt]
Local & Hamiltonian & 5 & --- & $[9.8,10.2]$ & $0.70$ \\
\addlinespace[5pt]
Non-local & Majorana coupling & 16 & 1.5 & $[9.8,10.2]$ & $\simeq0.39$ \\
\addlinespace[5pt]
Non-local & Hamiltonian & 8 & 1.5 & $[9.8,10.2]$ & $\simeq0.057$ \\
\addlinespace[5pt]
Petersen & Majorana coupling & 16 & 1.5 & $[9.8,10.2]$ & $1.0$ \\
\addlinespace[5pt]
Petersen & Hamiltonian & 8 & 2 & $[9.8,10.2]$ & $\simeq0.64$ \\
\noalign{\vskip 4pt\hrule height 0.7pt}
\end{tabular*}
\end{table}

Table~\ref{table-maj-site-weak-param} lists the parameters
used for the weak-site Majorana benchmarks.
  Table~\ref{Table-weak-site-maj-summary} summarises the mixing angle results for these three geometries in the Majorana scenario.

\begin{table}[tbp]
\centering
\small
\setlength{\arrayrulewidth}{0.4pt}
\setlength{\tabcolsep}{6pt}
\renewcommand{\arraystretch}{1.25}
\caption{Weak-site Majorana mixing patterns for the chosen
benchmarks. Both flavour channels produce peaked distributions
in all three geometries. Peak locations and widths depend
on the input flavour matrices and model parameters.}
\label{Table-weak-site-maj-summary}
\begin{tabularx}{\linewidth}{@{}>{\raggedright\arraybackslash}p{0.27\linewidth} *{3}{>{\centering\arraybackslash}X}@{}}
\noalign{\hrule height 0.7pt\vskip 5pt}
\textbf{Mixing Type} & \textbf{Local} & \textbf{Non-local} & \textbf{Petersen} \\
\noalign{\vskip 3pt\hrule height 0.4pt\vskip 5pt}
$W^{\alpha,\beta}$ Majorana Mixing & Peaked & Peaked & Peaked \\
\addlinespace[5pt]
$Y^{\alpha,\beta}_{Ham}$ Hamiltonian Mixing & Peaked & Peaked & Peaked \\
\noalign{\vskip 4pt\hrule height 0.7pt}
\end{tabularx}
\end{table}

\subsection{Hopping Disorder}
Here too, we study the impact of three specific geometries—(a) local, (b) Petersen, and (c) non-local—on the hierarchy of masses produced and the resulting mixing angles. 
Below, we present the results for the Petersen geometry. The corresponding results for the other two geometries are given in Appendix~\ref{app:weak_hopping}.

\subsubsection{Dirac}
In the Dirac scenario, we consider all new fields to be Dirac in nature, the same as in the above sections. Three different geometries are studied: local, non-local and Petersen with their respective Hamiltonians given by eq.(\ref{local-hamiltonian}), eq.(\ref{nonlocal-hamiltonian}) and eq.(\ref{petersen-Hamiltonian}). The Lagrangian for the 1-flavour case is given by eq.(\ref{ACS}). 
In this case, the non-diagonal terms of the Hamiltonians are taken from a random range $[-t, t]$. We also considered decaying hopping terms for beyond neighbouring couplings in all the geometries. For the flavour mixing, the three-generation Lagrangian is considered eq.(\ref{Dirac-Lagrangian-3Flavour}) with possible flavour mixing coming from the non-diagonal Yukawas $Y^{\alpha,\beta}_{yuk}$ and non-diagonal Hamiltonians $Y^{\alpha,\beta}_{Ham}$ in flavour basis. 
The flavour mixing angles and masses generated depend on the underlying theory space structure. They are studied below individually.

\paragraph{\textbf{Petersen}}
Both Petersen Dirac benchmarks use $N=16$, $b=3$,
and uniform site masses $\epsilon_i=10\,\mathrm{TeV}$.
The raw hopping couplings are sampled uniformly from
$[-t,t]$, with $t\simeq0.087\,\mathrm{TeV}$ for Yukawa
mixing and $t\simeq0.086\,\mathrm{TeV}$ for Hamiltonian
mixing. The allowed links receive the prescribed
distance weights.
\begin{figure}[ht]
    \begin{subfigure}{0.34\textwidth}
        \centering
        \includegraphics[width=1\textwidth]{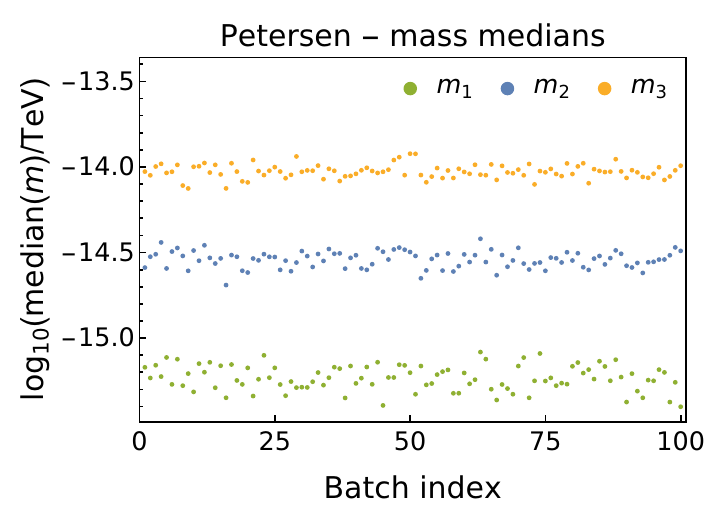}
        \caption{}
        \label{subfig:pet-hopp-yukmass}
    \end{subfigure}
    \hfill
    \begin{subfigure}{0.32\textwidth}
        \centering
        \includegraphics[width=1\textwidth]{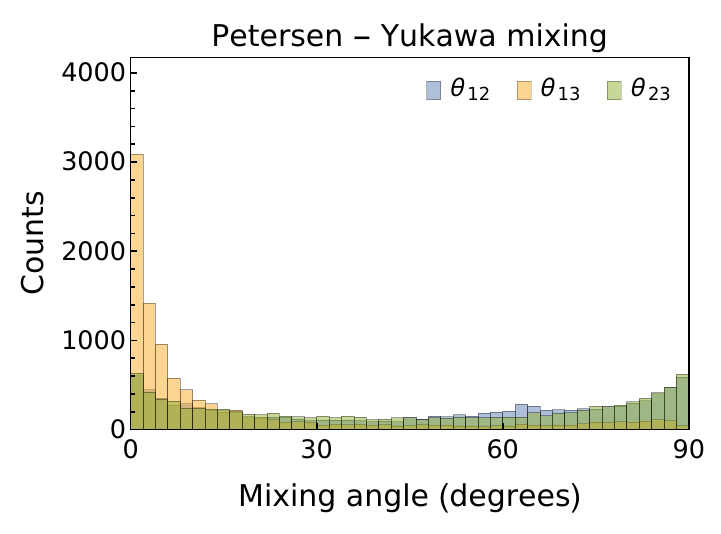}
        \caption{}
        \label{subfig:pet-hopp-yukangle}
    \end{subfigure}
    \hfill
    \begin{subfigure}{0.32\textwidth}
        \centering
        \includegraphics[width=1\textwidth]{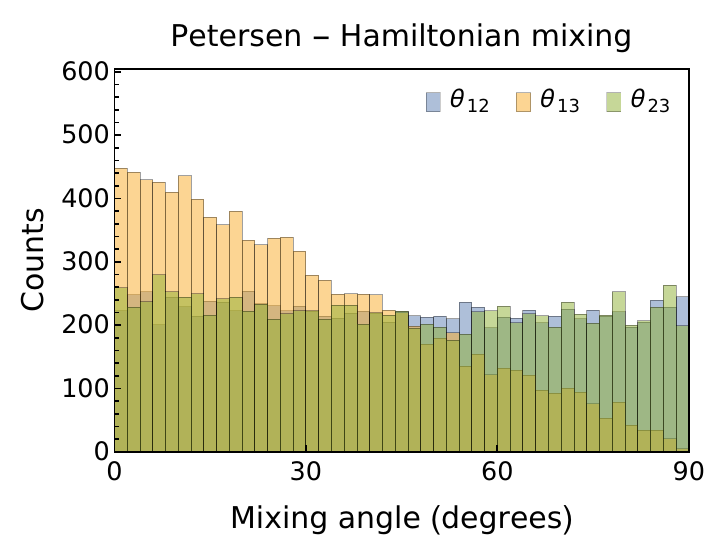}
        \caption{}
        \label{subfig:pet-hopp-hamangle}
    \end{subfigure}
    
    \caption{Median of the three lightest masses (in TeV, $\log_{10}$ scale) over 100 runs (left) and histograms of PMNS mixing angles for Yukawa mixing $Y^{\alpha,\beta}_{yuk}$ (middle) and Hamiltonian mixing $Y^{\alpha,\beta}_{Ham}$ (right), with matrices as in eq.~\eqref{31}. Petersen geometry, weak hopping disorder:
$\epsilon_i=10\,\mathrm{TeV}$, $b=3$, and $N=16$.
The raw hopping half-width is
$t\simeq0.087\,\mathrm{TeV}$ for Yukawa mixing and
$t\simeq0.086\,\mathrm{TeV}$ for Hamiltonian mixing.}
    \label{petersen-weak-coupling-dirac}
\end{figure}

Fig. \ref{petersen-weak-coupling-dirac} shows the median of three masses produced (a), the mixing angles due to Yukawa mixing $Y^{\alpha,\beta}_{yuk}$ (b) and Hamiltonian mixing $Y^{\alpha,\beta}_{Ham}$ (c). 
Although the hopping amplitudes are small compared
with the common site mass, their random magnitudes
and signs can substantially change the relative
propagation amplitudes between the boundaries.
Small hopping therefore does not by itself imply
a stable mixing pattern.

In the boundary-Yukawa channel, these fluctuating
amplitudes combine with the fixed
$Y_{yuk}^{\alpha,\beta}$ to produce endpoint-enhanced
distributions with broad tails. Both small and
large angles are populated, so the mixing cannot
be described as uniformly small.
In the Hamiltonian channel,
$Y_{Ham}^{\alpha,\beta}$ also couples flavours
within the heavy sector, producing broad
distributions for this benchmark.
Unlike the weak-site results, neither channel
retains narrow concentration around the selected
mixing angles under these hopping fluctuations.
Table~\ref{table-weak-hopping-dirac} $\&$ Table~\ref{Table-weak-hopp-dirac-summary} lists the parameters used for the numerical results and the summary of mixing angles for the three geometries.
\begin{table}[tbp]
\centering
\small
\setlength{\arrayrulewidth}{0.4pt}
\setlength{\tabcolsep}{6pt}
\renewcommand{\arraystretch}{1.25}
\caption{Parameters for the weak-hopping Dirac benchmarks.
Raw hopping couplings are sampled uniformly from the
listed intervals before geometric weighting.
The decay base $b$ applies only to non-local and Petersen
geometries. Hopping bounds are displayed to two significant
figures.}
\label{table-weak-hopping-dirac}
\begin{tabular*}{\linewidth}
{@{\extracolsep{\fill}}llcccc@{}}
\noalign{\hrule height 0.7pt\vskip 5pt}
\textbf{Geometry} & \textbf{Mixing}
& \textbf{N} & \textbf{b}
& \textbf{$\epsilon_i$ (TeV)}
& \textbf{Raw hopping (TeV)} \\
\noalign{\vskip 3pt\hrule height 0.4pt\vskip 5pt}
Local & Yukawa & 6 & --- & $10$ & $\simeq[-0.057,0.057]$ \\
\addlinespace[5pt]
Local & Hamiltonian & 9 & --- & $10$ & $\simeq[-0.36,0.36]$ \\
\addlinespace[5pt]
Non-local & Yukawa & 16 & 4 & $10$ & $\simeq[-0.063,0.063]$ \\
\addlinespace[5pt]
Non-local & Hamiltonian & 16 & 4 & $10$ & $\simeq[-0.064,0.064]$ \\
\addlinespace[5pt]
Petersen & Yukawa & 16 & 3 & $10$ & $\simeq[-0.087,0.087]$ \\
\addlinespace[5pt]
Petersen & Hamiltonian & 16 & 3 & $10$ & $\simeq[-0.086,0.086]$ \\
\noalign{\vskip 4pt\hrule height 0.7pt}
\end{tabular*}
\end{table}

\begin{table}[tbp]
\centering
\small
\setlength{\arrayrulewidth}{0.4pt}
\setlength{\tabcolsep}{6pt}
\renewcommand{\arraystretch}{1.25}
\caption{Weak-hopping Dirac mixing patterns for the chosen
benchmarks. Boundary-Yukawa distributions show endpoint
enhancement in the local and Petersen cases.
The non-local case has broad distributions with
small-angle enhancement in $\theta_{13}$.
Hamiltonian-channel distributions are broad in all
three geometries.}
\label{Table-weak-hopp-dirac-summary}
\begin{tabularx}{\linewidth}{@{}>{\raggedright\arraybackslash}p{0.27\linewidth} *{3}{>{\centering\arraybackslash}X}@{}}
\noalign{\hrule height 0.7pt\vskip 5pt}
\textbf{Mixing Type} & \textbf{Local} & \textbf{Non-local} & \textbf{Petersen} \\
\noalign{\vskip 3pt\hrule height 0.4pt\vskip 5pt}
$Y_{yuk}^{\alpha,\beta}$ Yukawa Mixing & Endpoint-enhanced with broad tails & Broad with small-angle enhancement & Endpoint-enhanced with broad tails \\
\addlinespace[5pt]
$Y^{\alpha,\beta}_{Ham}$ Hamiltonian Mixing & Broad & Broad & Broad \\
\noalign{\vskip 4pt\hrule height 0.7pt}
\end{tabularx}
\end{table}

\subsubsection{Majorana}
Finally, the Majorana neutrinos $\Psi$ are considered in the Lagrangian with Hamiltonians for different geometries. 
The Lagrangian for this case, too, remains the same as in the above section, eq.(\ref{craig})
but with the Hamiltonians having random hopping couplings $t_i$s instead of random $\epsilon_i$s.

\paragraph{\textbf{Petersen}}
Both Petersen Majorana benchmarks use uniform site
masses $\epsilon_i=10\,\mathrm{TeV}$ and a Majorana
mass prefactor of $10\,\mathrm{TeV}$.
For Majorana-coupling mixing we use $N=16$, $b=1.5$,
and raw hopping couplings sampled uniformly from
$[-t,t]$ with $t=1.0\,\mathrm{TeV}$.
For Hamiltonian mixing we use $N=8$, $b=2$,
and $t\simeq0.64\,\mathrm{TeV}$.
The coupling to $\Psi$ is fixed and flavour diagonal,
$t^{\alpha,\beta}=t\,\delta_{\alpha\beta}$,
with $t$ denoting the hopping half-width.
\begin{figure}[ht]
    \centering
    \begin{subfigure}{0.34\textwidth}
        \centering
        \includegraphics[width=1\textwidth]{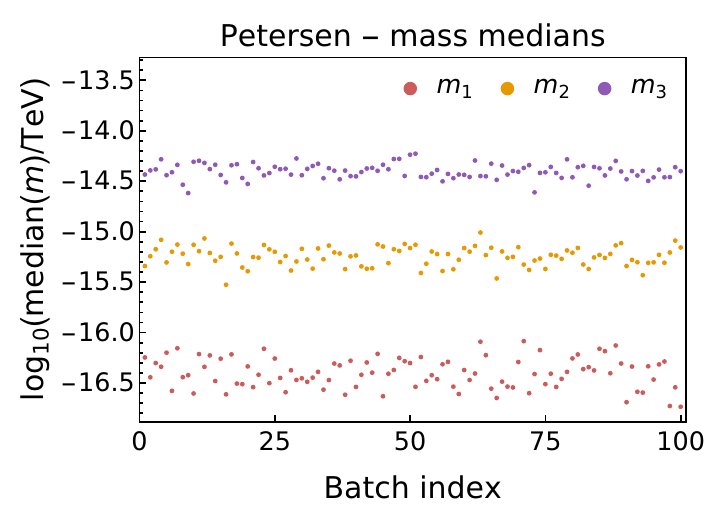}
        \caption{}
        \label{subfig:pet-hopp-mass-maj}
    \end{subfigure}
    \hfill
    \begin{subfigure}{0.32\textwidth}
        \centering
        \includegraphics[width=1\textwidth]{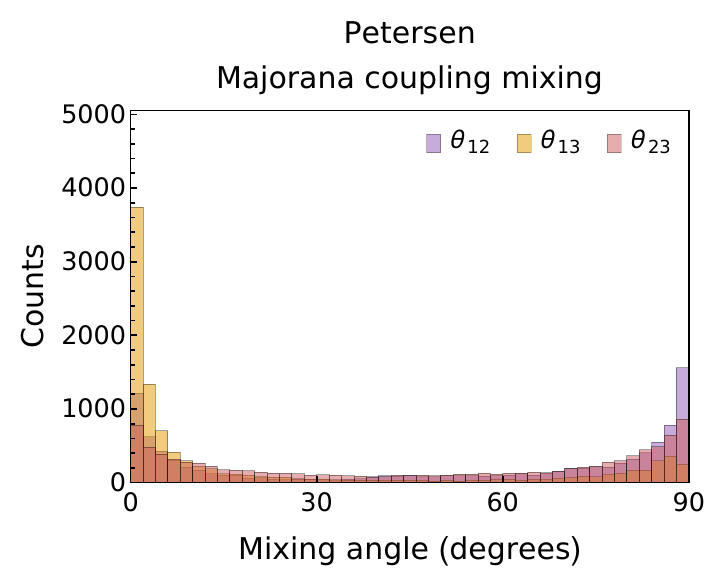}
        \caption{}
        \label{subfig:pet-hopp-yukangle-maj}
    \end{subfigure}
    \hfill
    \begin{subfigure}{0.32\textwidth}
        \centering
        \includegraphics[width=1\textwidth]{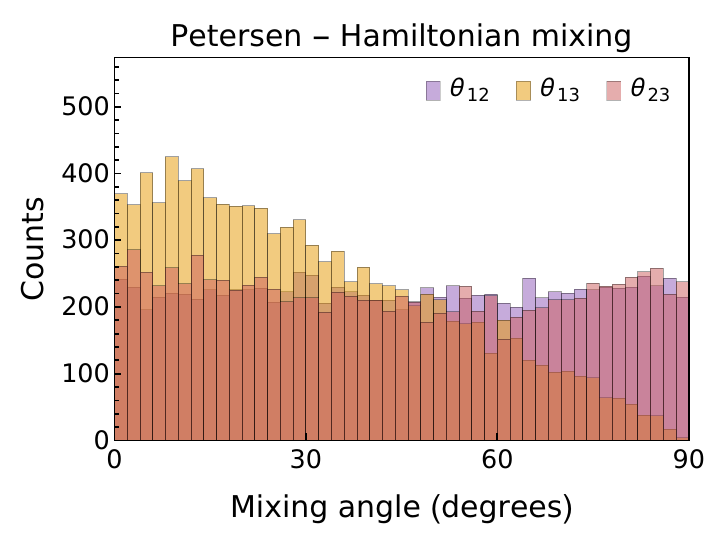}
        \caption{}
        \label{subfig:pet-hopp-hamangle-maj}
    \end{subfigure}
    
    \caption{Median of the three lightest masses (in TeV, $\log_{10}$ scale) over 100 runs (left) and histograms of PMNS mixing angles for Majorana coupling $W^{\alpha,\beta}$ (middle) and Hamiltonian mixing $Y^{\alpha,\beta}_{Ham}$ (right). The flavour matrices are given in
Eq.~\eqref{eq:petersen-weak-site-majorana-flavour}.
Both benchmarks have $\epsilon_i=10\,\mathrm{TeV}$.
The Majorana-coupling benchmark uses
$N=16$, $b=1.5$, and hopping half-width
$t=1.0\,\mathrm{TeV}$.
The Hamiltonian benchmark uses
$N=8$, $b=2$, and $t\simeq0.64\,\mathrm{TeV}$.} \label{weak_hopp_maj}
\end{figure}
Random hopping magnitudes and signs change the
relative propagation amplitudes between the
boundaries, despite remaining small compared
with the site mass. In the Majorana-coupling
channel, these amplitudes weight the flavour
structure supplied by $W^{\alpha,\beta}$ on
both sides of the effective light mass matrix.
Strongly unequal weights can favour flavour
alignment, providing intuition for the pronounced
endpoint enhancement. Intermediate angles remain
populated, so the mixing is not absent.
The Hamiltonian channel allows flavours to mix
within the heavy tower and produces broad
distributions. Neither channel retains the
narrow concentration of the corresponding
weak-site benchmark. The numerical results are shown in Fig.~\ref{weak_hopp_maj}.
The parameters are listed in
Table~\ref{table-maj-hopp-weak-param}.
The flavour matrices are unchanged from the Petersen
weak-site Majorana benchmarks in
Eq.~\eqref{eq:petersen-weak-site-majorana-flavour}.
\begin{table}[tbp]
\centering
\small
\setlength{\arrayrulewidth}{0.4pt}
\setlength{\tabcolsep}{5pt}
\renewcommand{\arraystretch}{1.25}
\caption{Parameters for the weak-hopping Majorana benchmarks.
Raw hopping couplings are sampled uniformly from the
listed intervals before geometric weighting.
The Majorana mass prefactor is $10\,\mathrm{TeV}$.
The fixed flavour-diagonal coupling to $\Psi$ equals
the hopping half-width $t$.
The decay base $b$ applies only to non-local and Petersen
geometries. Hopping bounds are displayed to two significant
figures.}
\label{table-maj-hopp-weak-param}
\begin{tabular*}{\linewidth}
{@{\extracolsep{\fill}}llcccc@{}}
\noalign{\hrule height 0.7pt\vskip 5pt}
\textbf{Geometry} & \textbf{Mixing}
& \textbf{N} & \textbf{b}
& \textbf{$\epsilon$ (TeV)}
& \textbf{Raw hopping (TeV)} \\
\noalign{\vskip 3pt\hrule height 0.4pt\vskip 5pt}
Local & Majorana coupling & 5 & --- & $10$
& $\simeq[-0.69,0.69]$ \\
\addlinespace[5pt]
Local & Hamiltonian & 5 & --- & $10$
& $[-0.70,0.70]$ \\
\addlinespace[5pt]
Non-local & Majorana coupling & 16 & 1.5 & $10$
& $\simeq[-0.39,0.39]$ \\
\addlinespace[5pt]
Non-local & Hamiltonian & 8 & 1.5 & $10$
& $\simeq[-0.057,0.057]$ \\
\addlinespace[5pt]
Petersen & Majorana coupling & 16 & 1.5 & $10$
& $[-1.0,1.0]$ \\
\addlinespace[5pt]
Petersen & Hamiltonian & 8 & 2 & $10$
& $\simeq[-0.64,0.64]$ \\
\noalign{\vskip 4pt\hrule height 0.7pt}
\end{tabular*}
\end{table}
Table~\ref{Table-weak-hopp-maj-summary} gives a summary of the resulting mixing angles for the three geometries considered. 
\begin{table}[tbp]
\centering
\small
\setlength{\arrayrulewidth}{0.4pt}
\setlength{\tabcolsep}{6pt}
\renewcommand{\arraystretch}{1.25}
\caption{Weak-hopping Majorana mixing patterns for the
chosen benchmarks. Majorana-coupling distributions
are endpoint-enhanced, with stronger concentration
in the local and Petersen examples.
Hamiltonian-channel distributions are broad
in all three geometries.}
\label{Table-weak-hopp-maj-summary}
\begin{tabularx}{\linewidth}{@{}>{\raggedright\arraybackslash}p{0.27\linewidth} *{3}{>{\centering\arraybackslash}X}@{}}
\noalign{\hrule height 0.7pt\vskip 5pt}
\textbf{Mixing Type} & \textbf{Local} & \textbf{Non-local} & \textbf{Petersen} \\
\noalign{\vskip 3pt\hrule height 0.4pt\vskip 5pt}
$W^{\alpha,\beta}$ Majorana Mixing & Strong endpoint enhancement & Endpoint enhancement with broad tails & Strong endpoint enhancement \\
\addlinespace[5pt]
$Y^{\alpha,\beta}_{Ham}$ Hamiltonian Mixing & Broad & Broad & Broad \\
\noalign{\vskip 4pt\hrule height 0.7pt}
\end{tabularx}
\end{table}

\subsection{Quantifying mixing-angle concentration}
\label{sec:angle-concentration}

Table~\ref{tab:mixing-angle-concentration} quantifies the contrast
visible in the angle histograms. For the benchmarks studied,
weak site disorder yields narrow distributions for all three
mixing angles, whereas strong site and weak hopping disorder
produce substantially broader distributions.
The weak-site flavour matrices were fitted to obtain approximately
observed mixing angles and subsequently transferred to the
weak-hopping ensemble without refitting.

\begin{table*}[t]
\centering
\caption{
Concentration of the mixing-angle distributions for the benchmarks
studied. The central 68\% width is
$w_{68}(\theta)=Q_{84}(\theta)-Q_{16}(\theta)$.
Each entry gives the minimum--maximum width across the twelve
benchmarks in that regime, covering three geometries, two neutrino
constructions and two flavour-mixing channels.
All widths are in degrees and are evaluated after numerical-quality
selection and the illustrative 1\% diagonal row-norm cut.
}
\label{tab:mixing-angle-concentration}

\begingroup
\renewcommand{\arraystretch}{1.4}
\begin{tabular*}{\textwidth}
{@{\extracolsep{\fill}}lccc@{}}
\hline
\textbf{Disorder regime}
& $\boldsymbol{w_{68}(\theta_{12})}$
& $\boldsymbol{w_{68}(\theta_{13})}$
& $\boldsymbol{w_{68}(\theta_{23})}$ \\
\hline
Strong site
& $49.61$--$83.74$
& $39.78$--$85.21$
& $58.43$--$89.00$ \\

Weak site
& $1.36$--$4.98$
& $0.54$--$1.47$
& $1.19$--$3.50$ \\

Weak hopping
& $60.44$--$85.23$
& $17.55$--$72.68$
& $55.05$--$84.94$ \\
\hline
\end{tabular*}
\endgroup
\end{table*}

\section{Conclusions and Outlook}
\label{sec:conclusion}

Disordered theory spaces offer two routes to a small
boundary-to-boundary mass: localisation under strong site
disorder, and GIM-like cancellations in the boundary propagator
when an approximately uniform diagonal dominates. The graph-path expansion in Eq.~\eqref{eq:resolvent-paths} identifies the ingredients controlling the latter: boundary placement, shortest paths, link weights and the diagonal scale. Spectral near-degeneracy by itself does not determine the suppression.

The mixing results depend on flavour input as well as disorder. In the strong-site benchmarks, Hamiltonian flavour mixing
produces broad distributions. The boundary channels show
endpoint enhancement or, for some Dirac benchmarks, an
interior peak, demonstrating that strong mass suppression
does not require small mixing angles. Weak site disorder can preserve peaked distributions for suitable fixed flavour matrices. For the weak-hopping benchmarks, the Hamiltonian channel
instead gives broad distributions, while the boundary channels
show geometry-dependent patterns with varying endpoint
enhancement. These benchmarks do not retain the narrow
mixing-angle concentration found with weak site disorder. This supports the possibility of structured mixing in a disordered theory space, but does not prove that randomness generates the observed pattern or that a particular graph is statistically preferred. Table~\ref{tab:summary} summarises the benchmark trends.
\begin{table}[tbp]
\centering
\small
\setlength{\arrayrulewidth}{0.4pt}
\setlength{\tabcolsep}{6pt}
\renewcommand{\arraystretch}{1.25}
\caption{Qualitative benchmark trends across the three geometries.}
\label{tab:summary}
\begin{tabularx}{\linewidth}{@{}>{\raggedright\arraybackslash}p{0.18\linewidth} >{\raggedright\arraybackslash}p{0.32\linewidth} >{\raggedright\arraybackslash}X@{}}
\noalign{\hrule height 0.7pt\vskip 5pt}
\textbf{Regime} & \textbf{Mass mechanism} & \textbf{Mixing} \\
\noalign{\vskip 3pt\hrule height 0.4pt\vskip 5pt}
Strong site & Localisation with geometry-dependent boundary suppression & Broad Hamiltonian distributions. Boundary channels show endpoint
or interior enhancement for the chosen inputs \\
\addlinespace[5pt]
Strong hopping & No sufficient suppression in the examples & No viable mass benchmark established \\
\addlinespace[5pt]
Weak site & Inverse-matrix cancellation and graph paths & Peaked distributions in both flavour channels for the chosen inputs \\
\addlinespace[5pt]
Weak hopping & Inverse-matrix cancellation with dominant diagonal & Broad Hamiltonian channel. Boundary mixing depends on geometry \\
\noalign{\vskip 4pt\hrule height 0.7pt}
\end{tabularx}
\end{table}
A quantitative comparison with neutrino anarchy requires
a specified probability measure on the flavour couplings
and consistent selection criteria. We do not perform such
a comparison here, so the present benchmarks do not establish
a statistical preference over Haar-distributed mixing.
A fuller phenomenological study should compare the geometries
under common flavour assumptions and evaluate the light-neutrino
masses, mixing angles and constraints on active--heavy mixing
for the same parameter points.

\textbf{Acknowledgments}
We thank Jason Kumar, Rajesh Narayan, Subroto Mukherjee, Alejandro Ibarra, Pinaki Muzumdar, Deepshikha Nagar, Diptiman Sen, Gowri Kurup and Anshuman Maharana for helpful discussions. SKV is supported by SERB, DST, Govt. of India Grants MTR/2022/000255, “Theoretical aspects of some physics beyond standard models”, CRG/2021/007170 “Tiny Effects from Heavy New Physics “and IoE funds from IISC. AS thanks CSIR, Govt. of India for SRF fellowship No. 09/0079(15487)/2022-EMR-I. All the codes used in this analysis will be made public through \href{https://github.com/AadarshSingh0/Neutrino-textures-using-Randomness}{GitHub \faGithub}.

\appendix

\section{Graph-Theoretic Properties of the Generalized Petersen Family}
\label{app:petersen}

The standard Petersen graph, denoted $GP(5,2)$, is a highly symmetric member
of the generalized Petersen family $GP(n,k)$. This family provides a useful
framework for constructing sparse graphs with controlled non-local
connectivity. In the main text and numerical sections, we use the term
``Petersen geometry'' as shorthand for the generalized Petersen-type
theory-space graphs used in our scans, rather than for the standard
Petersen graph $GP(5,2)$ alone.
More precisely, the numerical analysis uses the even-$n$ family
$GP(n,n/2)$. This family interpolates between the nearest-neighbour local
chain and the fully connected non-local graph by introducing a sparse set of
non-nearest-neighbour links. This distinction is important: the standard
Petersen graph $GP(5,2)$ has special symmetry properties, such as high
regularity and a large automorphism group, which are not shared by generic
members of the $GP(n,n/2)$ family.
The standard generalized Petersen family $GP(n,k)$ with
$1\leq k<n/2$ is cubic, or $3$-regular, and has $3n$ edges. The subfamily
$GP(n,n/2)$ studied in this work falls outside this range. It has $2n$
vertices, $5n/2$ edges, and is degree-heterogeneous (degrees $2$ and $3$): outer vertices have degree
$3$, while inner vertices have degree $2$. See, for example,
Refs.~\cite{weisstein2019generalizedpetersengraph,alspach1983classification,krnc2018characterization,bannai1978hamiltonian}.

\begin{enumerate}
    \item \textbf{Definition and size.}
    The standard generalized Petersen graph $GP(n,k)$, with
    $1\leq k<n/2$, has $2n$ vertices, $3n$ edges, and is cubic
    ($3$-regular). For the even-$n$ family $GP(n,n/2)$ used in our scans,
    the inner cycle degenerates into an antipodal matching. The graph has
    $2n$ vertices, $5n/2$ edges, and is degree-heterogeneous (degrees $2$ and $3$).

    \item \textbf{Connectivity of the numerical family.}
    For $k=n/2$ with even $n=N/2$, equivalently $N$ divisible by four, the graph $GP(n,n/2)$ remains connected:
    the outer $n$-cycle connects all outer vertices, each spoke connects
    one outer vertex to one inner vertex, and the antipodal inner matching
    connects each inner vertex to its antipodal partner. Together these
    links yield a single connected component.

    \item \textbf{Symmetry: standard Petersen vs.\ generalized family.}
    The standard Petersen graph $GP(5,2)$ is highly symmetric. In particular,
    it is vertex-transitive and edge-transitive, and its automorphism group
    is isomorphic to $S_5$. The $GP(n,n/2)$
    family used in our scans does not in general share the full symmetry of
    $GP(5,2)$.

    \item \textbf{Hamiltonicity.}
    The standard Petersen graph $GP(5,2)$ has no Hamiltonian cycle
    (i.e.\ no cycle visiting each vertex exactly once), although it does
    contain Hamiltonian paths. More generally, Hamiltonicity of $GP(n,k)$
    depends on $(n,k)$ and is not guaranteed. A complete classification is
    given in Ref.~\cite{alspach1983classification}.

    \item \textbf{Colouring.}
    The standard Petersen graph $GP(5,2)$ has chromatic number $3$, while
    the chromatic number of a general $GP(n,k)$ can depend on $(n,k)$.

    \item \textbf{Further standard Petersen facts.}
    The standard Petersen graph $GP(5,2)$ is strongly regular with
    parameters $\mathrm{srg}(10,3,0,1)$, has girth $5$, and diameter $2$.

    \item \textbf{Sparse non-locality.}
    The generalized Petersen-type construction includes links connecting
    vertices separated by a fixed number of steps on the outer cycle. In the
    $GP(n,n/2)$ family, these links provide a sparse and controlled form of
    non-local connectivity, intermediate between the local chain and the
    fully connected non-local graph.
\end{enumerate}

\begin{figure}[ht]
    \centering
    \begin{subfigure}{0.48\textwidth}
        \centering
        \includegraphics[width=\linewidth]{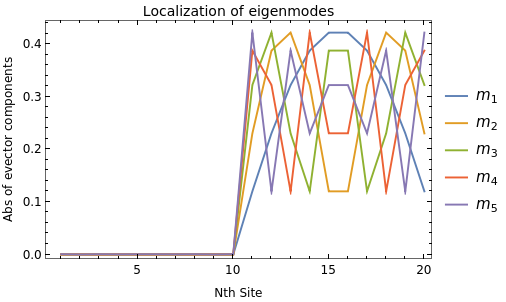}
        \caption{}
        \label{subfig:peter-first5}
    \end{subfigure}
    \begin{subfigure}{0.48\textwidth}
        \centering
        \includegraphics[width=\linewidth]{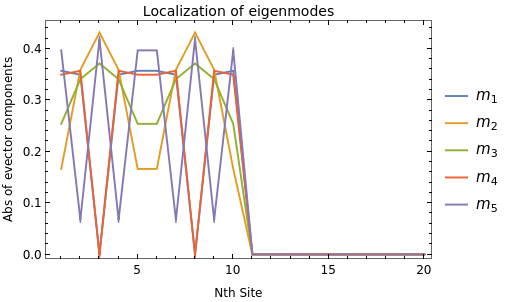}
        \caption{}
        \label{subfig:peter-last5}
    \end{subfigure}
    
    \caption{Mass modes of the $GP(n,n/2)$ generalized Petersen-type geometry
in the uniform-site limit: first five modes (left panel) and last five modes
(right panel), shown for a 20-site lattice.}
    \label{Petersen_spectrum}
\end{figure}

\noindent
In the uniform (no-disorder) limit of our $GP(n,n/2)$ Hamiltonian, we observe
a choice of eigenbasis with support predominantly on different site subsets, as illustrated in Fig.~\ref{Petersen_spectrum}.

\FloatBarrier

\section{Neutrino Oscillation Inputs}
\label{neutrino_numbers}
The atmospheric splitting convention is $\ell=1$ for NO and
$\ell=2$ for IO. The experimental CP-phase values are included
for completeness, although the real benchmark matrices
considered here conserve CP.
The weak-site fitting targets were taken from NuFIT~6.1
(normal ordering, IC23 without Super-Kamiokande atmospheric
data). The NuFIT~6.0 (2024) values below are retained as
experimental reference inputs.
Table~\ref{tab:nufit60} summarises the mixing angles, the CP-violating phase, and the mass-squared splittings for both normal ordering (NO) and inverted ordering (IO), quoting the best-fit values with $1\sigma$ uncertainties as well as the corresponding $3\sigma$ ranges \cite{esteban2025nufit}. 

\begin{table}[H]
\centering
\small
\setlength{\arrayrulewidth}{0.4pt}
\setlength{\tabcolsep}{6pt}
\renewcommand{\arraystretch}{1.25}
\caption{NuFIT 6.0 (2024) oscillation parameters using IC19 without Super-Kamiokande atmospheric data.}
\label{tab:nufit60}
\begin{tabularx}{\linewidth}{@{}>{\raggedright\arraybackslash}p{0.22\linewidth} *{4}{>{\centering\arraybackslash}X}@{}}
\noalign{\hrule height 0.7pt\vskip 5pt}
& \multicolumn{2}{c}{\textbf{Normal Ordering}} & \multicolumn{2}{c}{\textbf{Inverted Ordering}} \\
\noalign{\vskip 3pt}
\textbf{Parameter} & bfp $\pm 1\sigma$ & 3$\sigma$ range & bfp $\pm 1\sigma$ & 3$\sigma$ range \\
\noalign{\vskip 3pt\hrule height 0.4pt\vskip 5pt}
$\sin^2\theta_{12}$ 
 & $0.307^{+0.012}_{-0.011}$ 
 & $0.275 \to 0.345$
 & $0.308^{+0.012}_{-0.011}$ 
 & $0.275 \to 0.345$ \\
$\theta_{12} / {}^\circ$
 & $33.68^{+0.73}_{-0.70}$ 
 & $31.63 \to 35.95$
 & $33.68^{+0.73}_{-0.70}$ 
 & $31.63 \to 35.95$ \\
\addlinespace[5pt]
$\sin^2\theta_{23}$ 
 & $0.561^{+0.012}_{-0.015}$ 
 & $0.430 \to 0.596$
 & $0.562^{+0.012}_{-0.015}$ 
 & $0.437 \to 0.597$ \\
$\theta_{23} / {}^\circ$
 & $48.5^{+0.7}_{-0.9}$ 
 & $41.0 \to 50.5$
 & $48.6^{+0.7}_{-0.9}$ 
 & $41.4 \to 50.6$ \\
\addlinespace[5pt]
$\sin^2\theta_{13}$ 
 & $0.02195^{+0.00054}_{-0.00058}$ 
 & $0.02023 \to 0.02376$
 & $0.02224^{+0.00056}_{-0.00057}$ 
 & $0.02053 \to 0.02397$ \\
$\theta_{13} / {}^\circ$
 & $8.52^{+0.11}_{-0.11}$ 
 & $8.18 \to 8.87$
 & $8.58^{+0.11}_{-0.11}$ 
 & $8.24 \to 8.91$ \\
\addlinespace[5pt]
$\delta_{\rm CP} / {}^\circ$
 & $177^{+19}_{-20}$ 
 & $96 \to 422$
 & $285^{+25}_{-28}$ 
 & $201 \to 348$ \\
\addlinespace[5pt]
$\Delta m^2_{21}$ $(10^{-5}\,\text{eV}^2)$
 & $7.49^{+0.19}_{-0.19}$ 
 & $6.92 \to 8.05$
 & $7.49^{+0.19}_{-0.19}$ 
 & $6.92 \to 8.05$ \\
\addlinespace[5pt]
$\Delta m^2_{3\ell}$ $(10^{-3}\,\text{eV}^2)$
 & $+2.534^{+0.025}_{-0.023}$ 
 & $+2.463 \to +2.606$
 & $-2.510^{+0.024}_{-0.025}$ 
 & $-2.584 \to -2.438$ \\
\noalign{\vskip 4pt\hrule height 0.7pt}
\end{tabularx}
\end{table}

\FloatBarrier
\section{Numerical Results: Strong Site Disorder}
\label{app:strong_site}
In this appendix, we present the explicit numerical results for the local and non-local geometries, for both Dirac and Majorana neutrino scenarios with large disorder in site elements $\epsilon_i$s. 

\subsection{Dirac}
\begin{figure}
    \begin{subfigure}{0.48\textwidth}
        \centering
        \includegraphics[width=\linewidth]{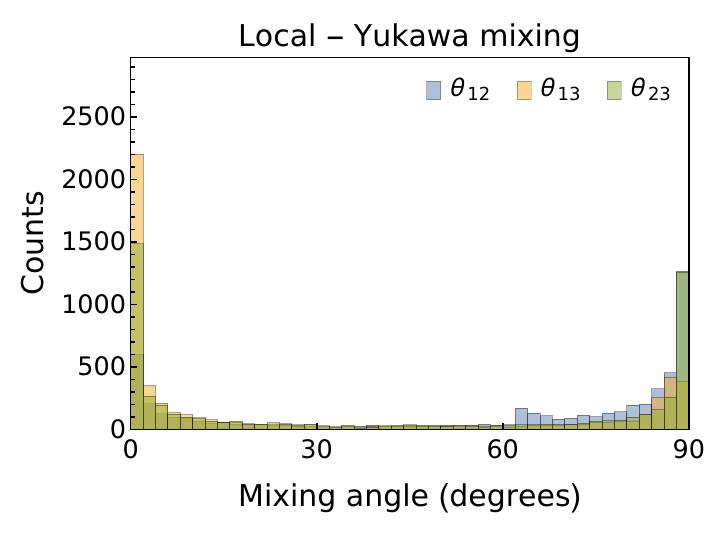}
        \caption{}
        \label{subfig:yuk}
    \end{subfigure}
    \hfill
    \begin{subfigure}{0.48\textwidth}
        \centering
        \includegraphics[width=\linewidth]{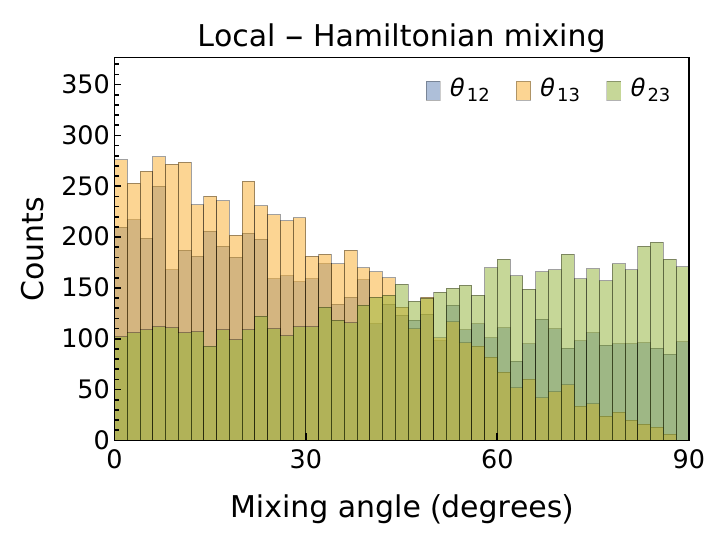}
        \caption{}
        \label{subfig:ham}
    \end{subfigure}
    
    \caption{Histogram of PMNS mixing angles in the local geometry (Dirac, strong site disorder): Yukawa mixing $Y^{\alpha,\beta}_{yuk}$ (left) and Hamiltonian mixing $Y^{\alpha,\beta}_{Ham}$ (right), with matrices as in eq.~\eqref{18}. Parameters: $W=5\,\mathrm{TeV}$, $N=12$,
$t=0.1\,\mathrm{TeV}$, and
$\epsilon_i\in[-2W,2W]$.}
    \label{dirac-strong-site-local}
\end{figure}

\paragraph{\textbf{Local}}
For the displayed strong-site Dirac benchmarks, the local
geometry gives the strongest suppression of the light masses.
Propagation between its endpoints must traverse the full
nearest-neighbour chain, whereas the other geometries provide
additional weighted paths. These paths can weaken the
suppression, although their effect depends on the link strengths
and disorder rather than the number of links alone. The total Lagrangian for a more realistic three-generation scenario is given by 
\begin{equation}
 \mathcal L_{\rm mass}=-\begin{pmatrix}\bar\nu_L&\bar L\end{pmatrix}
 \begin{pmatrix}0&B_L\\B_R&\mathbb H\end{pmatrix}
 \begin{pmatrix}\nu_R\\R\end{pmatrix}+\mathrm{h.c.}
 \label{Dirac-total-lagrangian}
\end{equation}
This is a broken-phase mass parametrisation with the boundary and gauge-matching assumptions of Sec.~\ref{sec:gauge}.  
For this scenario, we consider $N=12$ with a large randomness range, specifically $\epsilon_i \in [-2W,2W]$, and $t_i = t$. The parameters are set as $W = 5\,\text{TeV}$ and $t = 0.1\,\text{TeV}$. In the model, flavour-diagonal left-handed SM neutrino Yukawa couplings are considered alongside non-diagonal right-handed neutrino Yukawa couplings, $Y_{yuk}^{\alpha,\beta}$ (referred to as ``Yukawa mixing" in this work). A specific configuration is chosen randomly of O(1) $3 \times 3$ matrix. For concreteness we chose Yukawas as in eq.~\eqref{18}. The mixing generated for this set of parameters is shown in Fig.~\ref{dirac-strong-site-local} (left). The right plot in Fig.~\ref{dirac-strong-site-local} illustrates the case with diagonal $Y_{yuk}^{\alpha,\beta}$ couplings but with off-diagonal flavour matrices $Y^{\alpha,\beta}_{Ham}$, referred to as ``Hamiltonian mixing" in this work.
\begin{figure}
    \begin{subfigure}{0.48\textwidth}
        \centering
        \includegraphics[width=\linewidth]{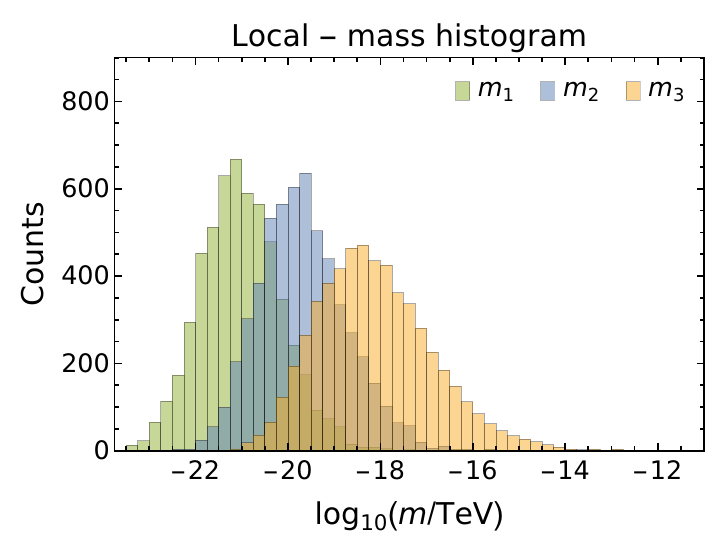}
        \caption{}
        \label{subfig:local-logmass}
    \end{subfigure}
    \hfill
    \begin{subfigure}{0.48\textwidth}
        \centering
        \includegraphics[width=\linewidth]{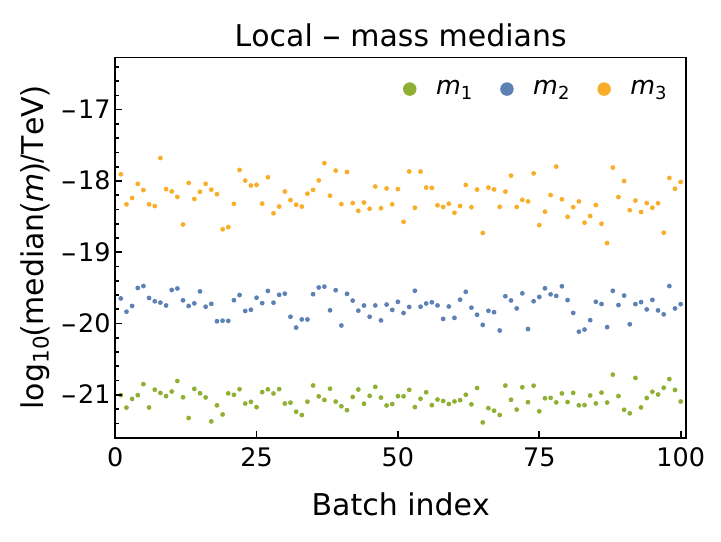}
        \caption{}
        \label{subfig:local-mass}
    \end{subfigure}
    
    \caption{Histograms of $\log_{10}$(mass/TeV) over various runs (left) and median of the three lightest masses over 100 runs (right) with Yukawa mixing. Local geometry, strong site disorder:
$W=5\,\mathrm{TeV}$, $t=0.1\,\mathrm{TeV}$,
$\epsilon_i\in[-2W,2W]$, and $N=12$.}
    \label{local-Dirac-strong}
\end{figure}

The off-diagonal elements of $Y_{Ham}^{\alpha, \beta}$ mixes the left-handed chiral field $L_{\alpha}$ with the right-handed chiral fields $R_{\beta}$ of different flavours.  
The masses produced with these parameters are given in Fig.~\ref{local-Dirac-strong}. 

In the local geometry, strong site disorder localizes the heavy
modes and suppresses propagation along the nearest-neighbour
chain. In the boundary-Yukawa channel
($Y_{yuk}^{\alpha,\beta}$), the heavy flavour sectors remain
separate, but their propagation amplitudes fluctuate across
disorder realisations. Their relative sizes, together with
the non-diagonal boundary Yukawa matrix, determine the flavour
structure of $m_D^{\mathrm{eff}}=-B_LH^{-1}B_R$.
Strongly unequal propagation amplitudes can favour alignment
with particular flavour directions, providing intuition for
the endpoint enhancement in
Fig.~\ref{dirac-strong-site-local}(a).
For this benchmark, $\theta_{12}$ is most strongly enhanced
near $90^\circ$, $\theta_{13}$ near $0^\circ$, and $\theta_{23}$
near both endpoints. Intermediate angles are also populated. The suppression of
the masses therefore does not imply vanishing mixing angles.

In the Hamiltonian channel, non-diagonal
$Y_{Ham}^{\alpha,\beta}$ allows flavours to mix during
propagation through the chain. The effective mass matrix
then receives contributions connecting different flavours
within the heavy sector, even with flavour-diagonal boundary
couplings. For the chosen benchmark, this produces the broad
distributions in Fig.~\ref{dirac-strong-site-local}(b):
$\theta_{12}$ and $\theta_{13}$ favour smaller angles, whereas
$\theta_{23}$ favours larger angles. These patterns describe
the selected flavour matrix and disorder ensemble.

\begin{figure}[ht]
    \begin{subfigure}{0.48\textwidth}
        \centering
        \includegraphics[width=\linewidth]{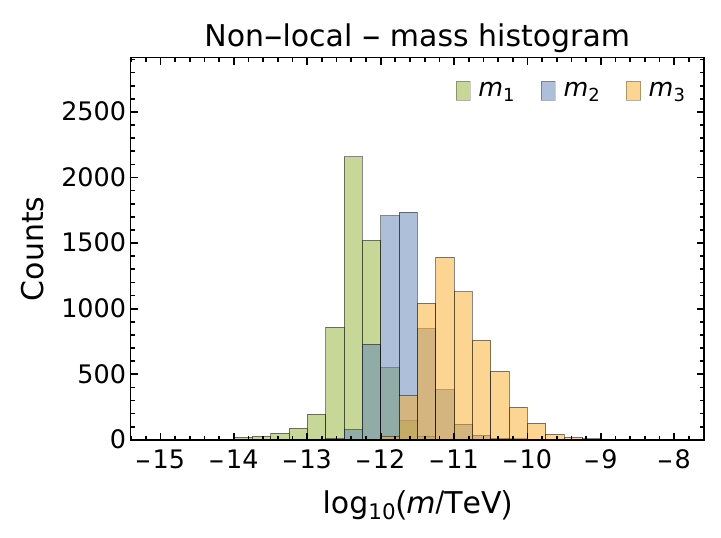}
        \caption{}
        \label{subfig:nonlocal-logmass}
    \end{subfigure}
    \hfill
    \begin{subfigure}{0.48\textwidth}
        \centering
        \includegraphics[width=\linewidth]{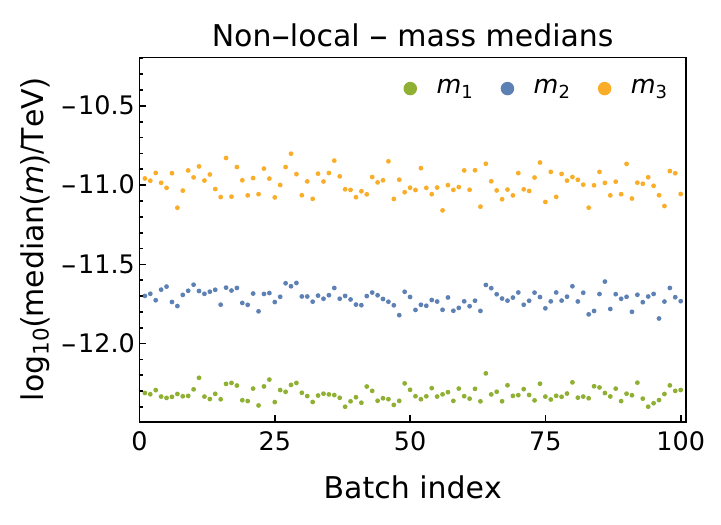}
        \caption{}
        \label{subfig:nonlocal-mass}
    \end{subfigure}
    
    \caption{Histograms of $\log_{10}$(mass/TeV) over various runs (left) and median of the three lightest masses over 100 runs (right). Non-local geometry, strong site disorder (Dirac),
with Yukawa mixing:
$W=5\,\mathrm{TeV}$, $b=5$, $N=12$,
$t=0.1\,\mathrm{TeV}$, and
$\epsilon_i\in[-2W,2W]$.}
    \label{nonlocal_masses}
\end{figure}

\begin{figure}
    \begin{subfigure}{0.48\textwidth}
        \centering
        \includegraphics[width=\linewidth]{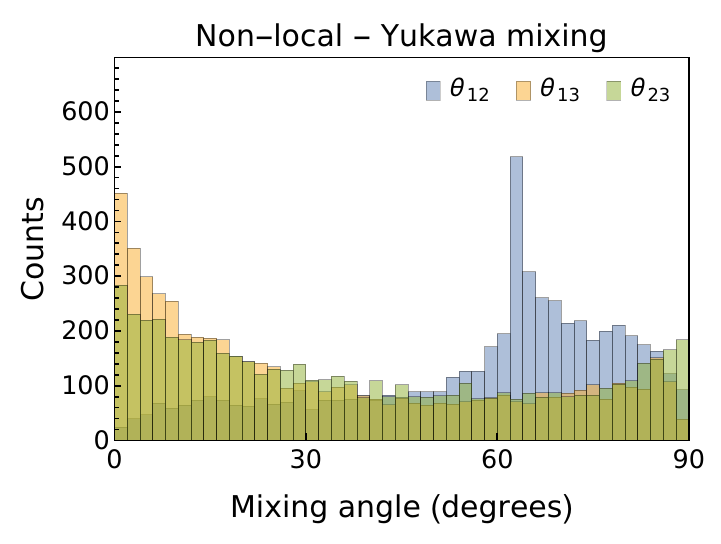}
        \caption{}
        \label{subfig:nonlocal-yukangle}
    \end{subfigure}
    \hfill
    \begin{subfigure}{0.48\textwidth}
        \centering
        \includegraphics[width=\linewidth]{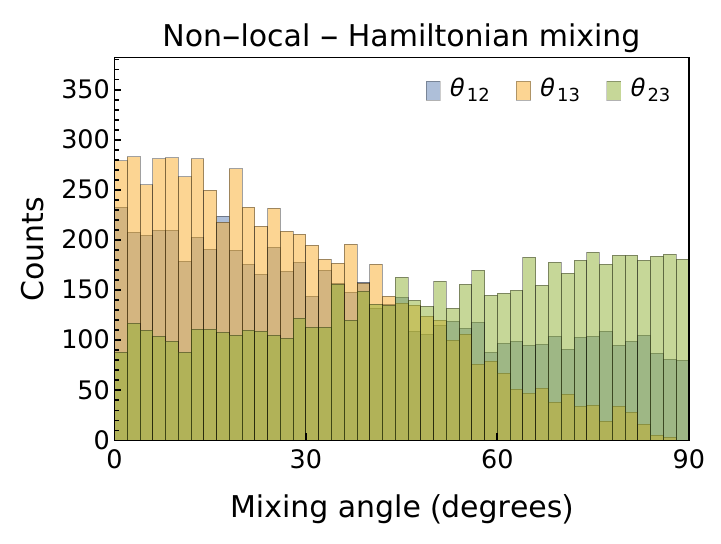}
        \caption{}
        \label{subfig:nonlocal-hamangle}
    \end{subfigure}
    
    \caption{Histogram of mixing angles for several runs in the non-local theory space for Yukawa mixing $Y^{\alpha,\beta}_{yuk}$ (left) and Hamiltonian mixing $Y^{\alpha,\beta}_{Ham}$ (right) as in eq.~\eqref{18}, with $W=5\,\mathrm{TeV}$, $b=5$, $N=12$,
$t=0.1\,\mathrm{TeV}$, and
$\epsilon_i\in[-2W,2W]$.} \label{nonlocal_maj_stringsite}
\end{figure}

\paragraph{\textbf{Non-Local}}
The non-local geometry considered here is the fully connected theory space, i.e, all fermions $\{L_i, R_i\}$ have non-zero coupling strengths to all other fermions $\{L_j, R_j\}$ of the model. The Hamiltonian for this geometry is given by
\begin{align}
    \mathcal{H}^{non-local}_{i,j} &= \epsilon_i\delta_{i,j} + \frac{t}{b^{|i-j|}}\left(1-\delta_{i, j}\right) \label{nonlocal-hamiltonian}
\end{align}
and the corresponding Lagrangian is given by eq.~\eqref{nonlocalDirac}.
The additional links provide alternative paths between the
boundaries and can increase the boundary propagator even
when the individual modes remain strongly concentrated. Even though the efficiency is smaller as compared to other geometries, we can still find natural fundamental parameters $O(1)$ to generate eV masses from the TeV scale required for neutrinos in this geometry.
For this non-local geometry, both Dirac benchmarks use
$N=12$, with site masses sampled uniformly from
$\epsilon_i\in[-2W,2W]$, where $W=5\,\mathrm{TeV}$.
The hopping prefactor is fixed at $t=0.1\,\mathrm{TeV}$
and the decay base is $b=5$. The resultant masses are shown in Fig.~\ref{nonlocal_masses}. The analysis for the mixing angles proceeds in two parts, analogous to previous geometries. First, we examine scenarios with non-diagonal right-handed neutrino Yukawa couplings, $Y_{yuk}^{\alpha,\beta}$, for a random $3 \times 3$ matrix. Second, we consider cases with diagonal $Y_{yuk}^{\alpha,\beta}$ couplings along with non-diagonal Hamiltonian mixing, $Y^{\alpha,\beta}_{Ham}$ given by Eq.~\eqref{18}. The results are shown in Fig.~\ref{nonlocal_maj_stringsite}.

In the non-local geometry, the additional links provide
alternative paths between the boundaries, modifying the
propagation amplitudes relative to the nearest-neighbour
chain. These links connect sites within each flavour sector
and do not themselves introduce flavour mixing when
$Y_{Ham}^{\alpha,\beta}$ is diagonal. In the boundary-Yukawa
channel, mixing instead arises from the non-diagonal
$Y_{yuk}^{\alpha,\beta}$, combined with the different
propagation amplitudes of the heavy flavour sectors.
The resulting angles depend on this relative flavour
structure, rather than simply on the overall suppression
of propagation.

For the chosen benchmark,
Fig.~\ref{nonlocal_maj_stringsite}(a) shows a pronounced
interior peak near $63^\circ$ in $\theta_{12}$.
The $\theta_{13}$ and $\theta_{23}$ distributions favour
smaller angles but retain substantial weight at intermediate
and large angles. The interior peak reflects
the chosen input flavour structure together with propagation
through the heavy sector. It should not be interpreted as
a universal prediction of non-local connectivity.

In the Hamiltonian channel, non-diagonal
$Y_{Ham}^{\alpha,\beta}$ allows flavour conversion within
the heavy sector itself. Disorder changes the amplitudes
connecting different flavours, producing broad mixing-angle
distributions for this benchmark, as shown in
Fig.~\ref{nonlocal_maj_stringsite}(b).
These are qualitatively similar to the local and Petersen
results, with $\theta_{12}$ and $\theta_{13}$ favouring
smaller angles and $\theta_{23}$ favouring larger angles.

\subsection{Majorana}
\paragraph{\textbf{Local}}
For the local geometry scenario, we will use the Hamiltonian eq.(\ref{local-hamiltonian}) in the Lagrangian eq.(\ref{craig}). 
Due to strong disorder in the diagonal terms $\epsilon_i$'s as compared to off-diagonal terms $ t_i$'s (hopping terms), the Anderson mechanism kicks in and gives localized modes. Localisation suppresses boundary propagation in both cases,
while the Majorana construction introduces additional
seesaw suppression.

\begin{figure}[ht]
    \begin{subfigure}{0.48\textwidth}
        \centering
        \includegraphics[width=\linewidth]{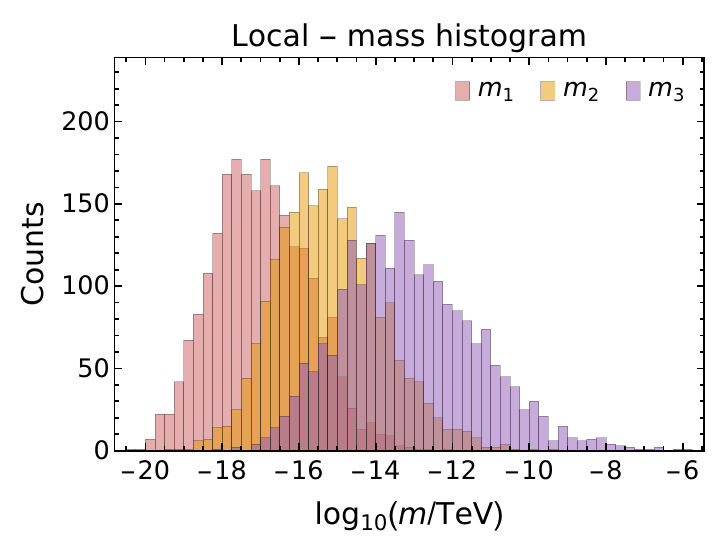}
        \caption{}
        \label{subfig:local-logmass-maj}
    \end{subfigure}
    \hfill
    \begin{subfigure}{0.48\textwidth}
        \centering
        \includegraphics[width=\linewidth]{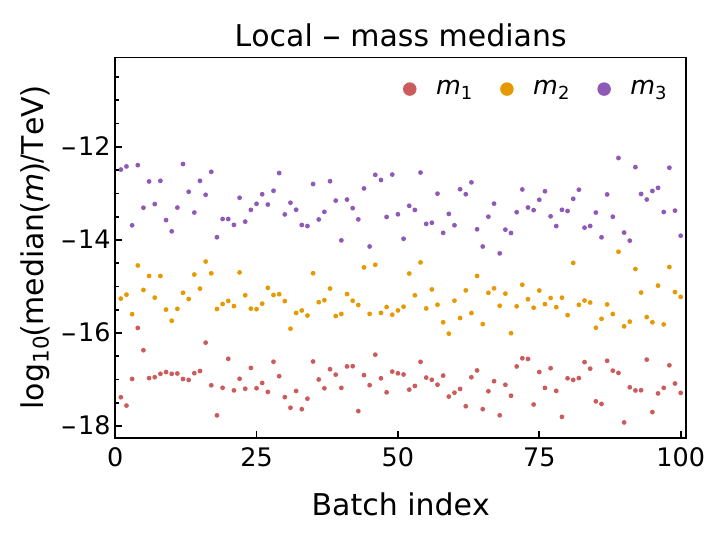}
        \caption{}
        \label{subfig:local-mass-maj}
    \end{subfigure}
    
    \caption{Histograms of $\log_{10}$(mass/TeV) over various runs (left) and median of the three lightest masses over 100 runs (right). Local geometry, strong site disorder (Majorana),
with Majorana-coupling mixing:
$\epsilon_i\in[-5,5]\,\mathrm{TeV}$,
$W=5\,\mathrm{TeV}$, $N=8$, and
$t\simeq0.33\,\mathrm{TeV}$.}
    \label{local-Majorana-strong}
\end{figure}

\begin{figure}[ht]
    \begin{subfigure}{0.48\textwidth}
        \centering
        \includegraphics[width=\linewidth]{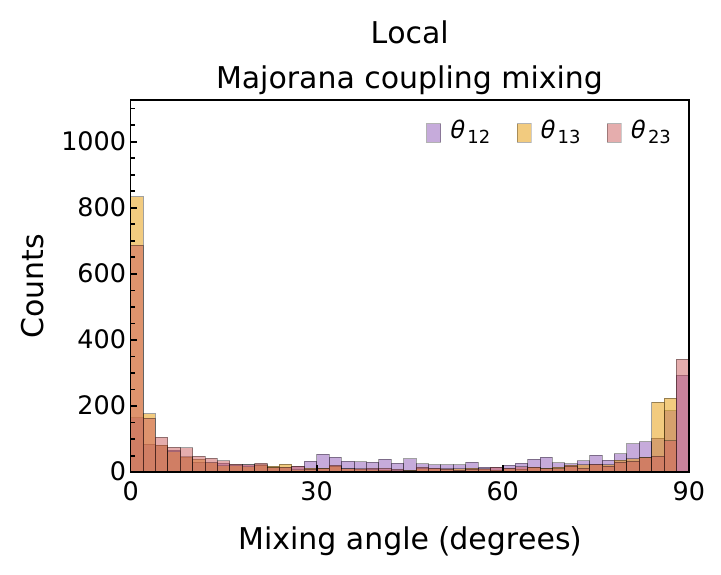}
        \caption{}
        \label{subfig:local-yukangle-maj}
    \end{subfigure}
    \hfill
    \begin{subfigure}{0.48\textwidth}
        \centering
        \includegraphics[width=\linewidth]{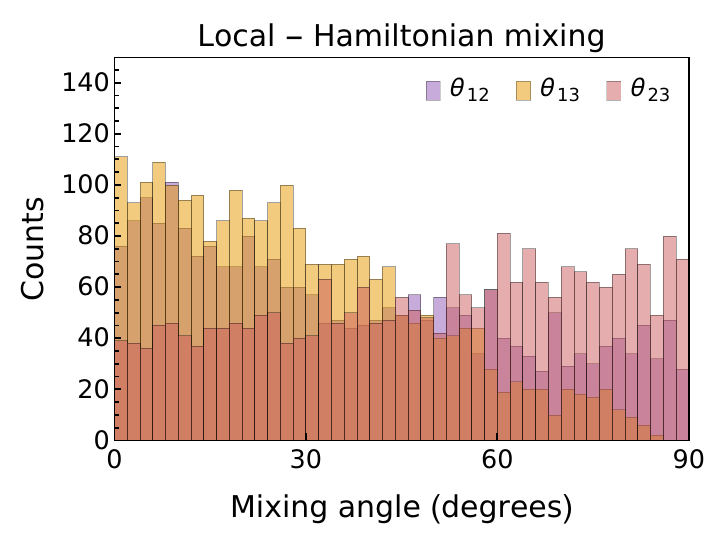}
        \caption{}
        \label{subfig:local-hamangle-maj}
    \end{subfigure}
    
    \caption{Histogram of PMNS mixing angles in the local geometry (Majorana, strong site disorder): Majorana coupling $W^{\alpha,\beta}$ (left) and Hamiltonian mixing $Y^{\alpha,\beta}_{Ham}$ (right), with $\mathbf W=\frac{W}{2}(Y_{\rm yuk}+Y_{\rm yuk}^{T})$
in the Majorana-coupling channel and $Y_{\rm Ham}$
as in eq.~\eqref{18} in the Hamiltonian channel.
Parameters: $\epsilon_i\in[-5,5]\,\mathrm{TeV}$,
$W=5\,\mathrm{TeV}$, $N=8$, and
$t\simeq0.33\,\mathrm{TeV}$.}
    \label{maj-strong-site-local}
\end{figure}

For this local Majorana scenario, both mixing channels use
$N=8$, with site masses sampled uniformly from
$\epsilon_i\in[-5,5]\,\mathrm{TeV}$ and fixed hopping
$t\simeq0.33\,\mathrm{TeV}$.
The Majorana mass prefactor is $W=5\,\mathrm{TeV}$.
The coupling to $\Psi$ is fixed and flavour diagonal,
$t^{\alpha,\beta}=(0.20\,\mathrm{TeV})\delta_{\alpha\beta}$,
and the active boundary prefactor is $0.173\,\mathrm{TeV}$.
The flavour matrices follow the strong-site Petersen
Majorana prescription, using the matrices in Eq.~\eqref{18}.
In the Majorana-coupling channel,
$\mathbf W=\frac{W}{2}(Y_{\rm yuk}+Y_{\rm yuk}^{T})$
and $Y_{\rm Ham}=I_3$.
In the Hamiltonian channel, $\mathbf W=W I_3$
and $Y_{\rm Ham}$ is given in eq.(\ref{18}). The results are illustrated in Fig.~\ref{maj-strong-site-local}. Furthermore, Fig.~\ref{local-Majorana-strong} provides specific neutrino masses for this local Majorana case: the left panel displays the distribution of the three smallest masses, while the right panel shows the median of masses over 100 runs.

In the Majorana-coupling channel, the heavy tower flavours
propagate separately and communicate through the
non-diagonal $W^{\alpha,\beta}$.
Strong localisation along the nearest-neighbour chain
can produce very unequal propagation amplitudes.
These amplitudes weight the Majorana flavour structure
on both sides of the effective light mass matrix,
which can favour alignment with particular flavour
directions. This provides intuition for the strong
endpoint enhancement observed in this benchmark.
The distributions nevertheless retain intermediate-angle
populations, so the mixing cannot be described as
absent or negligible. Compared with the non-local
and Petersen benchmarks, the local distributions
show stronger endpoint concentration.

In the Hamiltonian channel, non-diagonal
$Y_{Ham}^{\alpha,\beta}$ allows flavours to mix during
propagation through the heavy sector, even when
$W^{\alpha,\beta}$ is diagonal. Disorder changes both
the propagation amplitudes and their flavour composition.
For the chosen benchmark, the resulting distributions
are broad, with $\theta_{12}$ and $\theta_{13}$ favouring
smaller angles and $\theta_{23}$ favouring larger angles. These findings are illustrated in the accompanying Fig.~\ref{maj-strong-site-local}.

\begin{figure}[ht]
    \begin{subfigure}{0.48\textwidth}
        \centering
        \includegraphics[width=\linewidth]{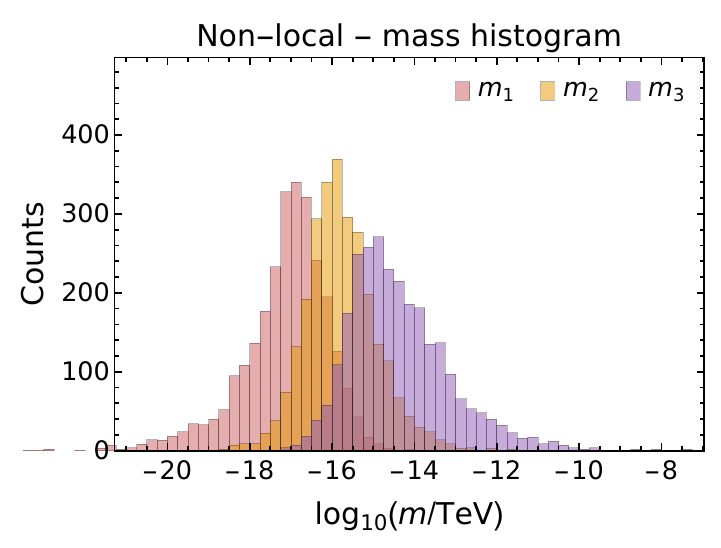}
        \caption{}
        \label{subfig:nonlocal-logmass-maj}
    \end{subfigure}
    \hfill
    \begin{subfigure}{0.48\textwidth}
        \centering
        \includegraphics[width=\linewidth]{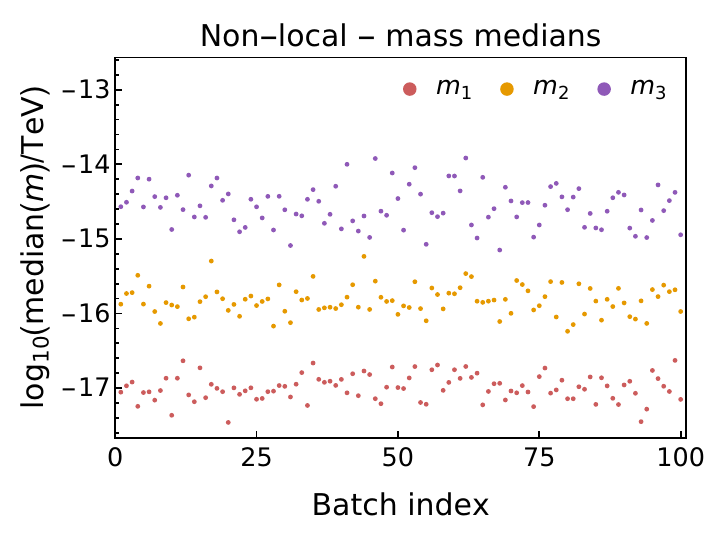}
        \caption{}
        \label{subfig:nonlocal-mass-maj}
    \end{subfigure}
    
    \caption{Histograms of $\log_{10}$(mass/TeV) over various runs (left) and median of the three lightest masses over 100 runs (right). Non-local geometry, strong site disorder (Majorana),
with Majorana-coupling mixing:
$\epsilon_i\in[-5,5]\,\mathrm{TeV}$,
$W=5\,\mathrm{TeV}$, $b=5$, $N=8$, and
$t\simeq0.33\,\mathrm{TeV}$.}
    \label{nonlocal_masses_maj_ss}
\end{figure}

\begin{figure}
    \begin{subfigure}{0.48\textwidth}
        \centering
        \includegraphics[width=\linewidth]{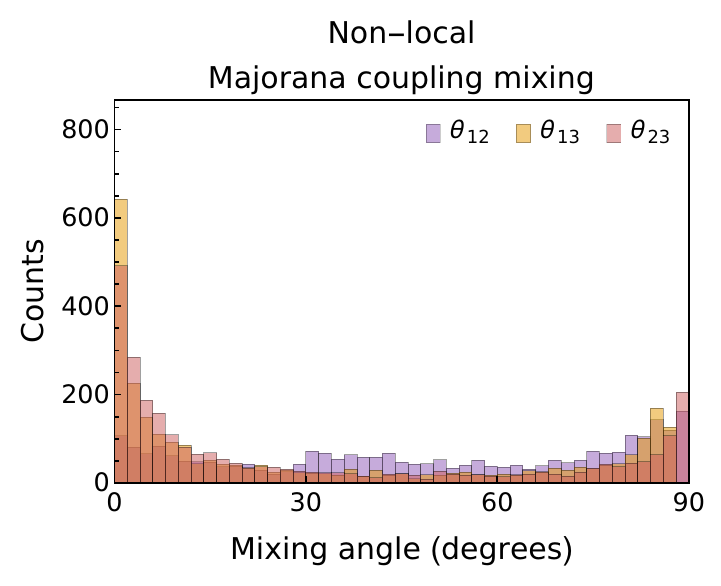}
        \caption{}
        \label{subfig:nl-yukangle-maj}
    \end{subfigure}
    \hfill
    \begin{subfigure}{0.48\textwidth}
        \centering
        \includegraphics[width=\linewidth]{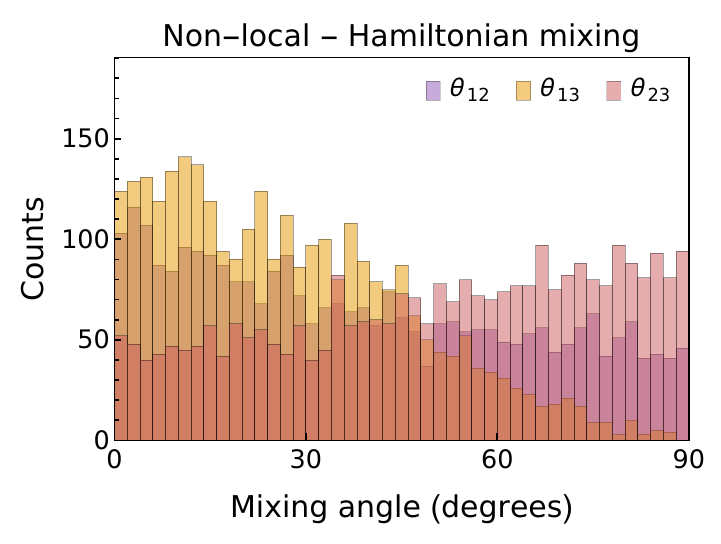}
        \caption{}
        \label{subfig:nl-hamangle-maj}
    \end{subfigure}
    
    \caption{Histogram of mixing angles for several runs in the non-local theory space for Majorana mixing $W^{\alpha,\beta}$ (left) and Hamiltonian mixing $Y^{\alpha,\beta}_{Ham}$ (right) with $\mathbf W=\frac{W}{2}(Y_{\rm yuk}+Y_{\rm yuk}^{T})$
in the Majorana-coupling channel and $Y_{\rm Ham}$
as in eq.~\eqref{18} in the Hamiltonian channel.
Parameters: $\epsilon_i\in[-5,5]\,\mathrm{TeV}$,
$W=5\,\mathrm{TeV}$, $b=5$, $N=8$, and
$t\simeq0.33\,\mathrm{TeV}$.}
    \label{maj-strong-site-nl}
\end{figure}

\paragraph{\textbf{Non-Local}}
We now extend our analysis of the non-local geometry to the Majorana neutrino model, used within the Lagrangian of eq.~(\ref{craig}). Both non-local Majorana benchmarks use $N=8$,
with site masses sampled uniformly from
$\epsilon_i\in[-5,5]\,\mathrm{TeV}$, decay base $b=5$,
and fixed hopping prefactor $t\simeq0.33\,\mathrm{TeV}$.
The Majorana mass prefactor, boundary couplings and
flavour matrices are the same as in the local
strong-site Majorana benchmarks described above. The light-mass distributions are shown in Fig.~\ref{nonlocal_masses_maj_ss}.
Their scales depend on the weighted propagation paths and
the Majorana mass matrix. Weaker suppression of the light
masses does not by itself imply less localised heavy modes.
 Following our established methodology, we investigate two mixing scenarios: first, non-diagonal right-handed Majorana couplings ($W^{\alpha,\beta}$), whose dimensionless flavour-matrix entries are
$\mathcal{O}(1)$ Fig.~\ref{maj-strong-site-nl} (left), and second, diagonal $W^{\alpha,\beta}$ couplings combined with off-block Hamiltonian mixing ($Y^{\alpha,\beta}_{Ham}$) Fig.~\ref{maj-strong-site-nl} (right).

In the non-local geometry, the additional links provide
alternative paths between the boundaries and modify
the relative propagation amplitudes of the flavour
sectors. In the Majorana-coupling channel, these sectors
remain separate within the heavy tower and communicate
through the non-diagonal $W^{\alpha,\beta}$.
Their propagation amplitudes weight the Majorana
flavour structure on both sides of the effective
light mass matrix. Geometry can therefore affect
the mixing pattern through these relative weights,
rather than simply through the overall mass suppression.

For the chosen benchmark, the Majorana-coupling
distributions show endpoint enhancement together
with broad intermediate-angle populations.
The endpoint concentration is weaker than in the
local benchmark and qualitatively resembles the
Petersen result.

In the Hamiltonian channel, non-diagonal
$Y_{Ham}^{\alpha,\beta}$ allows flavours to mix
within the heavy tower itself. Disorder changes
both the strength and flavour composition of
propagation. The resulting distributions are broad,
with $\theta_{12}$ and $\theta_{13}$ favouring
smaller angles and $\theta_{23}$ favouring larger
angles. Their qualitative similarity to the local
and Petersen benchmarks indicates less pronounced
geometry dependence in this channel.

\section{Numerical Results: Weak Site Disorder}
\label{app:weak_site}
We provide in this appendix the detailed numerical analysis for the local and non-local geometries, considering both Dirac and Majorana neutrino cases with weak disorder in site elements $\epsilon_i$s.

\subsection{Dirac}
\begin{figure}[ht]
    \centering
    \begin{subfigure}{0.33\textwidth}
        \centering
        \includegraphics[width=1\textwidth]{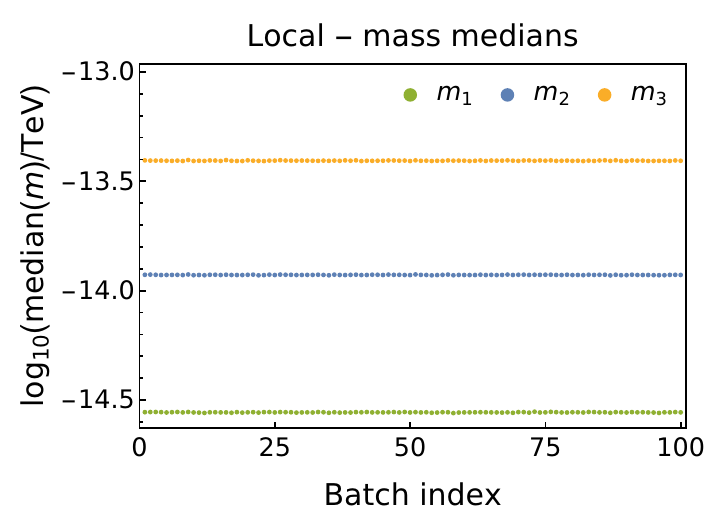}
        \caption{}
        \label{subfig:local-weak-mass}
    \end{subfigure}
    \hfill
    \begin{subfigure}{0.325\textwidth}
        \centering
        \includegraphics[width=1\textwidth]{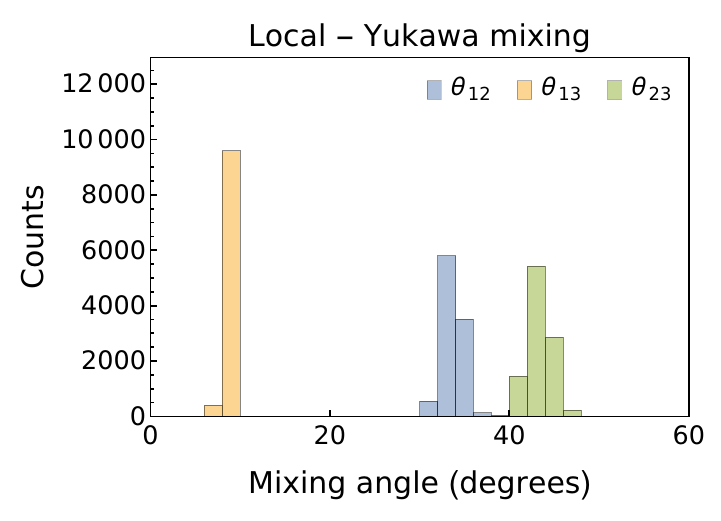}
        \caption{}
        \label{subfig:local-weak-yukangle}
    \end{subfigure}
    \hfill
    \begin{subfigure}{0.325\textwidth}
        \centering
        \includegraphics[width=1\textwidth]{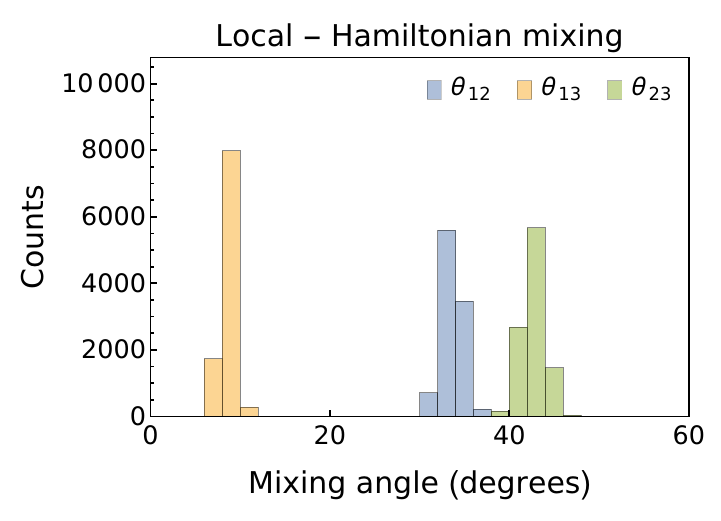}
        \caption{}
        \label{subfig:local-weak-hamangle}
    \end{subfigure}
    
    \caption{Median of the three lightest masses (in TeV, $\log_{10}$ scale) over 100 runs (left) and histograms of PMNS mixing angles for Yukawa mixing $Y^{\alpha,\beta}_{yuk}$ (middle) and Hamiltonian mixing $Y^{\alpha,\beta}_{Ham}$ (right). The flavour matrices are given in
Eq.~\eqref{eq:local-weak-site-dirac-flavour}.
Local geometry, weak site disorder:
$\epsilon_i\in[9.8,10.2]\,\mathrm{TeV}$.
The Yukawa-mixing benchmark uses $N=6$ and
$t\simeq0.057\,\mathrm{TeV}$.
The Hamiltonian-mixing benchmark uses $N=9$ and
$t\simeq0.36\,\mathrm{TeV}$.}
    \label{local-weak-site-dirac}
\end{figure}

\paragraph{\textbf{Local}}
The uniform nearest-neighbour chain provides a reference
for understanding this regime. Its single-flavour
eigenvalues follow a cosine spectrum centred on the
common site mass. When the spectral splittings are
small compared with this mass, cancellations between
mode contributions suppress propagation between
distinct boundary sites. This is the GIM-like
suppression discussed above. Site disorder perturbs
these contributions, but weak variation relative
to the common mass does not by itself guarantee
that the eigenvectors are only weakly perturbed.
The stability of the resulting flavour structure
must therefore be assessed from the mixing distributions. Now, once diagonal entries are randomised with tiny amplitudes, the spectrum distribution is also perturbed but follows the same overall mass distribution pattern. 
Both local Dirac benchmarks use site masses sampled
uniformly from $\epsilon_i\in[9.8,10.2]\,\mathrm{TeV}$.
The Yukawa-mixing benchmark uses $N=6$ and fixed hopping
$t\simeq0.057\,\mathrm{TeV}$.
The Hamiltonian-mixing benchmark uses $N=9$ and
$t\simeq0.36\,\mathrm{TeV}$.
The site-disorder half-width is fixed at
$0.2\,\mathrm{TeV}$ independently of the hopping.
The respective flavour matrices are
\begin{equation}
\begin{aligned}
Y_{\rm yuk} &\simeq
\begin{pmatrix}
0.14 & -0.050 & 0.050 \\
0.33 & 0.27 & -0.50 \\
0.34 & 1.5 & 1.5
\end{pmatrix},
\hspace{2cm}
Y_{\rm Ham} &\simeq
\begin{pmatrix}
1.7 & -1.2 & 0.70 \\
0.79 & 0.30 & -0.46 \\
0.25 & 0.052 & 0.19
\end{pmatrix}.
\end{aligned}
\label{eq:local-weak-site-dirac-flavour}
\end{equation}
Fig.~\ref{local-weak-site-dirac} demonstrates the scale of mass produced (left) and the mixing angles generated for both Yukawa $Y_{yuk}^{\alpha,\beta}$ (middle) and Hamiltonian mixing $Y^{\alpha,\beta}_{Ham}$ (right).
As can be seen from Fig.~\ref{local-weak-site-dirac}, both Yukawa and Hamiltonian mixing give rise to non-anarchical mixing.

Both flavour channels produce narrowly peaked
mixing-angle distributions for the chosen local
benchmarks. In the boundary-Yukawa channel, the
relative propagation amplitudes of the separate
heavy flavour sectors combine with the fixed
$Y_{yuk}^{\alpha,\beta}$ to determine the light
mixing matrix. In the Hamiltonian channel,
$Y_{Ham}^{\alpha,\beta}$ introduces flavour mixing
within the heavy sector itself.

The narrow distributions indicate that the
directions which diagonalise the effective light
mass matrix vary little across these disorder
realisations. This explains how a stable mixing
pattern can coexist with small masses generated
by suppressed boundary propagation.
The peak locations depend on the selected flavour
matrices and model parameters, rather than being
fixed by weak disorder alone.
Table~\ref{table-weak-site-dirac} lists the parameters,
and Table~\ref{Table-weak-site-summary} summarises
the mixing patterns across the three geometries.

\paragraph{\textbf{Non-Local}}
The non-local links modify both the heavy spectrum
and the paths connecting the boundary sites.
Their strengths depend on the hopping prefactor
and decay base, so connectivity alone does not
determine the efficiency of mass suppression.
For the chosen benchmarks, the common site mass
dominates the heavy spectrum, allowing cancellations
between mode contributions to suppress boundary
propagation. The resulting light masses depend
on this boundary propagator, rather than on the
spectral spread alone. The parameters used for the numerical results in this non-local geometry scenario are mentioned in Table~\ref{table-weak-site-dirac}. Both non-local Dirac benchmarks use $N=16$, $b=4$,
and site masses sampled uniformly from
$\epsilon_i\in[9.8,10.2]\,\mathrm{TeV}$.
The fixed hopping prefactor is
$t\simeq0.063\,\mathrm{TeV}$ for Yukawa mixing and
$t\simeq0.064\,\mathrm{TeV}$ for Hamiltonian mixing.
The site-disorder half-width is fixed at
$0.2\,\mathrm{TeV}$ independently of the hopping.
The respective flavour matrices are
\begin{equation}
\begin{aligned}
Y_{\rm yuk} &\simeq
\begin{pmatrix}
0.39 & -0.24 & 0.13 \\
0.53 & 0.54 & -0.60 \\
0.30 & 1.4 & 1.4
\end{pmatrix}, \hspace{2cm}
Y_{\rm Ham} &\simeq
\begin{pmatrix}
1.6 & -0.99 & 0.58 \\
0.59 & 0.56 & -0.67 \\
0.10 & 0.34 & 0.33
\end{pmatrix}.
\end{aligned}
\label{eq:nonlocal-weak-site-dirac-flavour}
\end{equation}
\begin{figure}[ht]
    \begin{subfigure}{0.335\textwidth}
        \centering
        \includegraphics[width=1\textwidth]{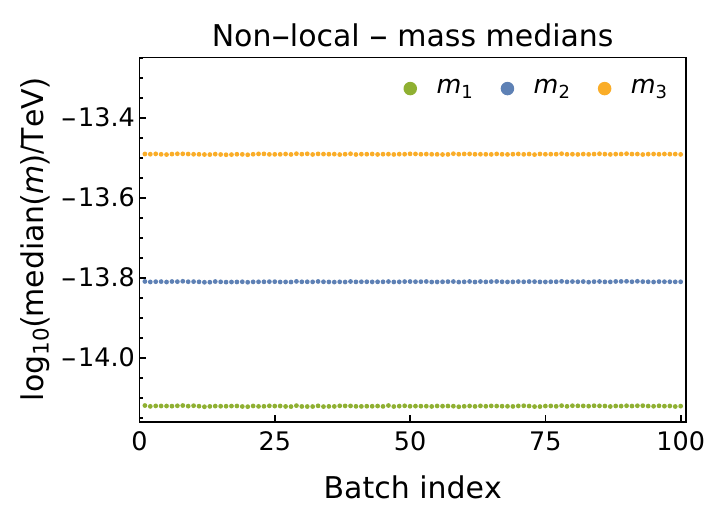}
        \caption{}
        \label{subfig:nonlocal-weak-mass}
    \end{subfigure}
    \hfill
    \begin{subfigure}{0.325\textwidth}
        \centering
        \includegraphics[width=1\textwidth]{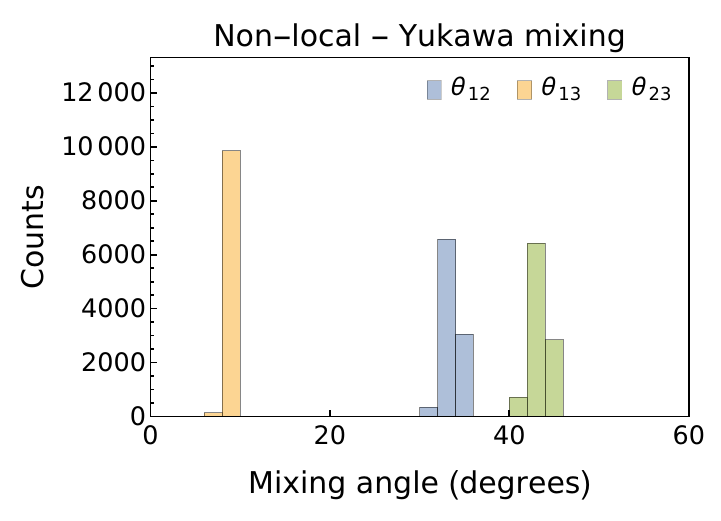}
        \caption{}
        \label{subfig:nonlocal-weak-yukangle}
    \end{subfigure}
    \hfill
    \begin{subfigure}{0.325\textwidth}
        \centering
        \includegraphics[width=1\textwidth]{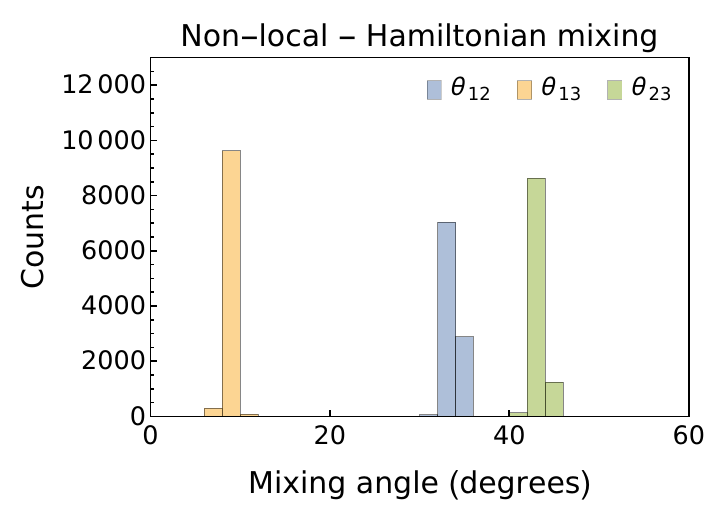}
        \caption{}
        \label{subfig:nonlocal-weak-hamangle}
    \end{subfigure}
    
    \caption{Median of the three lightest masses (in TeV, $\log_{10}$ scale) over 100 runs (left) and histograms of PMNS mixing angles for Yukawa mixing $Y^{\alpha,\beta}_{yuk}$ (middle) and Hamiltonian mixing $Y^{\alpha,\beta}_{Ham}$ (right). The flavour matrices are given in
Eq.~\eqref{eq:nonlocal-weak-site-dirac-flavour}.
Non-local geometry, weak site disorder:
$\epsilon_i\in[9.8,10.2]\,\mathrm{TeV}$,
$b=4$, and $N=16$.
The fixed hopping prefactor is
$t\simeq0.063\,\mathrm{TeV}$ for Yukawa mixing and
$t\simeq0.064\,\mathrm{TeV}$ for Hamiltonian mixing.}
    \label{non-local-weak-site-dirac}
\end{figure}

Fig. \ref{non-local-weak-site-dirac} demonstrates the median of three masses produced (left), the mixing angles due to Yukawa mixing $Y_{yuk}^{\alpha,\beta}$ (middle) and Hamiltonian mixing $Y^{\alpha,\beta}_{Ham}$ (right).
Both flavour channels produce narrowly peaked
mixing-angle distributions for these benchmarks.
In the boundary-Yukawa channel, the fixed
$Y_{yuk}^{\alpha,\beta}$ combines with the relative
propagation amplitudes of the separate heavy
flavour sectors. In the Hamiltonian channel,
$Y_{Ham}^{\alpha,\beta}$ mixes flavours within
the heavy sector itself. In either case, the
observed concentration indicates that weak site
fluctuations produce only small changes in the
directions that diagonalise the effective light
mass matrix. Small masses and stable mixing
angles can therefore coexist without requiring
the angles themselves to be small.
Their central values depend on the chosen
flavour matrices and model parameters.
Table~\ref{Table-weak-site-summary} summarises
the mixing patterns across the three geometries.

\subsection{Majorana}
\paragraph{\textbf{Local}}
The local geometry weak site disorder scenario with Majorana neutrinos has the same underlying working mechanism as the Dirac scenario, i.e. GIM-like cancellation. The Lagrangian and Hamiltonian for this scenario are given by eq.(\ref{craig}) and eq.(\ref{local-hamiltonian}), respectively. 
\begin{figure}[ht]
    \centering
    \begin{subfigure}{0.32\textwidth}
        \centering
        \includegraphics[width=1\textwidth]{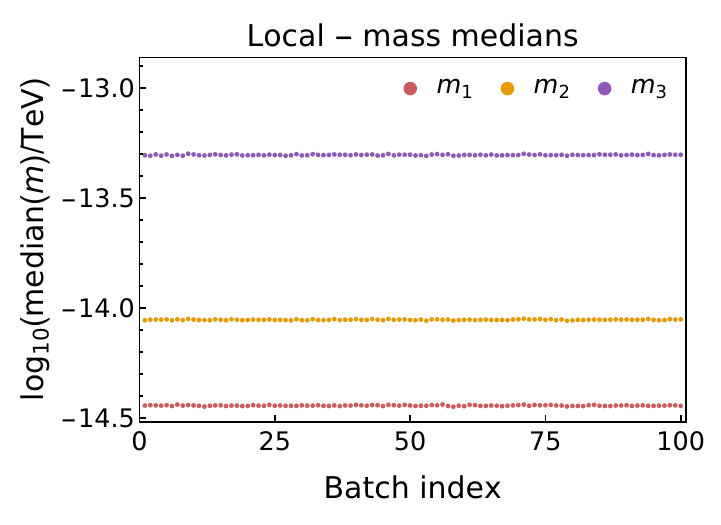}
        \caption{}
        \label{subfig:local-weak-hammass}
    \end{subfigure}
    \hfill
    \begin{subfigure}{0.32\textwidth}
        \centering
        \includegraphics[width=1\textwidth]{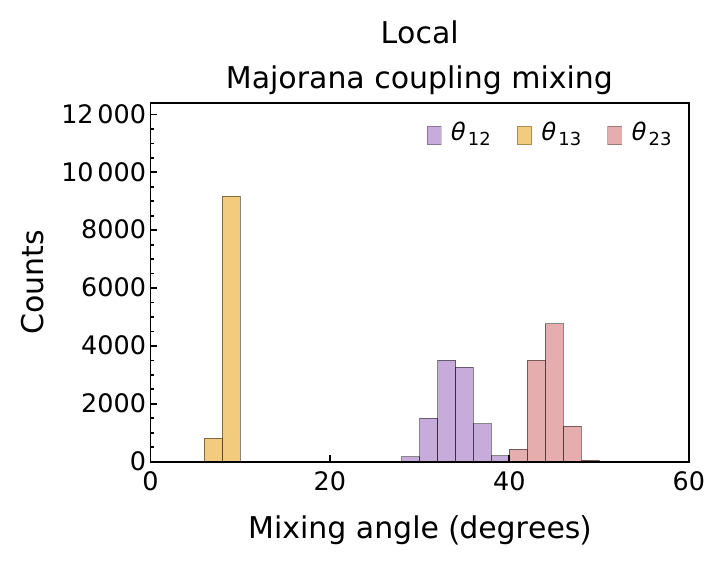}
        \caption{}
        \label{subfig:local-weak-yukangle-maj}
    \end{subfigure}
    \hfill
    \begin{subfigure}{0.32\textwidth}
        \centering
        \includegraphics[width=1\textwidth]{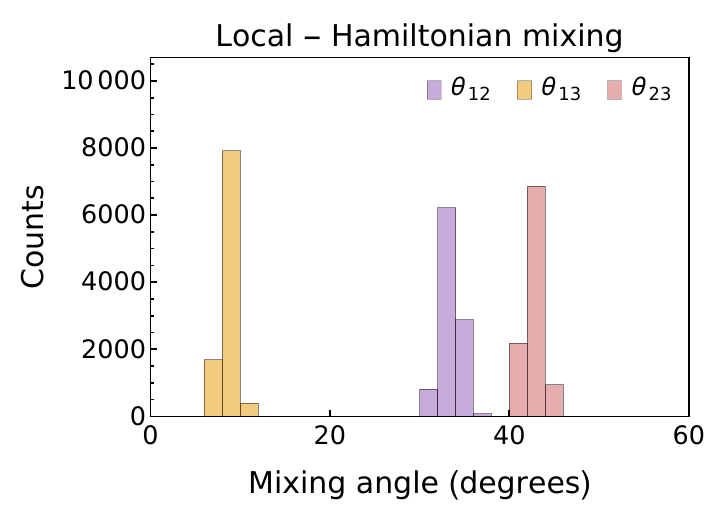}
        \caption{}
        \label{subfig:local-weak-hamangle-maj}
    \end{subfigure}
    
    \caption{Median of the three lightest masses (in TeV, $\log_{10}$ scale) over 100 runs (left) and histograms of PMNS mixing angles for Majorana coupling $W^{\alpha,\beta}$ (middle) and Hamiltonian mixing $Y^{\alpha,\beta}_{Ham}$ (right). The flavour matrices are given in
Eq.~\eqref{eq:local-weak-site-majorana-flavour}.
Local geometry, weak site disorder:
$\epsilon_i\in[9.8,10.2]\,\mathrm{TeV}$ and $N=5$.
The fixed hopping is $t\simeq0.69\,\mathrm{TeV}$ for
Majorana-coupling mixing and $t=0.70\,\mathrm{TeV}$
for Hamiltonian mixing.}
    \label{local-weak-site-majorana}
\end{figure}

Fig. \ref{local-weak-site-majorana} demonstrates the median of three masses produced (left), the mixing angles due to Majorana mixing $W^{\alpha,\beta}$(middle) and Hamiltonian mixing $Y^{\alpha,\beta}_{Ham}$ (right). The parameters are listed in
Table~\ref{table-maj-site-weak-param}.
Both local Majorana benchmarks use $N=5$ and site masses
sampled uniformly from
$\epsilon_i\in[9.8,10.2]\,\mathrm{TeV}$.
The fixed hopping is $t\simeq0.69\,\mathrm{TeV}$ for
Majorana-coupling mixing and $t=0.70\,\mathrm{TeV}$
for Hamiltonian mixing.
The site-disorder half-width is fixed at
$0.2\,\mathrm{TeV}$ independently of the hopping.
The coupling to $\Psi$ is fixed and flavour diagonal,
$t^{\alpha,\beta}=t\,\delta_{\alpha\beta}$.
The respective flavour matrices are
\begin{equation}
\begin{aligned}
\mathbf W &\simeq
(10\,\mathrm{TeV})
\begin{pmatrix}
1.6 & -0.54 & 0.22 \\
-0.54 & 0.77 & -0.50 \\
0.22 & -0.50 & 0.59
\end{pmatrix}, \hspace{2cm}
Y_{\rm Ham} &\simeq
\begin{pmatrix}
1.6 & -0.98 & 0.57 \\
0.57 & 0.53 & -0.62 \\
0.074 & 0.32 & 0.31
\end{pmatrix}.
\end{aligned}
\label{eq:local-weak-site-majorana-flavour}
\end{equation}
In the Majorana-coupling channel, $Y_{\rm Ham}=I_3$.
In the Hamiltonian channel,
$\mathbf W=(10\,\mathrm{TeV})I_3$. The plotted mass medians lie at the sub-eV scale,
while both flavour channels produce peaked
mixing-angle distributions. In the Majorana-coupling
channel, propagation through the separate heavy
flavour sectors weights the non-diagonal
$W^{\alpha,\beta}$. In the Hamiltonian channel,
flavours also mix within the tower through
$Y_{Ham}^{\alpha,\beta}$.
For these benchmarks, weak site fluctuations
produce only limited changes in the light mixing
matrix, preserving the concentration around the
values selected by the input flavour structure.
The Majorana-coupling distributions are somewhat
wider, particularly for $\theta_{12}$, but do not
indicate negligible mixing or inefficient transfer
of flavour mixing to the active sector.

\paragraph{\textbf{Non-Local}}
The additional non-local links modify propagation
between the boundaries and hence the effective
Majorana mass matrix. The resulting suppression
depends on the weighted links, boundary couplings
and Majorana mass scale, so connectivity alone
does not determine its efficiency. The one-flavour Hamiltonian and Lagrangian are given by eq.(\ref{nonlocal-hamiltonian}) and eq.(\ref{craig}), respectively. The parameters are listed in
Table~\ref{table-maj-site-weak-param}.
Both non-local Majorana benchmarks use $b=1.5$ and
site masses sampled uniformly from
$\epsilon_i\in[9.8,10.2]\,\mathrm{TeV}$.
The Majorana-coupling benchmark uses $N=16$ and fixed
hopping prefactor $t\simeq0.39\,\mathrm{TeV}$.
The Hamiltonian benchmark uses $N=8$ and
$t\simeq0.057\,\mathrm{TeV}$.
The site-disorder half-width is fixed at
$0.2\,\mathrm{TeV}$ independently of the hopping.
The coupling to $\Psi$ is fixed and flavour diagonal,
$t^{\alpha,\beta}=t\,\delta_{\alpha\beta}$.
The respective flavour matrices are
\begin{equation}
\begin{aligned}
\mathbf W &\simeq
(10\,\mathrm{TeV})
\begin{pmatrix}
1.8 & -0.57 & 0.23 \\
-0.57 & 0.97 & -0.50 \\
0.23 & -0.50 & 0.75
\end{pmatrix}, \hspace{2cm}
Y_{\rm Ham} &\simeq
\begin{pmatrix}
1.6 & -0.98 & 0.58 \\
0.56 & 0.54 & -0.64 \\
0.051 & 0.35 & 0.32
\end{pmatrix}.
\end{aligned}
\label{eq:nonlocal-weak-site-majorana-flavour}
\end{equation}
In the Majorana-coupling channel, $Y_{\rm Ham}=I_3$.
In the Hamiltonian channel,
$\mathbf W=(10\,\mathrm{TeV})I_3$.
\begin{figure}[ht]
    \centering
    \begin{subfigure}{0.32\textwidth}
        \centering
        \includegraphics[width=1\textwidth]{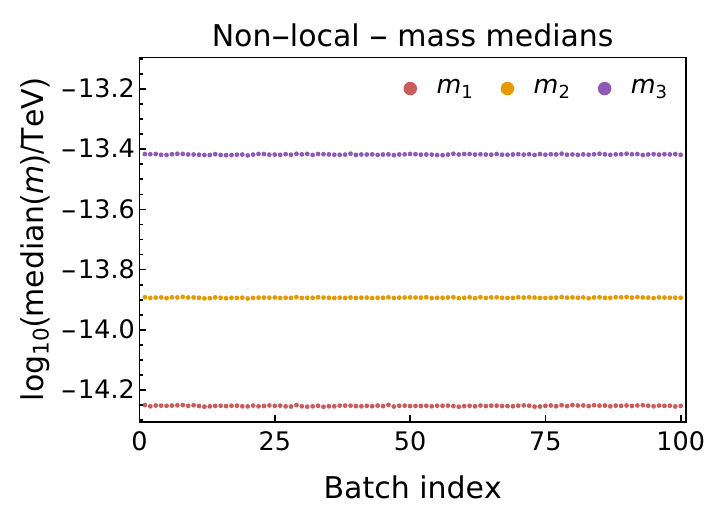}
        \caption{}
        \label{subfig:nl-weak-hammass}
    \end{subfigure}
    \hfill
    \begin{subfigure}{0.32\textwidth}
        \centering
        \includegraphics[width=1\textwidth]{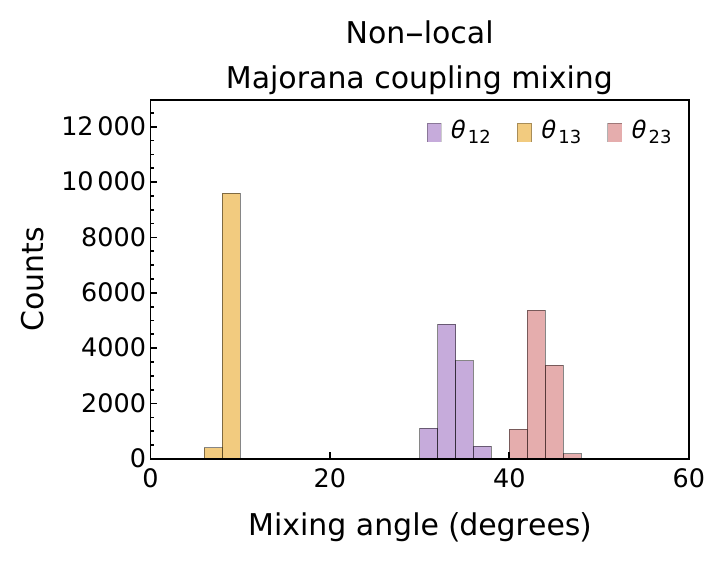}
        \caption{}
        \label{subfig:nl-weak-yukangle}
    \end{subfigure}
    \hfill
    \begin{subfigure}{0.32\textwidth}
        \centering
        \includegraphics[width=1\textwidth]{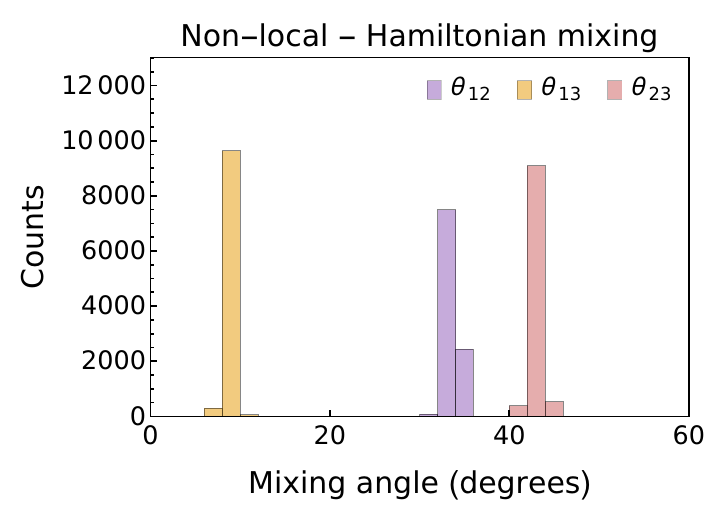}
        \caption{}
        \label{subfig:nl-weak-hamangle}
    \end{subfigure}
    
    \caption{Median of the three lightest masses (in TeV, $\log_{10}$ scale) over 100 runs (left) and histograms of PMNS mixing angles for Majorana coupling $W^{\alpha,\beta}$ (middle) and Hamiltonian mixing $Y^{\alpha,\beta}_{Ham}$ (right). The flavour matrices are given in
Eq.~\eqref{eq:nonlocal-weak-site-majorana-flavour}.
Non-local geometry, weak site disorder:
$\epsilon_i\in[9.8,10.2]\,\mathrm{TeV}$ and $b=1.5$.
The Majorana-coupling benchmark uses $N=16$ and
$t\simeq0.39\,\mathrm{TeV}$.
The Hamiltonian benchmark uses $N=8$ and
$t\simeq0.057\,\mathrm{TeV}$.}
    \label{maj-weak-site-nl}
\end{figure}

Fig.~\ref{maj-weak-site-nl} demonstrates the median of three masses produced (left), the mixing angles due to Majorana mixing $W^{\alpha,\beta}$(middle) and Hamiltonian mixing $Y^{\alpha,\beta}_{Ham}$ (right).
Both flavour channels produce peaked mixing-angle
distributions for the chosen benchmarks.
The Majorana-coupling channel combines the relative
propagation amplitudes with $W^{\alpha,\beta}$,
while the Hamiltonian channel introduces flavour
mixing within the heavy tower.
The narrow distributions indicate that the resulting
light mixing directions remain stable under the
sampled site fluctuations. Their central values
depend on the chosen inputs, and mixing is
not restricted to the Hamiltonian channel.

\section{Numerical Results: Weak Hopping Disorder}
\label{app:weak_hopping}
This appendix presents detailed numerical results for Dirac and Majorana neutrinos with weak hopping disorder $t_i$ in both local and non-local geometries.

\subsection{Dirac}
\paragraph{\textbf{Local}}
The local Hamiltonian eq.(\ref{local-hamiltonian}) is considered in the Lagrangian eq.(\ref{ACS}) with weak disorder in the off-diagonal/hopping terms (neighbouring interactions). 
Due to weak disorder, the mass spectrum is perturbed from the uniform case, the shape of the spectrum is dependent on the geometry. The perturbation in the eigenmass values is tiny enough that one would expect the GIM-like cancellation mechanism would work to generate small mass scales with diagonal terms still of the order of the fundamental scale of the theory. 
\begin{figure}[ht]
    \centering
    \begin{subfigure}{0.32\textwidth}
        \centering
        \includegraphics[width=1\textwidth]{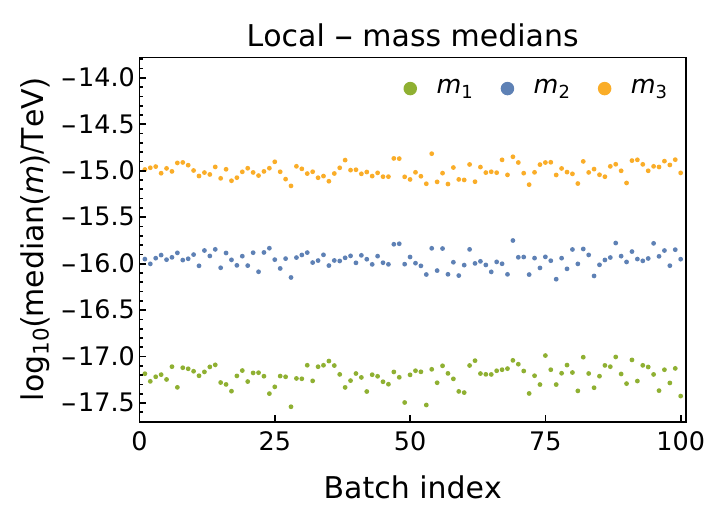}
        \caption{}
        \label{subfig:local-hopp-mass}
    \end{subfigure}
    \hfill
    \begin{subfigure}{0.32\textwidth}
        \centering
        \includegraphics[width=1\textwidth]{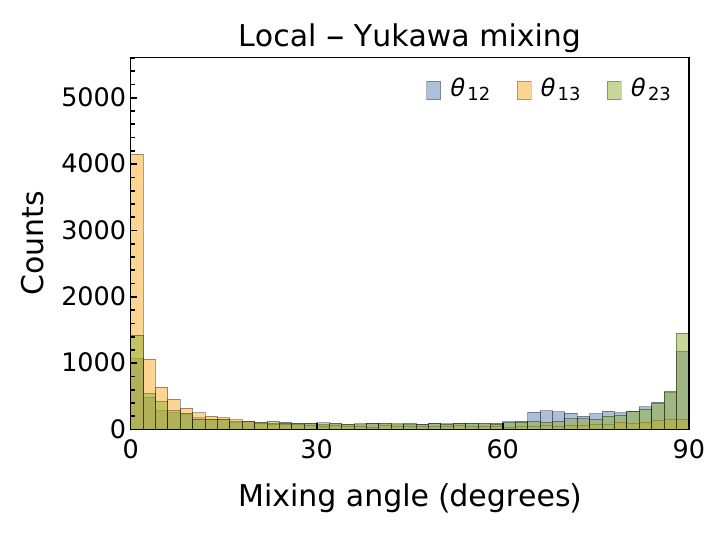}
        \caption{}
        \label{subfig:local-hopp-yukangle}
    \end{subfigure}
    \hfill
    \begin{subfigure}{0.32\textwidth}
        \centering
        \includegraphics[width=1\textwidth]{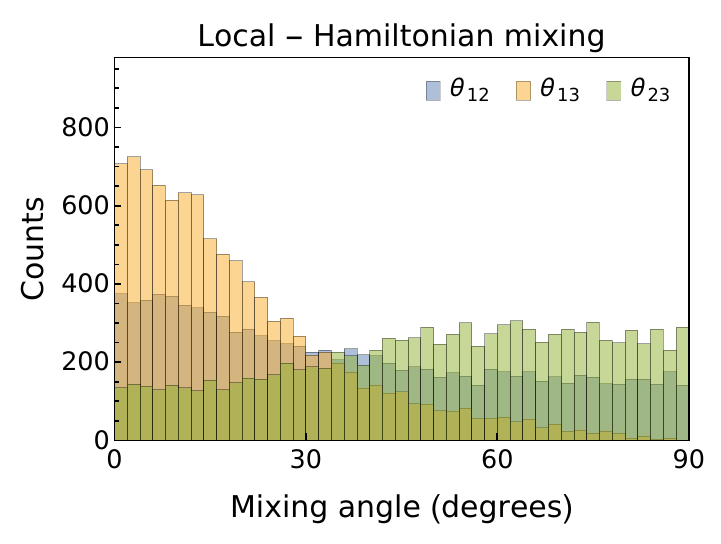}
        \caption{}
        \label{subfig:local-hopp-hamangle}
    \end{subfigure}
    \caption{Median of the three lightest masses (in TeV, $\log_{10}$ scale) over 100 runs (left) and histograms of PMNS mixing angles for Yukawa mixing $Y^{\alpha,\beta}_{yuk}$ (middle) and Hamiltonian mixing $Y^{\alpha,\beta}_{Ham}$ (right). The flavour matrices are given in
Eq.~\eqref{eq:local-weak-site-dirac-flavour}.
Local geometry, weak hopping disorder:
$\epsilon_i=10\,\mathrm{TeV}$, with nearest-neighbour
hopping couplings sampled uniformly from $[-t,t]$.
The Yukawa-mixing benchmark uses $N=6$ and
$t\simeq0.057\,\mathrm{TeV}$.
The Hamiltonian-mixing benchmark uses $N=9$ and
$t\simeq0.36\,\mathrm{TeV}$.}
    \label{local-weak-coupling-dirac}
\end{figure}

Fig.~\ref{local-weak-coupling-dirac} shows the median of three masses produced (left), the mixing angles due to Yukawa mixing $Y^{\alpha,\beta}_{yuk}$ (middle) and Hamiltonian mixing $Y^{\alpha,\beta}_{Ham}$ (right). Both local Dirac benchmarks use uniform site masses
$\epsilon_i=10\,\mathrm{TeV}$.
The nearest-neighbour hopping couplings are sampled
uniformly from $[-t,t]$.
The Yukawa-mixing benchmark uses $N=6$ and
$t\simeq0.057\,\mathrm{TeV}$.
The Hamiltonian-mixing benchmark uses $N=9$ and
$t\simeq0.36\,\mathrm{TeV}$.
The flavour matrices are unchanged from the local
weak-site Dirac benchmarks in
Eq.~\eqref{eq:local-weak-site-dirac-flavour}.
In the local chain, propagation between the boundaries
depends on successive nearest-neighbour hoppings.
Randomising these links can produce strongly unequal
propagation amplitudes across the heavy flavour
sectors, even when every hopping is small compared
with the site mass.
Combined with the fixed boundary Yukawa matrix,
these relative weights can favour flavour alignment,
providing intuition for the endpoint-enhanced
distributions. Intermediate and large angles remain
populated, so the mixing is not absent.
The Hamiltonian channel allows flavour mixing
within the chain and produces broad distributions
for the chosen benchmark.

\paragraph{\textbf{Non-Local}}
The non-local links provide additional paths between
the boundaries. Their weights and interference
affect the boundary propagator, so the number
of links alone does not determine the mass
suppression. For the displayed benchmark, the
light masses are less suppressed than in the
local and Petersen examples.
\begin{figure}[ht]
    \centering
    \begin{subfigure}{0.34\textwidth}
        \centering
        \includegraphics[width=1\textwidth]{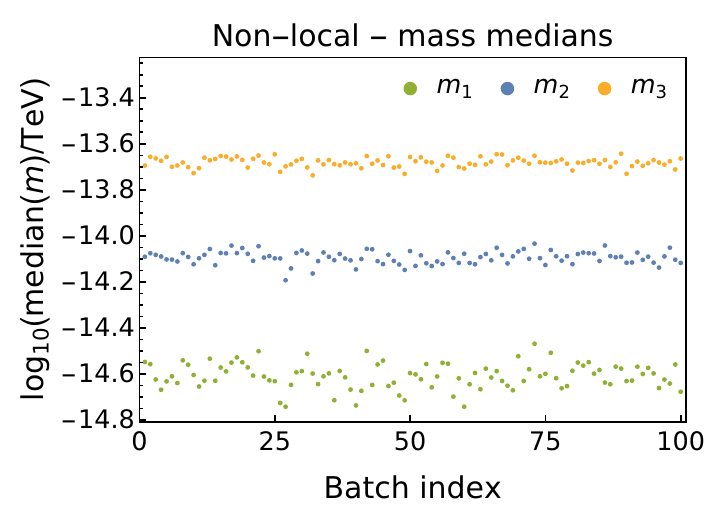}
        \caption{}
        \label{subfig:nl-hopp-mass}
    \end{subfigure}
    \hfill
    \begin{subfigure}{0.32\textwidth}
        \centering
        \includegraphics[width=1\textwidth]{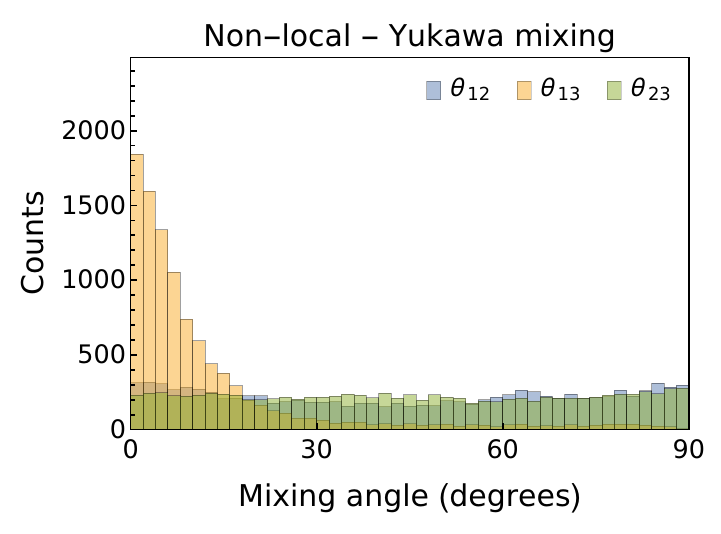}
        \caption{}
        \label{subfig:nl-hopp-yukangle}
    \end{subfigure}
    \hfill
    \begin{subfigure}{0.32\textwidth}
        \centering
        \includegraphics[width=1\textwidth]{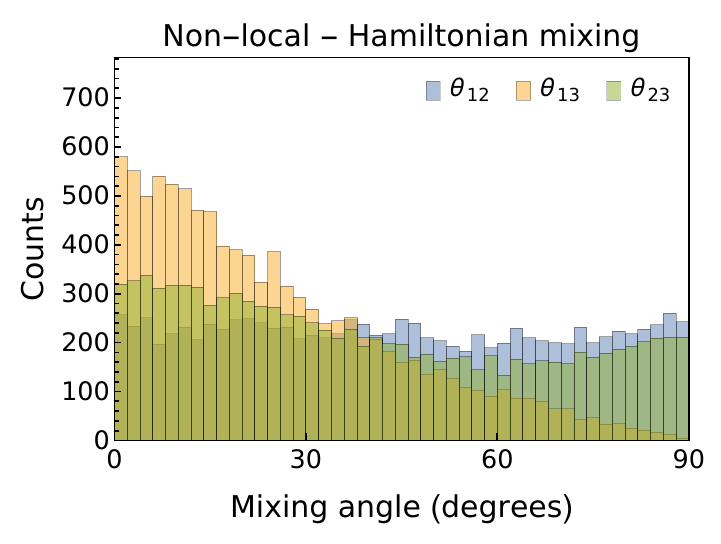}
        \caption{}
        \label{subfig:nl-hopp-hamangle}
    \end{subfigure}
    
    \caption{Median of the three lightest masses (in TeV, $\log_{10}$ scale) over 100 runs (left) and histograms of PMNS mixing angles for Yukawa mixing $Y^{\alpha,\beta}_{yuk}$ (middle) and Hamiltonian mixing $Y^{\alpha,\beta}_{Ham}$ (right). The flavour matrices are given in
Eq.~\eqref{eq:nonlocal-weak-site-dirac-flavour}.
Non-local geometry, weak hopping disorder:
$\epsilon_i=10\,\mathrm{TeV}$, $b=4$, and $N=16$.
Raw hopping couplings are sampled uniformly from
$[-t,t]$ before geometric weighting.
The hopping half-width is
$t\simeq0.063\,\mathrm{TeV}$ for Yukawa mixing and
$t\simeq0.064\,\mathrm{TeV}$ for Hamiltonian mixing.}
    \label{non-local-weak-coupling-dirac}
\end{figure}

Fig. \ref{non-local-weak-coupling-dirac} shows the median of three masses produced (left), the mixing angles due to Yukawa mixing $Y^{\alpha,\beta}_{yuk}$ (middle) and Hamiltonian mixing $Y^{\alpha,\beta}_{Ham}$ (right). Both non-local Dirac benchmarks use $N=16$, $b=4$,
and uniform site masses $\epsilon_i=10\,\mathrm{TeV}$.
The raw hopping couplings are sampled uniformly from
$[-t,t]$ before multiplication by $b^{-|i-j|}$.
The hopping half-width is
$t\simeq0.063\,\mathrm{TeV}$ for Yukawa mixing and
$t\simeq0.064\,\mathrm{TeV}$ for Hamiltonian mixing.
The flavour matrices are unchanged from the non-local
weak-site Dirac benchmarks in
Eq.~\eqref{eq:nonlocal-weak-site-dirac-flavour}.
Random hopping magnitudes and signs change the
contributions of different paths to boundary
propagation. In the boundary-Yukawa channel,
the resulting relative flavour weights combine
with the fixed $Y_{yuk}^{\alpha,\beta}$.
For this benchmark, $\theta_{12}$ and $\theta_{23}$
are broadly distributed, while $\theta_{13}$
is enhanced near small angles.
The distributions therefore differ from the
stronger endpoint concentration seen in the
local and Petersen benchmarks.
The Hamiltonian channel also produces broad
distributions, with flavour mixing occurring
within the heavy sector itself.
Neither channel retains the narrow concentration
observed in the corresponding weak-site benchmarks.

\subsection{Majorana}
\paragraph{\textbf{Local}}
In the local chain, boundary propagation depends
on successive nearest-neighbour hoppings.
Their random magnitudes and signs can generate
strongly unequal propagation amplitudes across
the heavy flavour sectors. These amplitudes
weight the Majorana flavour structure on both
sides of the effective light mass matrix,
which can favour flavour alignment.
For the chosen benchmark, the Majorana-coupling
channel produces strong endpoint enhancement
with non-zero intermediate-angle populations,
rather than absent mixing.
\begin{figure}[ht]
    \centering
    \begin{subfigure}{0.34\textwidth}
        \centering
        \includegraphics[width=1\textwidth]{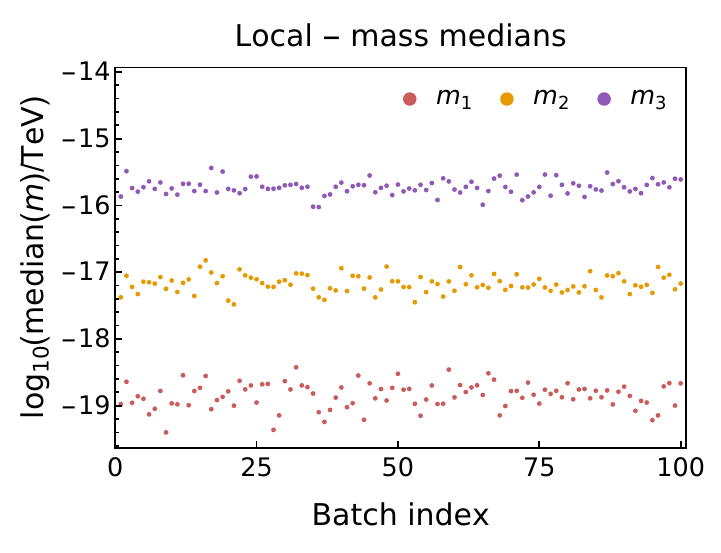}
        \caption{}
        \label{subfig:local-hopp-mass-maj}
    \end{subfigure}
    \hfill
    \begin{subfigure}{0.32\textwidth}
        \centering
        \includegraphics[width=1\textwidth]{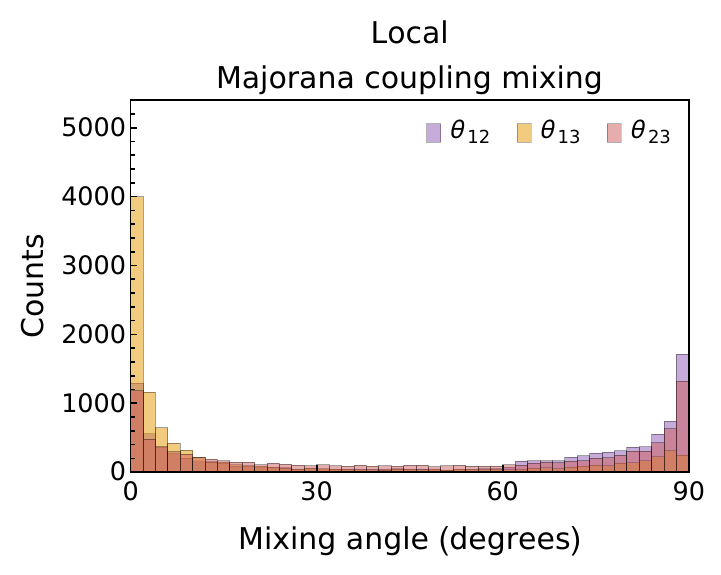}
        \caption{}
        \label{subfig:local-hopp-yukangle-maj}
    \end{subfigure}
    \hfill
    \begin{subfigure}{0.32\textwidth}
        \centering
        \includegraphics[width=1\textwidth]{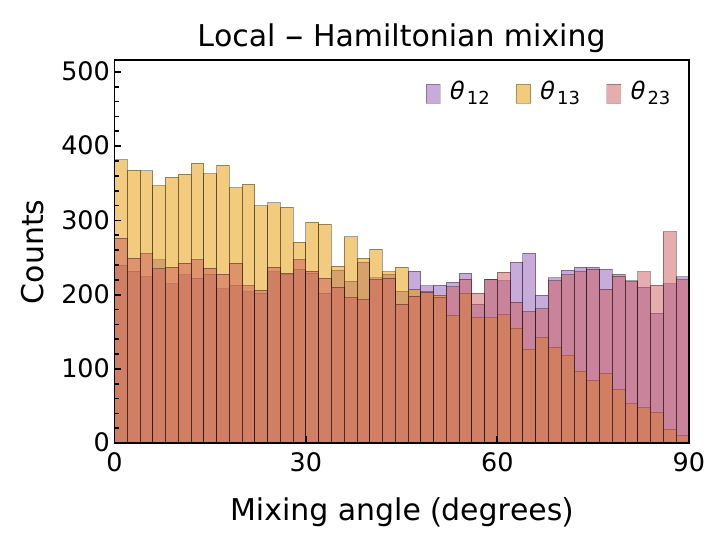}
        \caption{}
        \label{subfig:local-hopp-hamangle-maj}
    \end{subfigure}
    
    \caption{Median of the three lightest masses (in TeV, $\log_{10}$ scale) over 100 runs (left) and histograms of PMNS mixing angles for Majorana coupling $W^{\alpha,\beta}$ (middle) and Hamiltonian mixing $Y^{\alpha,\beta}_{Ham}$ (right). The flavour matrices are given in
Eq.~\eqref{eq:local-weak-site-majorana-flavour}.
Local geometry, weak hopping disorder:
$\epsilon_i=10\,\mathrm{TeV}$ and $N=5$.
Nearest-neighbour hopping couplings are sampled uniformly
from $[-t,t]$, with $t\simeq0.69\,\mathrm{TeV}$ for
Majorana-coupling mixing and $t=0.70\,\mathrm{TeV}$
for Hamiltonian mixing.} \label{local_maj_weak_hopp}
\end{figure}

Both local Majorana benchmarks use $N=5$,
uniform site masses $\epsilon_i=10\,\mathrm{TeV}$,
and a Majorana mass prefactor of $10\,\mathrm{TeV}$.
The nearest-neighbour hopping couplings are sampled
uniformly from $[-t,t]$.
The hopping half-width is $t\simeq0.69\,\mathrm{TeV}$
for Majorana-coupling mixing and $t=0.70\,\mathrm{TeV}$
for Hamiltonian mixing.
The coupling to $\Psi$ is fixed and flavour diagonal,
$t^{\alpha,\beta}=t\,\delta_{\alpha\beta}$.
The flavour matrices are unchanged from the local
weak-site Majorana benchmarks in
Eq.~\eqref{eq:local-weak-site-majorana-flavour}.
The parameters are listed in
Table~\ref{table-maj-hopp-weak-param}. The Hamiltonian channel allows flavours to mix
within the heavy tower and produces broad
mixing-angle distributions for this benchmark. Fig.~\ref{local_maj_weak_hopp} shows the mixing angles and neutrino masses generated.

\paragraph{\textbf{Non-local}}
The non-local links provide additional paths
between the boundaries. Random hopping magnitudes
and signs change their contributions to the
effective mass matrix, so the suppression cannot
be inferred from the number of random entries alone.
In the Majorana-coupling channel, the relative
propagation amplitudes weight the flavour structure
supplied by $W^{\alpha,\beta}$, producing
endpoint-enhanced distributions with broad tails.
The Hamiltonian channel also permits flavour
mixing within the heavy tower and produces
broad distributions for the chosen benchmark.
\begin{figure}[ht]
    \centering
    \begin{subfigure}{0.34\textwidth}
        \centering
        \includegraphics[width=1\textwidth]{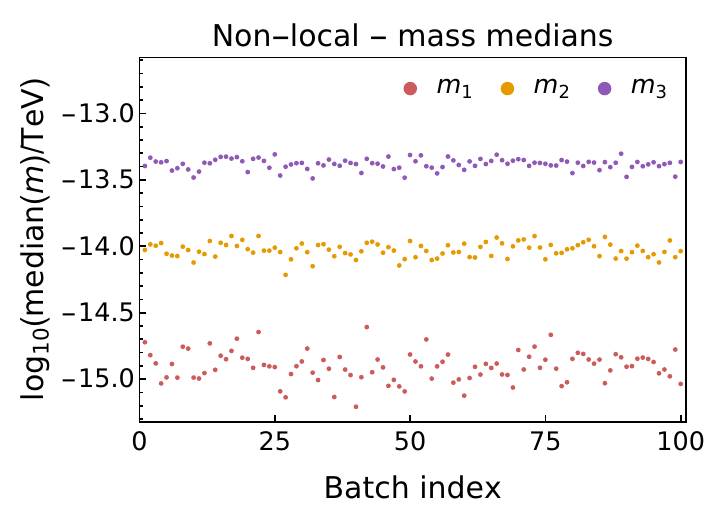}
        \caption{}
        \label{subfig:nl-hopp-mass-maj}
    \end{subfigure}
    \hfill
    \begin{subfigure}{0.32\textwidth}
        \centering
        \includegraphics[width=1\textwidth]{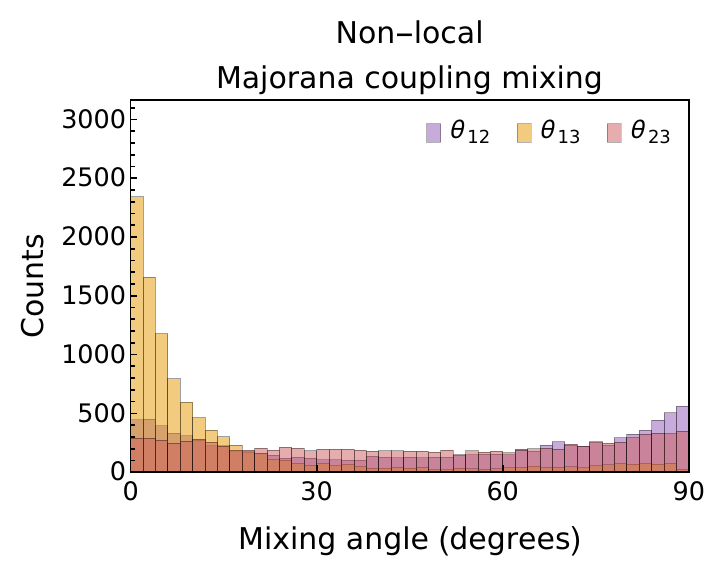}
        \caption{}
        \label{subfig:nl-hopp-yukangle-maj}
    \end{subfigure}
    \hfill
    \begin{subfigure}{0.32\textwidth}
        \centering
        \includegraphics[width=1\textwidth]{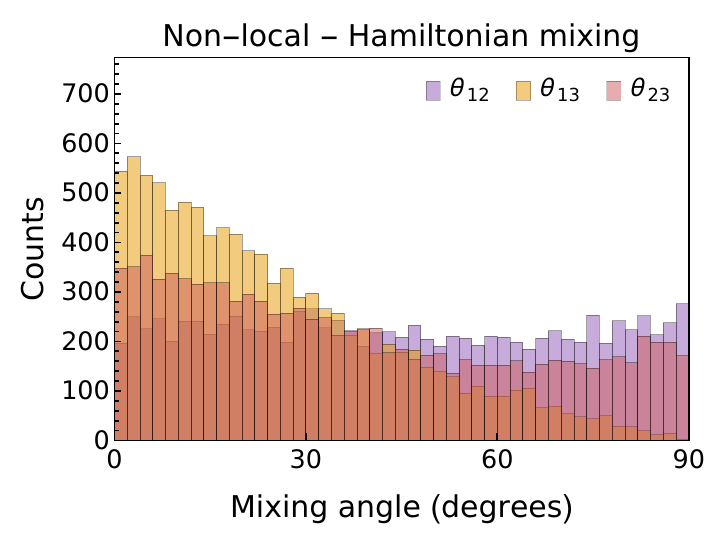}
        \caption{}
        \label{subfig:nl-hopp-hamangle-maj}
    \end{subfigure}
    
    \caption{Median of the three lightest masses (in TeV, $\log_{10}$ scale) over 100 runs (left) and histograms of PMNS mixing angles for Majorana coupling $W^{\alpha,\beta}$ (middle) and Hamiltonian mixing $Y^{\alpha,\beta}_{Ham}$ (right). The flavour matrices are given in
Eq.~\eqref{eq:nonlocal-weak-site-majorana-flavour}.
Non-local geometry, weak hopping disorder:
$\epsilon_i=10\,\mathrm{TeV}$ and $b=1.5$.
Raw hopping couplings are sampled uniformly from
$[-t,t]$ before geometric weighting.
The Majorana-coupling benchmark uses $N=16$ and
$t\simeq0.39\,\mathrm{TeV}$.
The Hamiltonian benchmark uses $N=8$ and
$t\simeq0.057\,\mathrm{TeV}$.} \label{nonlocal_maj_weak_hopp}
\end{figure}

Both non-local Majorana benchmarks use $b=1.5$,
uniform site masses $\epsilon_i=10\,\mathrm{TeV}$,
and a Majorana mass prefactor of $10\,\mathrm{TeV}$.
The raw hopping couplings are sampled uniformly from
$[-t,t]$ before multiplication by $b^{-|i-j|}$.
The Majorana-coupling benchmark uses $N=16$ and
hopping half-width $t\simeq0.39\,\mathrm{TeV}$.
The Hamiltonian benchmark uses $N=8$ and
$t\simeq0.057\,\mathrm{TeV}$.
The coupling to $\Psi$ is fixed and flavour diagonal,
$t^{\alpha,\beta}=t\,\delta_{\alpha\beta}$.
The flavour matrices are unchanged from the non-local
weak-site Majorana benchmarks in
Eq.~\eqref{eq:nonlocal-weak-site-majorana-flavour}.
The parameters are listed in
Table~\ref{table-maj-hopp-weak-param}. Fig.~\ref{nonlocal_maj_weak_hopp} shows the numerical results for neutrino masses and mixing angles generated. Compared with the local and Petersen benchmarks,
the Majorana-coupling distributions have more
intermediate-angle weight, while the Hamiltonian
distributions remain qualitatively similar.
These comparisons describe the selected benchmarks
and do not establish geometry-independent mixing.

\section{Numerical Sampling and Benchmark Ensembles}
\label{sec:rng}

This appendix describes the sampling and numerical
procedure for the three-flavour neutrino benchmarks.
Each benchmark starts from $10^4$ random realisations,
arranged in 100 batches of 100.
The flavour matrices, boundary couplings, site counts
and geometric parameters are fixed within each ensemble.
The displayed fitted parameters are rounded to two
significant figures, while the calculations use
the unrounded inputs.

For strong site disorder, the Dirac benchmarks sample
$\epsilon_i$ uniformly from $[-10,10]\,\mathrm{TeV}$,
whereas the Majorana benchmarks use
$[-5,5]\,\mathrm{TeV}$.
The hopping prefactors are fixed at
$0.1\,\mathrm{TeV}$ and $1/3\,\mathrm{TeV}$,
respectively.
For weak site disorder, all benchmarks sample
$\epsilon_i$ uniformly from
$[9.8,10.2]\,\mathrm{TeV}$.
The site-disorder half-width is therefore
$0.2\,\mathrm{TeV}$ independently of the
benchmark-specific fixed hopping.
For weak hopping disorder, the site masses are fixed
at $\epsilon_i=10\,\mathrm{TeV}$ and the raw hopping
couplings are sampled uniformly from $[-t,t]$,
with the half-width $t$ specified for each benchmark.
The local geometry retains only nearest-neighbour
links. The non-local and Petersen geometries apply
the stated distance weights to their allowed links.

The calculations use Mathematica with the
Mersenne Twister pseudorandom generator.
Random inputs are generated as exact rational numbers
on a grid with spacing $10^{-24}$ before rescaling
to the specified intervals.
Within each symmetric tower block, each undirected
edge is sampled once and assigned to both matrix entries.
Successive realisations use seeds $s_0+r-1$,
where $r=1,\ldots,10^4$.
The initial seeds are $s_0=20261009$ for strong site
disorder, $s_0=950001$ for weak site disorder, and
$s_0=970001$ for weak hopping disorder.
The full mass matrices are evaluated with
80-digit working precision.
Dirac masses and left-handed mixing vectors are
obtained by singular-value decomposition.
For the real symmetric Majorana matrices, an
eigendecomposition is used, with phases assigned
to negative-eigenvalue modes so that the physical
masses are nonnegative.
The three lightest physical masses are ordered
as $m_1\leq m_2\leq m_3$.

The plotted ensembles retain numerically resolved
samples with positive light masses.
Numerical resolution requires relative residuals
below $10^{-8}$ and at least ten digits of reported
precision in the light masses.
The active-row selection requires
\[
0.99\leq (UU^\dagger)_{\alpha\alpha}
\leq 1+10^{-30},
\]
where $U$ is the active $3\times3$ block of the
three light eigenvectors.
The upper tolerance accommodates numerical rounding.
Angle histograms additionally exclude samples
with undefined angles or light-mass relative
separations at or below $10^{-12}$.
Mass histograms use retained samples, and mass
medians are calculated within each original batch
after selection.
Empty batches are omitted.

All inputs, seeds and numerical settings will be
released with the code.
These ensembles describe the response to disorder
at fixed flavour benchmarks.
The weak-hopping cases reuse the flavour matrices
and geometric parameters of their corresponding
weak-site cases.

\section*{References}
\bibliographystyle{unsrt}
\bibliography{bib}

\end{document}